\documentclass[11pt,a4paper]{article}
\pdfoutput=1
\usepackage{jstyle}
\usepackage{amsmath, amsfonts, amssymb, siunitx}
\usepackage{graphicx, hyperref, color, verbatim}
\usepackage[dvipsnames]{xcolor}
\usepackage{float}
\usepackage{caption}
\usepackage{subcaption}

\title{Aspects of higher-point functions in BCFT$_d$}

\author[]{Junding Chen${}^{a,b}$,}
\author[]{Xinan Zhou${}^c$}
\affiliation[a]{CAS Key Laboratory of Theoretical Physics, Institute of Theoretical Physics, Chinese Academy of Sciences, Beijing 100190, China}
\affiliation[b]{School of Physical Sciences, University of Chinese Academy of Sciences, No.19A Yuquan Road, Beijing 100049, China}
\affiliation[c]{Kavli Institute for Theoretical Sciences,
University of Chinese Academy of Sciences, Beijing 100190, China. }

\abstract{We study three-point correlation functions of scalar operators in conformal field theories with boundaries and interfaces. We focus on two cases where there are one bulk and two boundary operators (B$\partial\partial$), or two bulk and one boundary operators (BB$\partial$). We perform a detailed analysis of the conformal blocks in different OPE channels. In particular, we obtain the bulk channel conformal blocks of the BB$\partial$ three-point functions for arbitrary exchanged spins in a series expansion with respect to the radial coordinates. We also study examples of such three-point functions in the simplest holographic dual where the $AdS_{d+1}$ space contains a brane filling an $AdS_d$ subspace. Such a setup arises in top-down models with probe branes and is also relevant for the functional approach to boundary and interface CFT correlators. We systematically study the Witten diagrams in this setup both in position space and in Mellin space. We also discuss in detail how to decompose these Witten diagrams into conformal blocks. }

\emailAdd{chenjunding@itp.ac.cn,
xinan.zhou@ucas.ac.cn}

\begin{document}
\maketitle
\tableofcontents

\newpage

\section{Introduction}
Boundary conformal field theories (BCFTs) have a wide range of applications in theoretical physics. Examples include surface phenomena in statistical mechanics \cite{Cardy:1984bb}, string worldsheet description of D-branes, formal quantum field theory ({\it e.g.}, \cite{Gaiotto:2008sa}) and holography \cite{Karch:2000gx,Aharony:2003qf}. In recent years, BCFTs have also emerged as a useful playground for developing analytic techniques for the conformal bootstrap program (see \cite{Rychkov:2016iqz,Simmons-Duffin:2016gjk,Poland:2018epd,Bissi:2022mrs} for reviews).\footnote{It should also be noted that in two dimensional rational CFTs, the study of boundary conditions also leads to analytic insights into the full theory \cite{Cardy:1989ir,Cardy:1991tv}.} For example, the analytic approach developed in \cite{Kaviraj:2018tfd,Mazac:2018biw} for BCFT two-point functions provided an important stepping stone for extending the functional method for boundary-free CFTs from the special case of 1D \cite{Mazac:2016qev,Mazac:2018mdx,Mazac:2018ycv} to higher spacetime dimensions \cite{Paulos:2019gtx,Mazac:2019shk}. The advantage for considering BCFTs is largely due to the fact that the partially broken conformal symmetry, as the result of introducing the boundary, allows correlators to develop nontrivial spacetime dependence at lower points. The simplest nontrivial correlators in CFTs without boundaries are four-point functions which have two cross ratios when the spacetime dimension $d\geq 2$. However, in BCFTs the first interesting case is two-point functions of bulk operators which have just one cross ratio. The latter has much simpler kinematics while still is capable of teaching us important lessons about the former. On the other hand, we  should note that currently most analytic studies of BCFT correlators are at this level of two-point functions ({\it e.g.}, \cite{McAvity:1995zd,Liendo:2012hy,Gliozzi:2015qsa,Liendo:2016ymz,Hogervorst:2017kbj,Bissi:2018mcq,Kaviraj:2018tfd,Mazac:2018biw,Dey:2020lwp,Giombi:2020rmc,Dey:2020jlc,Giombi:2021cnr}). However, it is also important to go to higher points. This is particularly relevant in light of the recent activities in extending the power of the analytic conformal bootstrap to correlation functions beyond four points in CFTs without boundaries.\footnote{See \cite{Bercini:2020msp,Bercini:2021jti,Antunes:2021kmm,Kaviraj:2022wbw} for recent progress of extending the lightcone bootstrap to higher points and \cite{Goncalves:2019znr,Alday:2022lkk,Goncalves:2023oyx} for results of computing five-point holographic correlators using bootstrap methods. } It should be again beneficial to turn to the BCFT playground and first study the higher-point correlators there. 

In this paper, we study three-point functions of scalar operators in BCFTs.\footnote{Note that the kinematics of BCFTs is the same as CFTs with a co-dimension 1 defect or interface. Therefore, the discussions in this paper also apply to such defect and interface CFTs.} Depending on whether the operators are inserted in the bulk (B) or on the boundary ($\partial$), there are four possibilities: BBB, BB$\partial$, B$\partial\partial$ and $\partial\partial\partial$. The last case is kinematically trivial because it is the same as in CFTs without boundaries and are fixed by conformal symmetry up to a constant. The other three nontrivial cases have 3, 2, and 1 cross ratios respectively. In this work, we will focus on the simpler BB$\partial$ and B$\partial\partial$ three-point functions, leaving the more complicated BBB case for the future. Compared to the two-point function case, these three-point functions have much richer kinematics and encode more data. In addition to the boundary OPE coefficients already contained in two-point functions, the B$\partial\partial$ three-point functions also give three-point function coefficients of boundary operators. These can be extracted from the OPE limit where the bulk operator is taken to be near the boundary. For the BB$\partial$ three-point functions, there are two OPE channels. The first is the bulk channel where we first perform a bulk OPE for the two bulk operators to reduce the three-point function to a sum of B$\partial$ two-point functions which are fixed by symmetry. The novel feature compared to the bulk channel of the BB two-point function is that operators with spins can now appear in the exchange. The other channel is the boundary channel where one applies the boundary OPE independently to both bulk operators. This OPE process has two operator exchanges and resembles the OPE of a five-point function in CFTs without boundaries, except that in the BCFT case the exchange operators are scalars. Therefore, we see that the investigations in BCFTs are again relevant for the study of higher-point correlation functions in CFTs without boundaries, and the former provides a simplified setup for the latter. 

Ultimately, we expect that the investigations will lead to a functional method for BCFT three-point functions which is similar to that of the two-point functions \cite{Kaviraj:2018tfd,Mazac:2018biw}. Using the functionals one can systematically extract all the constraints from the crossing equation, which form the starting point of various analytic analyses. However, to achieve this goal there are still a lot of prerequisite results which should be obtained first. One necessary ingredient is the conformal blocks. Although results for conformal blocks have been obtained for the B$\partial\partial$ three-point functions \cite{Lauria:2020emq,Buric:2020zea} and the BB$\partial$ three-point functions in the boundary channel \cite{Lauria:2020emq,Behan:2020nsf,Buric:2020zea}, the BB$\partial$ bulk channel conformal blocks are largely unknown, except for the case where the exchanged operators are scalars (and also with special exchanged dimensions) \cite{Behan:2020nsf}. In this paper, we revisit the first two cases. In particular, we will extend the method of \cite{Ferrara:1972xe,Ferrara:1972ay,Ferrara:1972uq,Ferrara:1972kab,Dolan:2000ut,Dolan:2011dv,Simmons-Duffin:2012juh} based on the shadow formalism to obtain BB$\partial$ boundary channel conformal blocks from conformal partial waves. Moreover, we will also obtain the bulk channel conformal blocks of BB$\partial$ three-point functions for arbitrary spins in a series expansion, with the series coefficients given by a recursion relation. This is achieved with the help of a set of new coordinates which are close analogues of the radial coordinates  in \cite{Hogervorst:2013sma,Lauria:2017wav}. On the other hand, an important lesson from the functional method of BCFT two-point functions\footnote{Also from that of four-point functions in CFTs without boundaries \cite{Mazac:2018ycv,Mazac:2019shk} and two-point functions in CFTs on real projective space \cite{Giombi:2020xah}.} \cite{Kaviraj:2018tfd,Mazac:2018biw} is that the analytic functionals are closely related to tree-level exchange Witten diagrams in AdS space. Essentially, the latter serves as a generating function for the former where the action of the  functionals can be read off from the conformal block decomposition coefficients of the Witten diagrams.\footnote{Two other subjects closely related to the functional method are the Polyakov-Mellin bootstrap \cite{Gopakumar:2016wkt,Gopakumar:2016cpb,Penedones:2019tng,Gopakumar:2021dvg} and the CFT dispersion relation \cite{Caron-Huot:2017vep,Carmi:2019cub}. In fact, all three approaches are related. See \cite{Caron-Huot:2020adz} for an explain of their connections.} We will not establish the functional method in this paper. Nevertheless, we expect that a similar connection persists for three-point functions as well. Motivated by this, in this paper we also perform a detailed study of the  relevant higher-point Witten diagrams in the holographic BCFT model which gives rise to the analytic functionals in the two-point function case \cite{Kaviraj:2018tfd,Mazac:2018biw}. This model is a probe brane setup where the full $AdS_{d+1}$ space contains an $AdS_d$ brane with no backreaction. The $AdS_d$ subspace can be viewed as the holographic dual of the boundary and carries localized degrees of freedom.  There are also local vertices on the brane through which the $AdS_d$ fields  can interact with the bulk fields. We will study three-point Witten diagrams in this setup both in position space and in Mellin space \cite{Rastelli:2017ecj}. Moreover, using our results of conformal blocks we will also compute the conformal block decomposition coefficients of these Witten diagrams.  

The rest of the paper is organized as follows. In Section \ref{Kinematics of BCFT three-point functions} we review the embedding space and use this formalism to discuss the kinematics of BCFT correlators. We study the conformal blocks in Section \ref{BCFT conformal blocks}. In Section \ref{Bpartialpartial case}  we compute the B$\partial\partial$ conformal block, reproducing the result of \cite{Lauria:2020emq}. Then we compute the BB$\partial$ conformal blocks in the boundary channel in Section \ref{Boundary channel CB} and in the bulk channel in Section \ref{Bulk channel CB}. We start studying Witten diagrams  in Section \ref{Witten diagrams Bpp case}. In Section \ref{Witten diagrams Bpp case} we study Witten diagrams for the B$\partial\partial$ case and in Section \ref{Witten diagrams: The BBp case} we study Witten diagrams for the BB$\partial$ case. These two sections are in position space and we also study the conformal block decomposition of Witten diagrams. In Section \ref{Mellin space representation} we perform a complementary study of these diagrams in Mellin space where we obtain simple expressions for their Mellin amplitudes. We conclude in Section \ref{outlook} with a brief discussion of future directions. 

\section{Kinematics of BCFT correlators}
\label{Kinematics of BCFT three-point functions}
In this section we set the stage for studying BCFT correlators. We start with a brief review of the embedding space formalism and the kinematics of one- and two-point functions in Section \ref{Embedding space and lower-point functions} . Then in Section \ref{Bulk-boundary-boundary} and \ref{Bulk-bulk-boundary} we discuss the kinematics of the ${\rm B}\partial\partial$ and ${\rm BB}\partial$ three-point functions. Most of the discussions are review material but the introduction of the radial coordinates for ${\rm BB}\partial$ three-point functions is new.

\subsection{Embedding space and lower-point functions}
\label{Embedding space and lower-point functions}
In this paper, we will simplify the kinematical discussions by using the embedding space \cite{Dirac:1936fq}, which linearizes the action of the conformal symmetry. As in the case without boundaries, we represent each point $x^{\mu} \in \mathbb{R}^{d}$ by a null ray $P^A$, $A=1,2,\ldots,d+2$, in the embedding space $\mathbb{R}^{d+1,1}$, satisfying\footnote{We choose the signature to be $(-,+,+,+, \ldots,+)$.}
\begin{equation}
	\label{coordP}
	 P \cdot P=0\;, \quad P \sim \lambda P \;.
\end{equation}
These conditions reduce the number of independent components of $P^A$ from $d+2$ to $d$, which is necessary to correctly account for the $d$ degrees of freedom of $x^{\mu} \in \mathbb{R}^{d}$. Explicitly, we can use the rescaling freedom to parameterize the embedding vector as 
\begin{equation}
	\label{Poincaresection}
	P^{A}=\bigg(\frac{1+x^{2}}{2}, \frac{1-x^{2}}{2}, x^{\mu}\bigg)\;,\quad x^{\mu} \in \mathbb{R}^{d}\;.
\end{equation}
In embedding space, the conformal group is just the Lorentz group $SO(d+1,1)$ which linearly acts on the embedding vector $P^A$. However, the presence of a planar boundary partially breaks the conformal group to $SO(d,1)$. This is achieved in this formalism by introducing a fixed vector in embedding space
\begin{equation}
N_{b}=(0,0,0, \ldots, 1)\;,
\end{equation}
which is invariant under the $SO(d,1)$ rotations in the first $d+1$ directions. The conformal boundary is determined by the condition 
\begin{equation}
 P \cdot N_b=0\;,   
\end{equation}
which corresponds to $x^d=0$ in the parameterization (\ref{Poincaresection}).  Follow the convention in the literature, we will denote $x^{d}$ as $x_{\perp}$ from now on. 

 Let us now consider operators and their correlation functions in the presence of the boundary.\footnote{In this paper we mostly consider external scalar operators, although operators exchanged in the OPE will have spins in some cases.} Operators are defined on the null rays with the following homogeneity condition 
\begin{equation}
	\label{homon}
	\mathcal{O}_{\Delta}(\lambda P)=\lambda^{-\Delta} \mathcal{O}_{\Delta}(P)\;.
\end{equation}
In ordinary CFTs one-point functions do not exist  because of the lack of conformal invariants constructed with just $P^A$. In BCFTs, however, one can  write down $P\cdot N_b$ which is invariant under the residual conformal group. Together with the rescaling property (\ref{homon}), this fixes the one-point function up to an overall factor
\begin{equation}
	\label{1ptfunction}
	\langle \mathcal{O}_{\Delta}(P)\rangle=\frac{a_{\mathcal{O}}}{(P \cdot N_b)^{\Delta}}=\frac{a_{\mathcal{O}}}{x_{\perp}^{\Delta}}\;,
\end{equation}
where the one-point function coefficient $a_\mathcal{O}$ is new CFT data. In addition to the bulk operators described above, there are also boundary operators $\widehat{\mathcal{O}}_{\widehat{\Delta}}(\widehat{P})$ which live on the boundary, {\it i.e.}, $\widehat{P}\cdot N_b =0$. We have put hats on the labels to distinguish them from the bulk operators. Clearly, their one-point functions vanish. 

For two-point functions, we need to consider several possibilities. The simplest one is the boundary-boundary ($\partial \partial$) two-point function
\begin{equation}
	\langle \widehat{\mathcal{O}}_{\widehat{\Delta}}(\widehat{P}_1) \widehat{\mathcal{O}}_{\widehat{\Delta}}(\widehat{P}_2)\rangle=\frac{1}{(-2 \widehat{P}_1\cdot \widehat{P}_2)^{\widehat{\Delta}}}=\frac{1}{(\widehat{x}_{12}^2)^{\widehat{\Delta}}}\;,
\end{equation}
which is just the familiar two-point function in CFT$_{d-1}$. Obviously, this is the only invariant one can write down which is compatible with the rescalings.\footnote{We have also chosen a normalization for the boundary operators such that the two-point function has coefficient $1$.} 

The next simplest case is the bulk-boundary (B$\partial$) two-point function. It is easy to see that the correlator is also determined by conformal symmetry and rescaling
\begin{equation}
	\label{Bptwopoint}
	\langle \mathcal{O}_{\Delta}(P_1) \widehat{\mathcal{O}}_{\widehat{\Delta}}(\widehat{P}_2)\rangle=\frac{b_{\mathcal{O}\mathcal{\widehat{O}}}}{( P_1 \cdot N_b)^{\Delta-\widehat{\Delta}} (-2 P_1 \cdot \widehat{P}_2)^{\widehat{\Delta}}}=\frac{b_{\mathcal{O}\mathcal{\widehat{O}}}}{x_{1\perp}^{\Delta-\widehat{\Delta}}(\widehat{x}_{12}^2+x_{1\perp}^2)^{\widehat{\Delta}}}\;,
\end{equation}
where $\widehat{x}_{12}^2$ is the squared distance between $x_1$ and $x_2$ projected along the boundary and $x_{12}^2=\widehat{x}_{12}^2+x_{1\perp}^2$. The coefficient $b_{\mathcal{O}\mathcal{\widehat{O}}}$ is also a new piece of data for the BCFT. Later we will also encounter ${\rm B}\partial$ two-point functions where the bulk operator transforms as a symmetric traceless tensor. The spacetime dependence of such two-point functions are uniquely fixed as well. Written in terms of the $\mathbb{R}^d$ coordinates, the correlator reads
\begin{equation}\label{Bp2ptspinJ}
	\langle  \mathcal{O}_{\Delta}^{\mu_{1} \ldots \mu_{J}}(x_{1})\widehat{\mathcal{O}}_{\widehat{\Delta}}(\widehat{x}_2)\rangle=b_{\mathcal{O}\mathcal{\widehat{O}}}\frac{X^{\mu_1}X^{\mu_2}\cdots X^{\mu_J}-\text{traces}}{x_{1\perp}^{\Delta-\widehat{\Delta}}(\widehat{x}_{12}^2+x_{1\perp}^2)^{\widehat{\Delta}}}\;,
\end{equation}
where
\begin{equation}
	X^{\mu}=2x_{1\perp}\frac{x_1^{\mu}-\widehat{x}_2^{\mu}}{x_{12}^2}-B^{\mu}\;,\quad\quad B^{\mu}=(0,0,\ldots,0,1)\in \mathbb{R}^d\;,
\end{equation}
and $X^{\mu}$ satisfies $X^{\mu}X_{\mu}=1$.  

The first nontrivial case is the bulk-bulk (BB) two-point function which is not determined by conformal symmetry. This is because one can construct the following conformal cross ratio 
\begin{equation}
\label{2ptcross ratio}
\zeta= \frac{-2P_1 \cdot P_2}{(2P_1 \cdot N_b)(2P_2 \cdot N_b)}=\frac{x_{12}^2}{4x_{1\perp}x_{2\perp}}\;,
\end{equation}
which is invariant under both the residual conformal group and the rescaling of the embedding vectors. Therefore, conformal symmetry only reduces the two-point function to an arbitrary function of the cross ratio
\begin{equation}
	\label{2ptfunction}
	\langle\mathcal{O}_{\Delta_1}(P_1) \mathcal{O}_{\Delta_2} (P_2) \rangle=\frac{1}{(2P_1 \cdot N_b)^{\Delta_1}(2 P_2 \cdot N_b)^{\Delta_2}} \mathcal{G}(\zeta)\;.
\end{equation}
This function has interesting kinematical limits which are related to the operator product expansion (OPE) which can be further divided into two types. The first limit is $\zeta\to 0$, which corresponds to the standard OPE in CFTs without boundaries. In this case, we take both  bulk operators to be close to each other 
\begin{equation}
	\label{bulkOPE}
	\mathcal{O}_{\Delta_1}(x) \mathcal{O}_{\Delta_2}(0)=\sum_{\mathcal{O}} f_{\mathcal{O}\mathcal{O}_1\mathcal{O}_2}\Bigg( \frac{x_{\mu_{1}} \ldots x_{ \mu_{J}}}{\left(x^{2}\right)^{\frac{\Delta_{1}+\Delta_{2}-\Delta+J}{2}}}  \mathcal{O}_{\Delta}^{\mu_{1} \ldots \mu_{J}}(0)+(\text{descendants}) \Bigg)\;,
\end{equation}
where in the expansion the operator $\mathcal{O}_{\Delta}^{\mu_{1} \ldots \mu_{J}}(0)$ has dimension $\Delta$ and spin $J$, and $f_{\mathcal{O}\mathcal{O}_1\mathcal{O}_2}$ is the three-point function coefficient. This is usually referred to as the {\it bulk channel} OPE. Taking the OPE reduces the BB two-point function to a sum of one-point functions. Furthermore, generalizing the argument leading to (\ref{1ptfunction}) one can show that only the $J=0$ scalar operators have nonvanishing one-point functions. Therefore, the two-point function can be written as the expansion
\begin{equation}
    \mathcal{G}(\zeta)=\sum_{\mathcal{O}}\mu_{\mathcal{O}} g^{{\rm 2pt,B}}_\Delta(\zeta)\;,
\end{equation}
where $\mu_{\mathcal{O}}=f_{\mathcal{O}\mathcal{O}_1\mathcal{O}_2}a_{\mathcal{O}}$ and $g^{{\rm 2pt,B}}_\Delta(\zeta)$ is the bulk channel conformal block \cite{McAvity:1995zd}\footnote{The $2^{\Delta}$ factor in  front of $g^{{\rm 2pt,B}}_\Delta$, which is absent in \cite{McAvity:1995zd}, accounts for the different normalization of the one-point function (\ref{1ptfunction}). Similarly, the normalization of the  B$\partial$ two-point function (\ref{Bptwopoint}) introduces the $2^{\Delta_1+\Delta_2-2\widehat{\Delta}}$ factor in the boundary conformal block (\ref{twopointblockboundry}).}
\begin{equation}\label{bulkCB2pt}
    g^{{\rm 2pt,B}}_\Delta(\zeta)=2^{\Delta}\zeta^{\frac{\Delta-\Delta_1-\Delta_2}{2}}{}_2F_1\bigg(\frac{\Delta+\Delta_1-\Delta_2}{2},\frac{\Delta+\Delta_2-\Delta_1}{2};\Delta-\frac{d}{2}+1;-\zeta\bigg)\;.
\end{equation}
The conformal block captures the contribution of exchanging a primary operator and its descendants in the bulk channel and can be obtained by solving the conformal Casimir equation \cite{Liendo:2012hy}.
The other interesting limit is $\zeta\to\infty$ which corresponds to a bulk operator being taken to the boundary. This gives rise to the {\it boundary channel} OPE
\begin{equation}
	\label{bdyOPE}
	\mathcal{O}_{\Delta}(x)=\sum_{\widehat{\mathcal{O}}}\Bigg( b_{\mathcal{O}\widehat{\mathcal{O}}}\frac{ \widehat{\mathcal{O}}_{\widehat{\Delta}}(\widehat{x})}{x_{\perp}^{\Delta-\widehat{\Delta}}}+ (\text{descendants})\Bigg)\;,
\end{equation}
where a bulk operator is expressed as a sum of boundary operators and $\widehat{x}$ is the coordinates on the boundary. The OPE coefficient $b_{\mathcal{O}\widehat{\mathcal{O}}}$ is the same as the one that appears in the B$\partial$ two-point function (\ref{Bptwopoint}). The resulting B$\partial$ two-point functions with the other bulk operator are also fixed by conformal symmetry. The boundary channel OPE allows us to decompose the two-point function into boundary channel conformal blocks \cite{McAvity:1995zd}
\begin{equation}
    \mathcal{G}(\zeta)=\sum_{\widehat{\mathcal{O}}}\widehat{\mu}_{\widehat{\mathcal{O}}} g^{{\rm 2pt,\partial}}_{\widehat{\Delta}}(\zeta) \;,
\end{equation}
where $\widehat{\mu}_{\widehat{\mathcal{O}}}=b_{\mathcal{O}_1\widehat{\mathcal{O}}}b_{\mathcal{O}_2\widehat{\mathcal{O}}}$ and 
\begin{equation}
\label{twopointblockboundry}
    g^{{\rm 2pt,\partial}}_{\widehat{\Delta}}(\zeta)=2^{\Delta_1+\Delta_2-2\widehat{\Delta}}\zeta^{-\widehat{\Delta}}{}_2F_1\bigg(\widehat{\Delta},\widehat{\Delta}-\frac{d}{2}+1;2\widehat{\Delta}+2-d;-\frac{1}{\zeta}\bigg)\;.
\end{equation}
is the contribution of exchanging an operator in the boundary channel. The two different ways of decomposing the correlator gives rise to the following crossing equation
\begin{equation}
\sum_{\mathcal{O}}\mu_{\mathcal{O}} g^{{\rm 2pt,B}}_\Delta(\zeta)=\sum_{\widehat{\mathcal{O}}}\widehat{\mu}_{\widehat{\mathcal{O}}} g^{{\rm 2pt,\partial}}_{\widehat{\Delta}}(\zeta)\;, 
\end{equation}
which contains infinitely many constraints on the BCFT data. 

In the following, we  discuss three-point functions in BCFTs which are the main focus of this paper. Many kinematic properties of the BB two-point functions similarly extend to the higher-point case as well. Depending on the kind of operators involved, there are three nontrivial types of three-point functions: bulk-boundary-boundary (B$\partial\partial$), bulk-bulk-boundary (BB$\partial$) and bulk-bulk-bulk (BBB)\footnote{The boundary-boundary-boundary case is trivial because it is the same as the three-point function in CFTs without boundaries and is determined by conformal symmetry.}, which have correspondingly 1, 2, and 3 cross ratios.\footnote{In general, when the spacetime dimension is high enough, the number of cross ratios in a correlator with $n$ bulk operators and $m$ boundary operators is $\frac{n(n-1)}{2}+\frac{m(m-1)}{2}+n m-m$, see \cite{Rastelli:2017ecj} for details of the counting.} We will focus on the first two cases in this paper, and leave the more complicated BBB case for the future. In Section \ref{Bulk-boundary-boundary} and \ref{Bulk-bulk-boundary}, we will discuss the basic kinematics of these three-point functions and point out various OPE channels. The computation of the associated conformal blocks will be the subject of Section \ref{BCFT conformal blocks}.

\subsection{Three-point functions: B$\partial\partial$}
\label{Bulk-boundary-boundary}
Let us first consider the simpler case of the B$\partial\partial$ three-point function which has only one cross ratio 
\begin{equation}
	\label{1}
	\xi=\frac{(-2 \widehat{P}_{2} \cdot \widehat{P}_{3})(P_1 \cdot N_b)^{2}}{(-2 P_1 \cdot \widehat{P}_{2})(-2 P_1 \cdot \widehat{P}_{3})}\equiv\frac{(\widehat{x}_2-\widehat{x}_3)^2 x_{1\perp}^2}{(x_1-\widehat{x}_2)^2(x_1-\widehat{x}_3)^2}\;.
\end{equation}
By extracting a kinematic factor, we can write the correlator as a function of $\xi$
\begin{equation}
		\label{2}
		\langle  \mathcal{O}_{\Delta_{1}}(P_1)\widehat{\mathcal{O}}_{\widehat{\Delta}_{2}}(\widehat{P}_2)\widehat{\mathcal{O}}_{\widehat{\Delta}_{3}}(\widehat{P}_3)\rangle=\frac{(P_1\cdot N_b)^{-\Delta_1+\widehat{\Delta}_2+\widehat{\Delta}_3}}{(-2P_1\cdot \widehat{P}_{2})^{\widehat{\Delta}_2}(-2P_1\cdot \widehat{P}_{3})^{\widehat{\Delta}_{3}}} \mathcal{G}(\xi)\;.
\end{equation}
It turns out that the B$\partial\partial$ three-point function has only {\it one} OPE channel where one sends the bulk operator $\mathcal{O}_{\Delta_{1}}(P_1)$ to the boundary while keeping it away from $\widehat{\mathcal{O}}_{\widehat{\Delta}_{2}}(\widehat{P}_2) $ and $\widehat{\mathcal{O}}_{\widehat{\Delta}_{3}}(\widehat{P}_3)$. This corresponds to the limit $\xi\to 0$. Using the boundary OPE (\ref{bdyOPE}), the three-point function reduces to a sum of products of B$\partial$ two-point functions and $\partial\partial\partial$ three-point functions which are both fixed by conformal symmetry. This gives rise to the following conformal block decomposition  
\begin{equation}\label{BppCBdecom}
\langle  \mathcal{O}_{\Delta_{1}}(P_1)\widehat{\mathcal{O}}_{\widehat{\Delta}_{2}}(\widehat{P}_2)\widehat{\mathcal{O}}_{\widehat{\Delta}_{3}}(\widehat{P}_3)\rangle=\sum_{\widehat{\mathcal{O}}} b_{\mathcal{O}_1\widehat{\mathcal{O}}}\widehat{f}_{\widehat{\mathcal{O}}\widehat{\mathcal{O}}_2\widehat{\mathcal{O}}_3} G_{\widehat{\Delta}}(P_1,\widehat{P}_2,\widehat{P}_3)\;,
\end{equation}
where $\widehat{f}_{\widehat{\mathcal{O}}\widehat{\mathcal{O}}_2\widehat{\mathcal{O}}_3}$ is the three-point function coefficient of the boundary three-point function $\langle \widehat{\mathcal{O}}\widehat{\mathcal{O}}_2\widehat{\mathcal{O}}_3\rangle$. Let us also define the stripped conformal block $g_{\widehat{\Delta}}(\xi)$
\begin{equation}
	\label{strippedblock1}
	G_{\widehat{\Delta}}(P_1,\widehat{P}_2,\widehat{P}_3)= \frac{(P_1\cdot N_b)^{-\Delta_1+\widehat{\Delta}_2+\widehat{\Delta}_3}}{(-2P_1\cdot \widehat{P}_{2})^{\widehat{\Delta}_2}(-2P_1\cdot \widehat{P}_{3})^{\widehat{\Delta}_{3}}} g_{\widehat{\Delta}}(\xi)\;.
\end{equation}
We will give their explicit expressions in the next section. But for the moment, let us just notice that the conformal block has the following asymptotic behavior as $\xi\to 0$
\begin{equation}
	\label{asy1}
g_{\widehat{\Delta}}(\xi)\to \xi^{\frac{\widehat{\Delta}-\widehat{\Delta}_{2}-\widehat{\Delta}_3}{2}}\;, \quad \xi\to 0\;,
\end{equation}
which follows from the leading term in the OPE (\ref{bdyOPE}). One may wonder if one can obtain a different OPE channel by applying the bulk channel OPE (\ref{bulkOPE}) to the two boundary operators. However, this is the same as (\ref{BppCBdecom}) because it also reduces to the same product of a three-point function and a two-point function. 

\subsection{Three-point functions: BB$\partial$}
\label{Bulk-bulk-boundary}
The BB$\partial$ three-point function has more complicated kinematics compared to the B$\partial\partial$ case because they have two cross ratios. A convenient choice is 
\begin{equation}
	\begin{gathered}
	\label{BBpcrossratio}
	\xi_1=\frac{-2P_2\cdot P_3}{(P_2\cdot N_b)(P_3\cdot N_b)}\;, \quad \xi_2=\frac{(-2\widehat{P}_1\cdot P_2)(P_3\cdot N_b)}{ (-2\widehat{P}_1\cdot P_3)(P_2\cdot N_b)}\;,
	\end{gathered}
\end{equation}
which leads to relatively simple expressions in subsequent calculations. Extracting a kinematic factor, the three-point function can be written as
\begin{equation}\label{BBpartialstripped}
	\langle \widehat{\mathcal{O}}_{\widehat{\Delta}_1}(\widehat{P}_1) \mathcal{O}_{\Delta_2}(P_2) \mathcal{O}_{\Delta_3}(P_3)\rangle=\frac{(-2 \widehat{P}_1 \cdot P_3)^{-\widehat{\Delta}_1}}{(P_2\cdot N_b)^{\Delta_2}(P_3\cdot N_b)^{\Delta_3-\widehat{\Delta}_1}} \mathcal{G}^{\text{BB}\partial}(\xi_1,\xi_2)\;.
\end{equation}
Unlike the B$\partial\partial$ case, the BB$\partial$ three-point function has two distinct OPE channels. We will describe these two cases separately in the following. 

\vspace{0.3cm}
\noindent{\bf The bulk channel}
\vspace{0.2cm}

\noindent In the first case, we use the bulk channel OPE (\ref{bulkOPE}) for the two bulk operators. This reduces the three-point function to B$\partial$ two-point functions which are fixed by conformal symmetry. The OPE limit corresponds to taking $P_2\cdot P_3\to 0$, which in terms of the cross ratios is 
\begin{equation}
\label{bulkOPElimit}
	\xi_1\to0\;,\quad \xi_2 \to 1\;.
\end{equation}
We will refer to this channel as the {\it bulk channel}. This OPE leads to the following conformal block decomposition 
\begin{equation}
	\langle \widehat{\mathcal{O}}_{\widehat{\Delta}_1}(\widehat{P}_1) \mathcal{O}_{\Delta_2}(P_2) \mathcal{O}_{\Delta_3}(P_3)\rangle= \sum_{\mathcal{O}} b_{\mathcal{O}\widehat{\mathcal{O}}_1} f_{\mathcal{O}\mathcal{O}_2\mathcal{O}_3} G^{\text{B}}_{\Delta,J}(\widehat{P}_1,P_2,P_3)\;,
\end{equation}
where  $G^{\text{B}}_{\Delta,J}$ is the bulk channel conformal block. It should be emphasized that, unlike the bulk channel of the BB two-point function, operators with spins can be exchanged due to the nonvanishing B$\partial$ two-point function (\ref{Bp2ptspinJ}). For later convenience, we also define  the stripped conformal block $g^{\text{B}}_{\Delta,J}(\xi_1,\xi_2)$ from
\begin{equation}
	G^{\text{B}}_{\Delta,J}(\widehat{P}_1,P_2,P_3)=\frac{(-2 \widehat{P}_1 \cdot P_3)^{-\widehat{\Delta}_1}}{(P_2\cdot N_b)^{\Delta_2}(P_3\cdot N_b)^{\Delta_3-\widehat{\Delta}_1}} g^{\text{B}}_{\Delta,J}(\xi_1,\xi_2)\;,
\end{equation}
to manifest the dependence on the cross ratios. In the bulk channel OPE limit (\ref{bulkOPE}), it has the following asymptotic behavior
\begin{equation}
\label{asybulkblock}
    g^{\text{B}}_{\Delta,J}(\xi_1,\xi_2) \to \xi_1^{\frac{\Delta-\Delta_2-\Delta_3}{2}}\widehat{C}^{\nu}
_J(\omega)\;,
\end{equation}
with $\omega\equiv(\xi_2-1)/\sqrt{\xi_1}$ fixed. Here $\widehat{C}_{\ell}^{\nu}(x)$ is the normalized Gegenbauer polynomial.\footnote{The Gegenbauer polynomial is related to $\widehat{C}_{\ell}^{\nu}(x)$ by  
\begin{equation}
C_{\ell}^{\nu}(x)=
	\frac{2^{\ell} \Gamma(\ell+\nu)}{\Gamma(\ell+1) \Gamma(\nu)}\widehat{C}_{\ell}^{\nu}(x)=\frac{\Gamma(2\nu+\ell)}{\Gamma(\ell+1)\Gamma(2\nu)}{}_2 F_1\bigg(2\nu+\ell,-\ell,\nu+\frac{1}{2},\frac{1-x}{2}\bigg)\;.
 \end{equation}
 }

\vspace{0.3cm}
\noindent{\bf The boundary channel}
\vspace{0.2cm}

\noindent In the second case, we apply the boundary channel OPE (\ref{bdyOPE}) independently to both bulk operators. We can do this in steps. Applying the OPE first to $\mathcal{O}_{\Delta_2}(P_2)$ for example, we obtain a B$\partial\partial$ three-point function. According to the analysis in the previous subsection, the B$\partial\partial$ three-point functions also admit an OPE by applying (\ref{bdyOPE}) again to  $\mathcal{O}_{\Delta_3}(P_3)$. Such a double OPE limit with $P_2\cdot N_b\to0$, $P_3\cdot N_b\to0$ corresponds to taking $\xi_1\xi_2$ and $\xi_1/\xi_2$ as the independent variables and sending them to infinity. Let us denote this combination as $\zeta_1$ and $\zeta_2$, and this limit reads
\begin{equation}
\label{newcrossratiosbdy}
	\zeta_1=\sqrt{\xi_1 \xi_2} \to \infty\;,\quad \zeta_2=\sqrt{\xi_1/\xi_2} \to \infty\;.
\end{equation}
We will refer to this OPE channel as the {\it boundary channel}. It leads to the following conformal block decomposition
\begin{equation}
	\langle \widehat{\mathcal{O}}_{\widehat{\Delta}_1}(\widehat{P}_1) \mathcal{O}_{\Delta_2}(P_2) \mathcal{O}_{\Delta_3}(P_3)\rangle= \sum_{\widehat{\mathcal{O}}}\sum_{\widehat{\mathcal{O}}^{\prime}} b_{\mathcal{O}_{2} \widehat{\mathcal{O}}} b_{\mathcal{O}_{3} \widehat{\mathcal{O}}^{\prime}} \widehat{f}_{\widehat{\mathcal{O}}_1\widehat{\mathcal{O}}\widehat{\mathcal{O}}^{\prime}} G^{\partial}_{\widehat{\Delta},\widehat{\Delta}^{\prime}}(\widehat{P}_1,P_2,P_3)\;,
\end{equation}
where $G^{\partial}_{\widehat{\Delta},\widehat{\Delta}^{\prime}}$ is the boundary channel conformal block and  $\widehat{\mathcal{O}}_{\widehat{\Delta}}$ and $\widehat{\mathcal{O}}^{\prime}_{\widehat{\Delta}^{\prime}}$ are the boundary exchanged primary operators from boundary OPE of $\mathcal{O}_{\Delta_2}(P_2)$ and $\mathcal{O}_{\Delta_3}(P_3)$ respectively. Let us also define the stripped conformal block 
\begin{equation}
	G^{\partial}_{\widehat{\Delta},\widehat{\Delta}^{\prime}}(\widehat{P}_1,P_2,P_3)= \frac{(-2 \widehat{P}_1 \cdot P_3)^{-\widehat{\Delta}_1}}{(P_2\cdot N_b)^{\Delta_2}(P_3\cdot N_b)^{\Delta_3-\widehat{\Delta}_1}} g^{\partial}_{\widehat{\Delta},\widehat{\Delta}^{\prime}}(\zeta_1,\zeta_2)\;.
\end{equation}
In the OPE limit, it should have the following asymptotic behavior
\begin{equation}
	\label{asybdyblock}
	g_{\widehat{\Delta},\widehat{\Delta}^{\prime}}^{\partial}(\zeta_1,\zeta_2)\to  \zeta_1^{-\widehat{\Delta}}\zeta_2^{\widehat{\Delta}_1-\widehat{\Delta}^{\prime}}\;.
\end{equation}
The two different OPE channels give the following crossing equation for the BB$\partial$ three-point functions
\begin{equation}
 \sum_{\mathcal{O}} b_{\mathcal{O}\widehat{\mathcal{O}}_1} f_{\mathcal{O}\mathcal{O}_2\mathcal{O}_3} G^{\text{B}}_{\Delta,J}(\widehat{P}_1,P_2,P_3)=\sum_{\widehat{\mathcal{O}}}\sum_{\widehat{\mathcal{O}}^{\prime}} b_{\mathcal{O}_{2} \widehat{\mathcal{O}}} b_{\mathcal{O}_{3} \widehat{\mathcal{O}}^{\prime}} \widehat{f}_{\widehat{\mathcal{O}}_1\widehat{\mathcal{O}}\widehat{\mathcal{O}}^{\prime}} G^{\partial}_{\widehat{\Delta},\widehat{\Delta}^{\prime}}(\widehat{P}_1,P_2,P_3)\;.
\end{equation}

\vspace{0.3cm}
\noindent{\bf Radial coordinates}
\vspace{0.2cm}

\noindent The conformal blocks $g^{\text{B}}_{\Delta,J}$ and $g^{\partial}_{\widehat{\Delta},\widehat{\Delta}^{\prime}}$ will be studied in detail in Section \ref{BCFT conformal blocks}. However, before that let us introduce another convenient choice of the cross ratios which will be useful for the BB$\partial$ case. We will call them the {\it radial coordinates}, in close analogy with the cross ratios used for CFT four-point functions \cite{Pappadopulo:2012jk,Hogervorst:2013sma} and for two-point functions in defect CFTs \cite{Lauria:2017wav}. As was emphasized in these works, the radial coordinates have a clear geometric meaning, which makes them more suitable to discuss the OPE convergence and to compute conformal blocks in series expansions. We will see that this is also the case for BCFT three-point functions.

\begin{figure}
	\centering
	\includegraphics[width=0.4\linewidth]{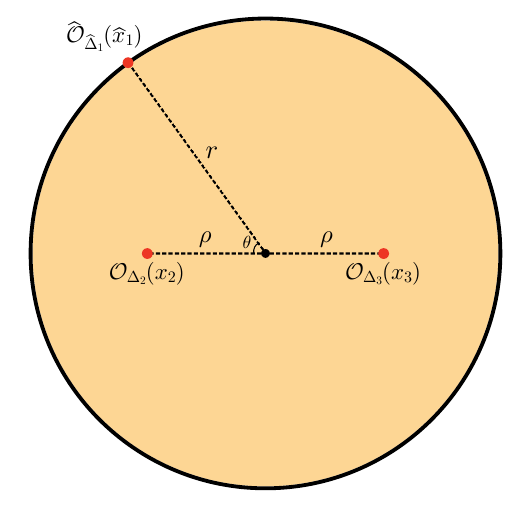}
	\caption{The desired configuration where all three operators lie within a sphere with radius $r$.  The operator   $\widehat{\mathcal{O}}_{\widehat{\Delta}_1}(\widehat{x}_1)$ is at its boundary and $\mathcal{O}_{\Delta_2}(x_2)$, $\mathcal{O}_{\Delta_3}(x_3)$ are located at antipodal points of another sphere with radius $\rho$.}
	\label{figsphere}
\end{figure}
\begin{figure}
	\centering
	\includegraphics[width=1.0\linewidth]{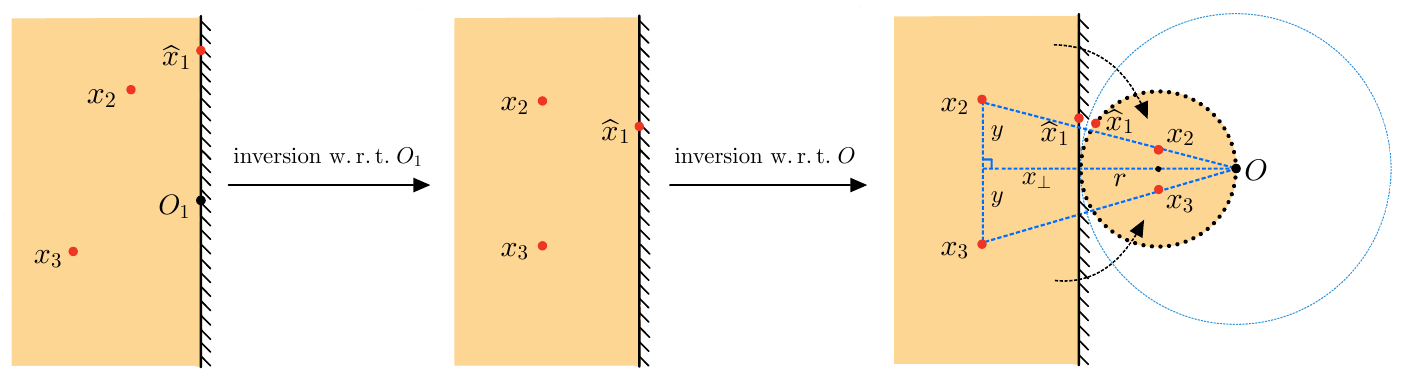}
	\caption{The sequence of conformal transformations that map a generic configuration of  $\widehat{\mathcal{O}}_{\widehat{\Delta}_1}(\widehat{x}_1)$, $\mathcal{O}_{\Delta_2}(x_2)$, $\mathcal{O}_{\Delta_3}(x_3)$ to the desired configuration. From left to the middle, we use a conformal inversion w.r.t.$\;$a point $O_1$ on the boundary to move $x_2$ and $x_3$ to positions with the same perpendicular distance $x_{\perp}$ to the boundary. The conformal inversion preserves the boundary. From middle to the right, we use another inversion w.r.t. a point $O$ located on the other side of the boundary to map the domain with $x_{\perp}>0$ to the interior of a ball with radius $r$. The point $O$ is located at the horizontal line that intersects the segment $\overline{x_2 x_3}$ in the middle. The radius of the ball $r$ is such that the image of $\overline{x_2 x_3}$ goes through the center of the ball. Obviously, $\widehat{\mathcal{O}}_{\widehat{\Delta}_1}(\widehat{x}_1)$  is mapped to the boundary of the ball.}
	\label{conformaltransformation}
\end{figure}

To define the radial coordinates, we will work in the original $\mathbb{R}^d$ space. The basic idea is to distribute the three operators $\widehat{\mathcal{O}}_{\widehat{\Delta}_1}(\widehat{x}_1)$, $\mathcal{O}_{\Delta_2}(x_2)$, $\mathcal{O}_{\Delta_3}(x_3)$ on concentric spheres so that we can conveniently apply the radial quantization. The desired configuration is depicted in Fig. \ref{figsphere}. To arrive at such a configuration from a generic one, we can use conformal transformations and Weyl transformation as we are in a CFT. The sequence of transformations are described in the caption of Fig. \ref{conformaltransformation}. To satisfy the condition mentioned in the caption, $r$ should be chosen as  
\begin{equation}
	r=\frac{\sqrt{x_{\perp}^2+y^2}}{2}\;,
\end{equation}
where $2y$ is the distance between $x_2$ and $x_3$ in the middle panel of Fig. \ref{conformaltransformation}. In that panel, $x_2$ and $x_3$ have the same perpendicular distance to the boundary which is denoted by $x_{\perp}$. The upshot is that $\widehat{\mathcal{O}}_{\widehat{\Delta}_1} (\widehat{x}_1)$ sits at the boundary of a sphere with radius $r$, and $\mathcal{O}_{\Delta_2}(x_2)$ and $\mathcal{O}_{\Delta_3}(x_3)$ are distributed symmetrically on another concentric sphere with radius $\rho$ (Fig. \ref{figsphere}).
These two radii are related by
\begin{equation}
	\frac{\rho}{r}=\frac{y}{2r+x_{\perp}}\;.
\end{equation}
This configuration motivates us to choose the following variables
\begin{equation}
	z\equiv \frac{\rho}{r}=\frac{y}{\sqrt{x_{\perp}^2+y^2}+x_{\perp}}\;,\quad x\equiv\cos \theta\;,
\end{equation}
where $z \in (0,1) \in \mathbb{R}^{+}$ and $x \in [-1,1] \in \mathbb{R}$. We see that $z$ represents the  ratio of radial distance of $\mathcal{O}_{\Delta_2}(x_2)$, $\mathcal{O}_{\Delta_3}(x_3)$ in the circle, and $x$ measures the angle between \big($\mathcal{O}_{\Delta_2}(x_2),\mathcal{O}_{\Delta_3}(x_3)$\big) and $\widehat{\mathcal{O}}_{\widehat{\Delta}_1}(\widehat{x}_1)$.  A simple calculation leads to the relation between the new variables $(z,x)$ and the cross ratios $(\xi_1,\xi_2)$
\begin{equation}
	\label{radialcoord1}
	\xi_1= \frac{16 z^2}{(1-z^2)^2}\;, \quad \xi_2 = \frac{1+2 x z+z^2}{1-2xz+z^2}\;,
\end{equation}
and the inverse transformation is
\begin{equation}
	\label{radialcoord2}
	z=\frac{\sqrt{\xi_1}}{\sqrt{\xi_1+4}+2}\;,\quad x=\sqrt{\frac{\xi_1+4}{\xi_1}}\frac{\xi_2-1}{\xi_2+1}\;.
\end{equation}
We will refer to $(z,x)$ as the radial coordinate of BB$\partial$ three-point function. We will use them in Section \ref{BCFT conformal blocks} to calculate the series expansion of BB$\partial$ bulk channel conformal blocks.

\section{BCFT conformal blocks}
\label{BCFT conformal blocks}
In this section, we compute the conformal blocks introduced in Section \ref{Kinematics of BCFT three-point functions} using a collection of different methods. In Section \ref{Bpartialpartial case} we obtain the B$\partial\partial$ conformal blocks by solving the Casimir equation and the result is given by (\ref{block}). These conformal blocks have already been obtained in the literature \cite{Lauria:2020emq,Buric:2020zea}, and are reproduced here with complete agreement. We discuss the BB$\partial$ conformal blocks in Section \ref{BBpartial case}. For the boundary channel, we will use the shadow formalism  \cite{Ferrara:1972xe,Ferrara:1972ay,Ferrara:1972uq,Ferrara:1972kab,Dolan:2000ut,Dolan:2011dv,Simmons-Duffin:2012juh} to extract the conformal block from the conformal partial wave which can be written as a Mellin-Barnes integral. The result is the series expansion (\ref{blockboundary}). The  boundary channel conformal blocks have  also been obtained in \cite{Buric:2020zea} using a totally different approach. Although a direct comparison with our results is difficult, we believe that they are fully equivalent, see the comments at the end of Section \ref{Boundary channel CB}. For the bulk channel, we will compute the  conformal blocks for arbitrary spin $J$ in the radial expansion by extending the techniques of \cite{Hogervorst:2013sma,Lauria:2017wav}. The final result for the conformal blocks takes the form (\ref{radial expansion}) in which the expansion coefficients can be computed recursively in principle to arbitrarily high orders. These results are new and have not appeared in the literature to the best of our knowledge. We also computed the $J=0$ case in (\ref{blockbulkscalar}) using the method based on conformal partial waves.

\subsection{The B$\partial\partial$ case}
\label{Bpartialpartial case}

An elegant way to compute conformal blocks is to use the quadratic conformal Casimir operator of which the conformal blocks are eigenfunctions \cite{Dolan:2003hv}. The eigen equation gives rise to a second order ODE,  which can be solved with appropriate boundary conditions. This method was extended to BCFT two-point functions in \cite{Liendo:2012hy} and reproduced the bulk and boundary channel conformal blocks originally obtained in \cite{McAvity:1995zd}. We will use the same strategy to compute the conformal block for the B$\partial\partial$ three-point function.

The quadratic Casimir of the boundary conformal group $SO(d,1)$ is defined as 
\begin{equation}
    \mathbf{C}=-\frac{1}{2}\widehat{L}_{ab}\widehat{L}^{ab}\;,
\end{equation}
where $\widehat{L}_{ab}$ are the Lorentz generators of $SO(d,1)$. Acting on a boundary primary state $| \widehat{\Delta},\ell \rangle$, the Casimir becomes its eigenvalue $\lambda_{\widehat{\Delta}, \ell} = \widehat{\Delta}(\widehat{\Delta}-d+1)+\ell(\ell+d-3)$. Let us denote as $\widehat{\mathbf{L}}_1$ the differential operators of the $SO(d,1)$ Lorentz generators acting on the field $\mathcal{O}_{\Delta_{1}}(P_1)$. In embedding space, they take the form
\begin{equation}
\widehat{\mathbf{L}}_{1,\widehat{a}\widehat{b}}= P_{1,\widehat{a}} \frac{\partial}{\partial P_{1,\widehat{b}}}-P_{1,\widehat{b}} \frac{\partial}{\partial P_{1,\widehat{a}}}\;,
\end{equation}
where $\widehat{a}, \widehat{b}=1,2,\ldots,d+1$ are the restriction of the full $SO(d+1,1)$ generators to the $SO(d,1)$ subgroup. We have  
\begin{equation}
\widehat{\mathbf{L}}_1\mathcal{O}_{\Delta_{1}}(P_1)|0\rangle=[\widehat{L}_{ab},\mathcal{O}_{\Delta_{1}}(P_1)]|0\rangle=\widehat{L}_{ab}\mathcal{O}_{\Delta_{1}}(P_1)|0\rangle\;,
\end{equation}
where it follows 
\begin{equation}
\mathbf{C}\mathcal{O}_{\Delta_{1}}(P_1)|0\rangle=-\frac{1}{2}\widehat{\mathbf{L}}_1^2\mathcal{O}_{\Delta_{1}}(P_1)|0\rangle\;.
\end{equation}
Let us define the boundary projector
\begin{equation}
	|\widehat{\mathcal{O}}| \equiv \sum_{\alpha, \beta=\mathcal{\widehat{O}},\widehat{\partial} \widehat{\mathcal{O}},\widehat{\partial}\widehat{\partial} \widehat{\mathcal{O}},\ldots}|\alpha\rangle \mathcal{N}_{\alpha \beta}^{-1}\langle\beta|\;,\quad \mathcal{N}_{\alpha \beta} \equiv\langle\alpha|\beta\rangle\;,
\end{equation}
which projects to the boundary conformal primary  $\widehat{\mathcal{O}}$ and its descendants. Then the conformal block of the B$\partial\partial$ three-point function is
\begin{equation}
	G_{\widehat{\Delta}}(P_1, \widehat{P}_2,\widehat{P}_3) \propto \langle  \mathcal{O}_{\Delta_{1}}(P_1)|\widehat{\mathcal{O}}|\widehat{\mathcal{O}}_{\widehat{\Delta}_{2}}(\widehat{P}_2)\widehat{\mathcal{O}}_{\widehat{\Delta}_{3}}(\widehat{P}_3)\rangle\;.
\end{equation}
Acting on it with the conformal Casimir, we have 
\begin{equation}
	\begin{gathered}
		-\frac{1}{2}\widehat{\mathbf{L}}_1^2\langle  \mathcal{O}_{\Delta_{1}}(P_1)|\widehat{\mathcal{O}}|\widehat{\mathcal{O}}_{\widehat{\Delta}_{2}}(\widehat{P}_2)\widehat{\mathcal{O}}_{\widehat{\Delta}_{3}}(\widehat{P}_3)\rangle=\langle  \mathcal{O}_{\Delta_{1}}(P_1)\mathbf{C}|\widehat{\mathcal{O}}|\widehat{\mathcal{O}}_{\widehat{\Delta}_{2}}(\widehat{P}_2)\widehat{\mathcal{O}}_{\widehat{\Delta}_{3}}(\widehat{P}_3)\rangle\\=\lambda_{\widehat{\Delta}, 0} \langle  \mathcal{O}_{\Delta_{1}}(P_1)|\widehat{\mathcal{O}}|\widehat{\mathcal{O}}_{\widehat{\Delta}_{2}}(\widehat{P}_2)\widehat{\mathcal{O}}_{\widehat{\Delta}_{3}}(\widehat{P}_3)\rangle\;.
	\end{gathered}
\end{equation}
In terms of the stripped conformal block, this gives the following differential equation 
\begin{equation}
	\label{casimirBpp}
	\left((\widehat{\Delta} _2-\widehat{\Delta }_3)^2-\widehat{\Delta} ^2\right) \xi  f(\xi )-2 \xi  \left((d+2 (\widehat{\Delta} +1) \xi -2 \widehat{\Delta}-3) f'(\xi )+2 (\xi -1) \xi  f''(\xi )\right)=0\;,
\end{equation}
where we have defined for convenience
\begin{equation}
	g_{\widehat{\Delta}}(\xi)\equiv \xi^{\frac{\widehat{\Delta}-\widehat{\Delta}_2-\widehat{\Delta}_3}{2}} f(\xi)\;.
\end{equation}
Combined with the asymptotic behavior $f(\xi)\to 1$  as $\xi \to 0$ determined by the OPE (\ref{asy1}), the Casimir equation (\ref{casimirBpp}) can be easily solved and we  get the following expression for the  B$\partial\partial$ conformal block (\ref{casimirBpp})\footnote{Conformal block of B$\partial\partial$ three-point function was also discussed in \cite{Karch:2018uft}, but their result has asymptotic behavior like a combination of conformal block and its shadow block.}
\begin{equation}
	\label{block}
	g_{\widehat{\Delta}}(\xi)=\xi^{\frac{\widehat{\Delta}-\widehat{\Delta}_{2}-\widehat{\Delta}_3}{2}} \, _2F_1\bigg(\frac{\widehat{\Delta}+\widehat{\Delta}_2-\widehat{\Delta}_3}{2},\frac{\widehat{\Delta}-\widehat{\Delta}_2+\widehat{\Delta}_3}{2};\frac{3-d}{2}+\widehat{\Delta};\xi \bigg)\;.
\end{equation}
The B$\partial\partial$ three-point function was also considered in \cite{Lauria:2020emq} (and also \cite{Buric:2020zea}), where the authors computed the B$\partial\partial$ conformal block.\footnote{They considered  defect CFTs with general co-dimension $q$ and the BCFT case corresponds to $q=1$.} In \cite{Lauria:2020emq} a different cross ratio $\hat{\chi}$ was used and it is related to $\xi$ by
\begin{equation}
\hat{\chi}=\frac{1}{\xi}-1\;.
\end{equation}
With the same choice of the prefactor as ours (\ref{2}), their result, given in (107) of \cite{Lauria:2020emq}, is
\begin{equation}
g_{\widehat{\Delta}}^{\text{theirs}}(\xi)=\bigg(\frac{1}{\xi}-1\bigg)^{-\kappa_{\widehat{\Delta}}} \xi^{-\Delta_2} {}_2 F_1 \bigg(\kappa_{\widehat{\Delta}},\frac{3-d}{2}+\kappa_{\widehat{\Delta}};\frac{3-d}{2}+\widehat{\Delta};\frac{\xi}{\xi-1}\bigg)\;,
\end{equation}
where $\kappa_{\widehat{\Delta}}=(\widehat{\Delta}+\widehat{\Delta}_2-\widehat{\Delta}_3)/2$. This is the same as our result  $\left(\ref{block}\right)$ up to a Pfaff transformation (\ref{pfaff}) of Gauss hypergeometric function.\footnote{The Pfaff transformation is defined as 
\begin{equation}
\label{pfaff}
		{ }_2 F_1(a, b ; c ; z)=(1-z)^{-a}{ }_2 F_1\Big(a, c-b ; c ; \frac{z}{z-1}\Big)\;, \quad z \notin(1, \infty)\;.
\end{equation}}

Let us also make a comment that the above conformal block is the same as the conformal block of two-point functions in CFTs on $\mathbb{RP}^{d-1}$ \cite{Nakayama:2016xvw}, upon identifying the cross ratios. This is in fact not a coincidence as the B$\partial\partial$ three-point function in BCFT$_d$ has the same symmetry as the two-point function in $\mathbb{RP}^{d-1}$ CFT.\footnote{From the point of view of the two operators on the boundary, the embedding vector $P_1$ further breaks the boundary conformal group to $SO(d-1)$. This has the same effect as introducing a time-like fixed vector $N_c$ which implements the $\mathbb{Z}_2$ quotient in $\mathbb{RP}^{d-1}$ and breaks the conformal group from $SO(d,1)$ to $SO(d)$ (see \cite{Giombi:2020xah} for details).} Moreover, the conformal Casimir operators in the Casimir equations are both with respect to the $SO(d,1)$ conformal group.

\subsection{The BB$\partial$ case}
\label{BBpartial case}
Let us now turn to the BB$\partial$ case which is more complicated. As we have discussed in Section \ref{Kinematics of BCFT three-point functions}, there are two kinds of conformal blocks which are associated with the bulk and boundary channel respectively. In the boundary channel, only scalar operators can be exchanged. However, in the bulk channel spinning operators can be exchanged as well. Moreover, the BB$\partial$ three-point functions have two cross ratios. As a result, the Casimir equations for the conformal blocks are second order PDEs which are much more difficult to solve than ODEs. Fortunately, there are other techniques in the literature which are particularly suitable for computing the BB$\partial$ conformal blocks. In Section \ref{Boundary channel CB}, we will calculate the boundary channel conformal block by constructing the conformal partial wave \cite{Dolan:2000ut,Simmons-Duffin:2012juh} (see also \cite{Dolan:2011dv}). In Section \ref{Bulk channel CB}, we will use the radial coordinates introduced in Section \ref{Kinematics of BCFT three-point functions} to calculate bulk channel conformal blocks in series expansions.

\subsubsection{Boundary channel}
\label{Boundary channel CB}
As is reviewed in Appendix \ref{BBpartial boundary conformal partial wave}, for four-point functions in a CFT without boundaries, a conformal partial wave (CPW)  of dimension $\Delta$ and spin $J$ is the linear combination of a conformal block with these quantum numbers and its ``shadow" with conformal dimension $d-\Delta$ and spin $J$, with appropriate coefficients such that the CPW is single-valued. Moreover, they can be expressed as simple conformal integrals using the shadow integral representation  \cite{Simmons-Duffin:2012juh} (see also \cite{Simmons-Duffin:2017nub}). This fact makes the CPWs easier to compute  and we can then extract from them the conformal blocks. Here we will generalize this strategy to the BCFT case. We find that similar to the four-point function case, the CPW is a linear combination of the boundary channel conformal blocks $G^\partial_{\delta,\delta'}$ where the two exchanged dimensions $(\delta,\delta')$ can be any combination of $(\Delta,\Delta')$ and their shadows
\begin{equation}
\Psi_{\widehat{\Delta},\widehat{\Delta}^{\prime}}^{\partial}=\widetilde{\mathcal{K}}_{\widehat{\Delta},\widehat{\Delta}^{\prime}}^{\widehat{\Delta}_1} G_{\widehat{\Delta},\widehat{\Delta}^{\prime}}^{\partial}+\mathcal{K}_{\widehat{\Delta},\widehat{\Delta}^{\prime}}^{\widehat{\Delta}_1} G_{\widetilde{\widehat{\Delta}},\widetilde{\widehat{\Delta}^{\prime}}}^{\partial}+ \mathcal{S}_{\widehat{\Delta},\widehat{\Delta}^{\prime}}^{\widehat{\Delta}_1} G_{\widetilde{\widehat{\Delta}},\widehat{\Delta}^{\prime}}^{\partial}+ \mathcal{S^{\prime}}_{\widehat{\Delta},\widehat{\Delta}^{\prime}}^{\widehat{\Delta}_1} G_{\widehat{\Delta},\widetilde{\widehat{\Delta}^{\prime}}}^{\partial}\;.
\end{equation}
Here $\widetilde{\widehat{\Delta}}\equiv d-1-\widehat{\Delta}$ is the boundary shadow dimension of $\widehat{\Delta}$ and similarly   $\widetilde{\widehat{\Delta}}^{\prime}=d-1-\widehat{\Delta}^{\prime}$. In Appendix \ref{BBpartial boundary conformal partial wave}, we will derive this identity along with the coefficients in front of the conformal blocks. Here we will use it as a starting point and focus on the computation of the CPW. From it we will  extract the conformal block $g^\partial_{\widehat{\Delta},\widehat{\Delta}'}$ in the form of a series expansion (\ref{blockboundary}).

Explicitly, the boundary channel CPW for the B$\partial\partial$ three-point function is defined by the following integral (\ref{bdyCPW})
\begin{equation}
\begin{aligned}	&\Psi_{\widehat{\Delta},\widehat{\Delta}^{\prime}}^{\partial}(P_i)\\&=\int D^{d-1} \widehat{P} D^{d-1} \widehat{P^{\prime}} \langle\langle \widehat{\mathcal{O}}_1(\widehat{P}_1)\widehat{\mathcal{O}}_{\widehat{\Delta}}(\widehat{P})\widehat{\mathcal{O}}_{\widehat{\Delta}^{\prime}}(\widehat{P^{\prime}})\rangle\rangle \langle\langle \widehat{\mathcal{O}}_{\widetilde{\widehat{\Delta}}}(\widehat{P}) \mathcal{O}_2(P_2)\rangle\rangle \langle\langle \widehat{\mathcal{O}}_{\widetilde{\widehat{\Delta}^{\prime}}}(\widehat{P^{\prime}}) \mathcal{O}_3(P_3)\rangle\rangle\;.
\end{aligned}
\end{equation}
Here we have written the integral in embedding space similar to \cite{Simmons-Duffin:2012juh} and the integrand is a product of a three-point function and two two-point functions of the operators or their shadows. We used the double angle brackets to denote the conformal invariant structures 
\begin{equation}
	\label{conformalinvariant3pt}
\langle \langle \widehat{\mathcal{O}}_{1} \widehat{\mathcal{O}}_2 \widehat{\mathcal{O}}_3\rangle \rangle= \frac{1}{(-2 \widehat{P}_1 \cdot \widehat{P}_2)^{\frac{\widehat{\Delta}_1+\widehat{\Delta}_2-\widehat{\Delta}_3}{2}} (-2 \widehat{P}_1 \cdot \widehat{P}_3)^{\frac{\widehat{\Delta}_1+\widehat{\Delta}_3-\widehat{\Delta}_2}{2}} (-2 \widehat{P}_2 \cdot \widehat{P}_3)^{\frac{\widehat{\Delta}_2+\widehat{\Delta}_3-\widehat{\Delta}_1}{2}}}\;,
\end{equation}
\begin{equation}
\label{conformalinvariant2pt}
    \langle\langle \mathcal{O}_{1}\widehat{\mathcal{O}}_{2}\rangle\rangle =\frac{1}{(P_1 \cdot N_b)^{\Delta_1-\widehat{\Delta}_2} (-2 P_1 \cdot \widehat{P}_2)^{\widehat{\Delta}_2}}\;.
\end{equation}
This gives the following integral
\begin{equation}
	\begin{aligned}
			\label{cpwboundary01}
&\Psi_{\widehat{\Delta},\widehat{\Delta}^{\prime}}^{\partial}\\&=\int D^{d-1} \widehat{P}  D^{d-1} \widehat{P^{\prime}} \frac{(P_2 \cdot N_b)^{\widetilde{\widehat{\Delta}}-\Delta_2}}{(-2 P_2 \cdot \widehat{P})^{\widetilde{\widehat{\Delta}}} }\frac{(P_3 \cdot N_b)^{\widetilde{\widehat{\Delta}^{\prime}}-\Delta_3}}{(-2 P_3 \cdot \widehat{P^{\prime}})^{\widetilde{\widehat{\Delta}^{\prime}}}} \frac{(-2\widehat{P} \cdot \widehat{P^{\prime}})^{\frac{-\widehat{\Delta}-\widehat{\Delta}^{\prime}+\widehat{\Delta}_1}{2}}}{(-2\widehat{P}_1 \cdot \widehat{P})^{\frac{\widehat{\Delta}_1+\widehat{\Delta}-\widehat{\Delta}^{\prime}}{2}} (-2\widehat{P}_1 \cdot \widehat{P^{\prime}})^{\frac{\widehat{\Delta}_1+\widehat{\Delta}^{\prime}-\widehat{\Delta}}{2}} } \;.
\end{aligned}
\end{equation}
To compute this integral, we will use the method of \cite{Dolan:2011dv}. To start, we compute the $\widehat{P^{\prime}}$ integral, which  can be easily done by using Feynman parameterization. We get
\begin{equation}
\begin{aligned} \label{Pprimeintegral1}
	&\int D^{d-1}\widehat{P^{\prime}} \frac{1}{ (-2\widehat{P}_1 \cdot \widehat{P^{\prime}})^{\frac{\widehat{\Delta}_1+\widehat{\Delta}^{\prime}-\widehat{\Delta}}{2}} (-2\widehat{P} \cdot \widehat{P^{\prime}})^{\frac{\widehat{\Delta}+\widehat{\Delta}^{\prime}-\widehat{\Delta}_1}{2}} (-2 P_3 \cdot \widehat{P^{\prime}})^{\widetilde{\widehat{\Delta}^{\prime}}}} \\ &=\frac{\pi^{h^{\prime}}\Gamma(h^{\prime})}{\prod_{i} \Gamma(a_{i})} \frac{(P_3 \cdot N_b)^{a_1+a_2-a_3} }{(-2 \widehat{P}_1 \cdot P_3)^{a_1}(-2 \widehat{P} \cdot P_3)^{a_2} } \int_{0}^{1} \prod_{i=1}^{3} d t_{i} t_{i}^{a_{i}-1} \frac{\delta(\sum_{i} t_{i}-1)}{\left(t_{3}^{2}+t_{1} t_{3}+t_{2} t_{3}+\xi_0^{-1} t_{1} t_{2}\right)^{h^{\prime}}}\\ &=\frac{\pi^{h^{\prime}}}{\prod_{i} \Gamma(a_{i})} \frac{(P_3 \cdot N_b)^{a_1+a_2-a_3} }{(-2 \widehat{P}_1 \cdot P_3)^{a_1}(-2 \widehat{P} \cdot P_3)^{a_2} }\\&\quad \times
		\int_{-i\infty}^{+i\infty} \frac{d\tau}{2\pi i} \;\Gamma(-\tau)\Gamma(a_3-h^{\prime}-\tau)\Gamma( a _ { 1 } + \tau ) \Gamma(a_{2}+\tau) \xi_0^{-\tau}\;,
\end{aligned}
\end{equation}
where $h^{\prime}\equiv\frac{d-1}{2}$ and
\begin{equation}
	a_1=\frac{\widehat{\Delta}_1+\widehat{\Delta}^{\prime}-\widehat{\Delta}}{2}\;, \quad a_2=\frac{-\widehat{\Delta}_1+\widehat{\Delta}+\widehat{\Delta}^{\prime}}{2}\;, \quad a_3=\widetilde{\widehat{\Delta}^{\prime}}\;,\quad\xi_0=\frac{(-2 \widehat{P}\cdot P_3) (-2 \widehat{P}_1 \cdot P_3)}{(P_3 \cdot N_b)^2 (-2 \widehat{P}_1 \cdot \widehat{P})}\;.
\end{equation}
In the last line, we used the Mellin-Barnes (MB) representation 
\begin{equation}
	\begin{split}
 \label{MBrep}
		\frac{1}{\left(A_{1}+\cdots+A_{n}\right)^{c}}={}&\frac{1}{(2 \pi i)^{n-1}}\frac{1}{\Gamma(c)} \int_{-i\infty}^{+i \infty} d \sigma_{1} \cdots d \sigma_{n-1} \Gamma(-\sigma_{1}) \cdots \Gamma(-\sigma_{n-1}) \\
{}&\times 	\Gamma(\sigma_{1}+\cdots+\sigma_{n-1}+c) A_{1}^{\sigma_{1}}  \cdots A_{n-1}^{\sigma_{n-1}} A_{n}^{-\sigma_{1}-\cdots-\sigma_{n-1}-c}\;,
	\end{split}
\end{equation}
and then integrated out the $t_i$ variables. We can similarly perform the $\widehat{P}$ integral and get 
\begin{equation}
\begin{aligned}
&\int D^{d-1} \widehat{P} \frac{1}{(-2 \widehat{P}_1 \cdot \widehat{P})^{\frac{\widehat{\Delta}_1+\widehat{\Delta}-\widehat{\Delta}^{\prime}}{2}-\tau}  (-2 P_2 \cdot \widehat{P})^{\widetilde{\widehat{\Delta}}} (-2 P_3 \cdot \widehat{P})^{\frac{\widehat{\Delta}+\widehat{\Delta}^{\prime}-\widehat{\Delta}_1}{2}+\tau}}
		\\ &=  \frac{\pi^{h^{\prime}}}{\prod_{i} \Gamma(b_{i})} \frac{(P_2 \cdot N_b)^{b_1-b_2} }{(-2 \widehat{P}_1 \cdot P_2)^{b_1}(P_3 \cdot N_b)^{b_3}}  \int_{-i \infty}^{i \infty} \prod_{i=1}^{3} \frac{d s_i}{(2 \pi i)} \,\xi_2^{-s_3} (\xi_1+1)^{-(\sum s_i)-h^{\prime}}  \\ & \quad\times \Gamma(b_1+s_3)\Gamma(b_2-h^{\prime}-s_2-s_3)\Gamma(b_3-h^{\prime}-s_1+s_2)\Gamma(-s_2)\Gamma(-s_3)\Gamma(\sum s_i+h^{\prime})\;, 
  \end{aligned}
\end{equation}
where
\begin{equation}
	b_1=\frac{\widehat{\Delta}_1+\widehat{\Delta}-\widehat{\Delta}^{\prime}}{2}-\tau\;, \quad b_2=\widetilde{\widehat{\Delta}}\;, \quad b_3=\frac{\widehat{\Delta}+\widehat{\Delta}^{\prime}-\widehat{\Delta}_1}{2}+\tau\;.
\end{equation}
This ultimately leads to the following MB representation for the CPW 
\begin{equation}
\begin{aligned}
&\psi_{\widehat{\Delta},\widehat{\Delta}^{\prime}}^{\partial}(\xi_1,\xi_2)= \frac{\pi^{2 h^{\prime}}}{ \prod \Gamma(a_i) \Gamma(b_2)}  \xi_2^{\frac{\widehat{\Delta}^{\prime}-\widehat{\Delta}-\widehat{\Delta}_1}{2}}  \int \prod_{i=1}^{3} \frac{d s_i}{2 \pi i} \frac{d\tau}{2\pi i}\xi_2^{-s_3+\tau} (\xi_1+1)^{-(\sum s_i)-h^{\prime}}\\& \times \bigg[ \frac{\Gamma(-\tau)\Gamma(a_3-h^{\prime}-\tau)\Gamma( a _ { 1 } + \tau ) }{\Gamma(\frac{\widehat{\Delta}_1+\widehat{\Delta}-\widehat{\Delta}^{\prime}}{2}-\tau) }  \Gamma(-s_2)\Gamma(-s_3)   \Gamma\big(\sum s_i+h^{\prime}\big) \Gamma\bigg(\frac{\widehat{\Delta}_1+\widehat{\Delta}-\widehat{\Delta}^{\prime}}{2}-\tau+s_3\bigg)
\\&\times\Gamma(h^{\prime}-\widehat{\Delta}-s_2-s_3)\Gamma\bigg(\frac{\widehat{\Delta}+\widehat{\Delta}^{\prime}-\widehat{\Delta}_1}{2}+\tau-h^{\prime}-s_1+s_2\bigg) \bigg]\;,
\end{aligned}
\end{equation}
where
\begin{equation}
\Psi_{\widehat{\Delta},\widehat{\Delta}^{\prime}}^{\partial}(P_i)= \frac{(-2 \widehat{P}_1 \cdot P_3)^{-\widehat{\Delta}_1}}{(P_2\cdot N_b)^{\Delta_2}(P_3\cdot N_b)^{\Delta_3-\widehat{\Delta}_1}}  \psi_{\widehat{\Delta},\widehat{\Delta}^{\prime}}^{\partial}(\xi_1,\xi_2)\;.
\end{equation}

The integrand of this integral representation has many poles in $s_i$ and $\tau$. However, we are only interested in a subset of them, namely the poles which are associated with the boundary conformal block $g_{\widehat{\Delta},\widehat{\Delta}^{\prime}}^{\partial}$. It is not difficult to identify these poles using (\ref{asybdyblock})
\begin{equation}
	s_1=\frac{\widehat{\Delta}+\widehat{\Delta}^{\prime}-\widehat{\Delta}_1}{2}+\tau-h^{\prime}+s_2+n_1\;, \quad s_2=n_2\;, \quad s_3=n_3\;, \quad \tau=n\;,
\end{equation}
where $n,n_i=0,1,2,\cdots$. Taking residues at these poles, we find that it is possible to first sum over $n_1$ and then over $n_2$. Note that the conformal block $g_{\widehat{\Delta},\widehat{\Delta}^{\prime}}^{\partial}$ appears in the  CPW with a coefficient $\widetilde{\mathcal{K}}_{\widehat{\Delta},\widehat{\Delta}^{\prime}}^{\widehat{\Delta}_1}$ which is given explicitly in (\ref{coeffkhat}). Taking this into account, we find 
\begin{equation}
	\begin{split}
		\label{blockboundary0}
&g_{\widehat{\Delta},\widehat{\Delta}^{\prime}}^{\partial}(\xi_1,\xi_2)=\sum_{m=0}^{\infty} \sum_{n=0}^{\infty} \bigg[(-1)^{m+n}\tilde{\xi}_1^{\,\frac{\widehat{\Delta}_1-\widehat{\Delta}-\widehat{\Delta}^{\prime}}{2}-m-n} \xi_2^{\frac{-\widehat{\Delta}_1-\widehat{\Delta}+\widehat{\Delta}^{\prime}}{2}-m+n} \\&\times  \frac{ \Gamma(h^{\prime}-m-\widehat{\Delta})\Gamma(h^{\prime}-n-\widehat{\Delta}^{\prime}) \Gamma\big(\frac{2(m-n)+\widehat{\Delta}-\widehat{\Delta}^{\prime}+\widehat{\Delta}_1}{2}\big) \Gamma\big(\frac{2(m+n)+\widehat{\Delta}+\widehat{\Delta}^{\prime}-\widehat{\Delta}_1}{2}\big) \Gamma\big(\frac{2n-\widehat{\Delta}+\widehat{\Delta}^{\prime}+\widehat{\Delta}_1}{2}\big)}{m! n! \Gamma(h^{\prime}-\widehat{\Delta}) \Gamma(h^{\prime}-\widehat{\Delta}^{\prime}) \Gamma\big(\frac{\widehat{\Delta}+\widehat{\Delta}^{\prime}-\widehat{\Delta}_1}{2}\big) \Gamma\big(\frac{-\widehat{\Delta}+\widehat{\Delta}^{\prime}+\widehat{\Delta}_1}{2}\big)\Gamma\big(\frac{-2n+\widehat{\Delta}-\widehat{\Delta}^{\prime}+\widehat{\Delta}_1}{2}\big)  } \\  &\times  {}_2F_1\Bigg(\frac{2(m+n)+\widehat{\Delta}+\widehat{\Delta}^{\prime}-\widehat{\Delta}_1}{4},\frac{2+2(m+n)+\widehat{\Delta}+\widehat{\Delta}^{\prime}-\widehat{\Delta}_1}{4};1-h^{\prime}+m+\widehat{\Delta};\frac{4}{\tilde{\xi}_1^{\,2}}\Bigg) \Bigg]\;,
	\end{split}
\end{equation}
where $\tilde{\xi}_1=\xi_1+2$. However, this representation is not most convenient for expansions in the boundary OPE limit where natural choice of cross ratios $\zeta_{1,2}\to \infty$. To make progress, we use the following  identity to rewrite the hypergeometric function
\begin{equation}
z^{-y}{ }_2 F_1\bigg(y,\frac{1-x}{2}+y ;-x+2 y+1 ;-\frac{4}{z}\bigg)=(z+2)^{-y}{ }_2 F_1\bigg(\frac{y+1}{2}, \frac{y}{2} ;-\frac{x}{2}+y+1 ; \frac{4}{(z+2)^2}\bigg) \;.
\end{equation}
This gives the following triple sum representation of the boundary channel conformal block
\begin{eqnarray}
\begin{aligned}
\label{blockboundary}
&g^{\partial}_{\widehat{\Delta},\widehat{\Delta}^{\prime}}(\zeta_1,\zeta_2)=\sum_{k,m,n=0}^{\infty}\bigg[ (-1)^{k+m+n}4^{k}  \zeta_1^{-k-2m-\widehat{\Delta}} \zeta_2^{-k-2n+\widehat{\Delta}_1-\widehat{\Delta}^{\prime}}\\&\quad\times\frac{\left(\frac{1-2h^{\prime}+2m+2\widehat{\Delta}}{2}\right)_k\left(\frac{-2n+\widehat{\Delta}-\widehat{\Delta}^{\prime}+\widehat{\Delta}_1}{2}\right)_m\left(\frac{-\widehat{\Delta}+\widehat{\Delta}^{\prime}+\widehat{\Delta}_1}{2}\right)_n\left(\frac{\widehat{\Delta}+\widehat{\Delta}^{\prime}-\widehat{\Delta}_1}{2}\right)_{k+m+n}}{k!m!n!\left(h^{\prime}-m-\widehat{\Delta}\right)_m\left(h^{\prime}-n-\widehat{\Delta}^{\prime}\right)_n\left(1-2h^{\prime}+2m+2\widehat{\Delta}\right)_k}\Bigg]\;,
    \end{aligned}
\end{eqnarray} 
as a series in inverse powers of $\zeta_{1,2}$. Although not manifest, it is not difficult to check that the conformal block (\ref{blockboundary}) is invariant under $P_2\leftrightarrow P_3$, $\widehat{\Delta} \leftrightarrow \widehat{\Delta}^{\prime}$.

As a consistency check, let us also consider the Casimir equations. The derivation of these equations is similar to the B$\partial\partial$ case, except that we now have two equations
\begin{eqnarray}
\label{bdyCasimir1}&& \bigg(\frac{1}{2} \widehat{\mathbf{L}}_{2}^{2}+\widehat{\Delta}(\widehat{\Delta}-d+1)\bigg) G_{\widehat{\Delta},\widehat{\Delta}^{\prime}}^{\partial}=0\;, \\
\label{bdyCasimir2}&& \bigg(\frac{1}{2} \widehat{\mathbf{L}}_{3}^{2}+\widehat{\Delta}^{\prime}(\widehat{\Delta}^{\prime}-d+1)\bigg) G_{\widehat{\Delta},\widehat{\Delta}^{\prime}}^{\partial}=0\;.
\end{eqnarray}
Rewritten in terms of the cross ratios (and omitting the upstairs label $\partial$), they read
\begin{eqnarray}
  \nonumber&& \zeta
   _1^{-2} \zeta _2^{-1}\left(\zeta _2 \left((d-2)\zeta_1+\zeta _2\right) g^{(0,1)}_{\widehat{\Delta},\widehat{\Delta}^{\prime}}(\zeta _1,\zeta _2)+\zeta_1 \left(\left(d \zeta_1 \left(\zeta_1 \zeta_2+1\right)-\zeta_2\right) g^{(1,0)}_{\widehat{\Delta},\widehat{\Delta}^{\prime}}(\zeta_1,\zeta_2)\right.\right.\\\nonumber&& \left.\left.+2 \zeta _1
   \zeta_2 g^{(1,1)}_{\widehat{\Delta},\widehat{\Delta}^{\prime}}(\zeta _1,\zeta _2)+\zeta_1 \left(2 \zeta_1+\left(\zeta_1^2-1\right) \zeta_2\right) g^{(2,0)}_{\widehat{\Delta},\widehat{\Delta}^{\prime}}(\zeta _1,\zeta_2)\right)+\zeta_2^3 g^{(0,2)}_{\widehat{\Delta},\widehat{\Delta}^{\prime}}(\zeta_1,\zeta_2)\right)\\&&+\widehat{\Delta}  (d-\widehat{\Delta} -1) g_{\widehat{\Delta},\widehat{\Delta}^{\prime}}^{\partial}(\zeta _1,\zeta_2)=0\;,\\ \nonumber&& 
 \zeta_2^{-2}\left(\zeta _2 (d-2 (\widehat{\Delta} _1+1))+2 \widehat{\Delta}_1 \zeta _1+\zeta_1\right) g^{(1,0)}_{\widehat{\Delta},\widehat{\Delta}^{\prime}}(\zeta_1,\zeta_2)+\zeta_1^{-1} \zeta_2^{-1} \left(\zeta_2
   (d-2 \widehat{\Delta} _1)\right.\\\nonumber&&\left.+\zeta_1(\zeta_2^2 (d-2 \widehat{\Delta} _1)+2 \widehat{\Delta} _1-1)\right)g^{(0,1)}_{\widehat{\Delta},\widehat{\Delta}^{\prime}}(\zeta_1,\zeta_2)+\zeta_1^2 \zeta_2^{-2} g^{(2,0)}_{\widehat{\Delta},\widehat{\Delta}^{\prime}}(\zeta_1,\zeta_2)+\left(\zeta_2^2+2\zeta_1^{-1} \zeta_2-1\right)\\\nonumber&&\times g^{(0,2)}_{\widehat{\Delta},\widehat{\Delta}^{\prime}}(\zeta_1,\zeta _2)+2
g^{(1,1)}_{\widehat{\Delta},\widehat{\Delta}^{\prime}}(\zeta_1,\zeta_2)+(\widehat{\Delta}_1-\widehat{\Delta}^{\prime}) 
   (-d+\widehat{\Delta}_1+\widehat{\Delta}^{\prime}+1)g_{\widehat{\Delta},\widehat{\Delta}^{\prime}}^{\partial}(\zeta _1,\zeta_2)=0\;,\\\label{cas2}&&
\end{eqnarray}
 where $g_{\widehat{\Delta},\widehat{\Delta}^{\prime}}^{(m,n)}(\zeta_1,\zeta_2)=\partial_{\zeta_1}^m \partial_{\zeta_2}^n g_{\widehat{\Delta},\widehat{\Delta}^{\prime}}^{\partial}(\zeta_1,\zeta_2)$. Although solving these PDEs is difficult, we can easily check that (\ref{blockboundary}) satisfies both equations order by order.

We can perform further checks by taking limits of (\ref{blockboundary}). Taking $\widehat{\Delta}_1 \to 0$ and $\widehat{\Delta}^{\prime} \to \widehat{\Delta}$, (\ref{blockboundary}) nicely reduces to the boundary channel conformal block for two-point functions $\left(\ref{twopointblockboundry}\right)$. Another interesting limit is to send $\xi_1 \to \infty$ keeping $\xi_2 / \xi_1 = \xi$ fixed, and set $\widehat{\Delta} \to \Delta_2$. This moves $\mathcal{O}_{2 }$ to the boundary, and we find that $\left(\ref{blockboundary}\right)$ reduces to B$\partial\partial$ conformal block (\ref{block}). A special “cylindrical” configuration of the boundary channel conformal block was also considered in \cite{Lauria:2020emq}, which corresponds to taking $\xi_2=1$ (or taking $\zeta_1=\zeta_2$) in our expression (\ref{blockboundary0}) (or in (\ref{blockboundary})). Under this condition, we have checked that our result correctly reproduced their (114), {\it i.e.},
\begin{equation}\label{gpartialcylind}
g^{\partial}_{\widehat{\Delta},\widehat{\Delta}^{\prime}}(\xi_1,1)=\xi_1^{-\widehat{\kappa}} 
{ }_4 F_3\Bigg(\begin{array}{c|c}
\frac{1}{2}-h^{\prime}+\frac{\widehat{\Delta}+\widehat{\Delta}^{\prime}}{2}, 1-h^{\prime}+\frac{\widehat{\Delta}+\widehat{\Delta}^{\prime}}{2},\widehat{\kappa} ,1-h^{\prime}+\widehat{\kappa}\\
1-h^{\prime}+\widehat{\Delta}, 1-h^{\prime}+\widehat{\Delta}^{\prime}, 1-2h^{\prime}+\widehat{\Delta}+\widehat{\Delta}^{\prime}
\end{array} -\frac{4}{\xi_1}\Bigg)\;,
\end{equation}
where $\widehat{\kappa}=(\widehat{\Delta}+\widehat{\Delta}^{\prime}-\widehat{\Delta}_1)/2$. Note that the boundary channel conformal block was also obtained as an Appell $F_4$ function in (4.55) of \cite{Buric:2020zea}. Both here and in \cite{Buric:2020zea}, it was checked that the conformal block satisfies the Casimir equations with the correct boundary conditions. Moreover, \cite{Buric:2020zea} also checked their result reduces to (\ref{gpartialcylind}) in the ``cylindrical'' limit. Therefore, our expression (\ref{blockboundary}) for the boundary channel conformal block should be equivalent to (4.55) of \cite{Buric:2020zea}. However, it would be interesting to directly prove the equivalence by massaging the triple sum representation (\ref{blockboundary}).

\subsubsection{Bulk channel}
\label{Bulk channel CB}
To compute the BB$\partial$ bulk channel conformal blocks, it is more convenient to use the radial coordinates introduced in Section \ref{Kinematics of BCFT three-point functions}. Using conformal symmetry we already showed that the insertion locations of 
$\widehat{\mathcal{O}}_1$, $\mathcal{O}_2$, $\mathcal{O}_3$ can be arranged into the configuration    Fig. \ref{figsphere}. We now further perform a Weyl transformation into a cylinder $(\tau,{\bf n})$, where $\tau=-\log z$ is the time along the cylinder and $\mathbf{n}$ is the direction of the operator insertion in  Fig. \ref{figsphere} with $x^{\mu} = z \mathbf{n}$. Using radial quantization, the bulk channel conformal block can be written as an overlap between states  \cite{Hogervorst:2013sma}
 \begin{equation}
 \label{radialquantization1}
 g_{\Delta,J}^{\text{B}}(z, x)=(4z)^{-\Delta_2-\Delta_3}\langle \widehat{0}|\widehat{\mathcal{O}}_{1,\text{cyl.}} (\mathbf{n}_1) |\mathcal{O}_{\Delta,J}|e^{-\tau D} \mathcal{O}_{2,\text{cyl.}}(\mathbf{n}_2) \mathcal{O}_{3,\text{cyl.}}(-\mathbf{n}_2) | 0\rangle\;,
 \end{equation}
where $\langle\widehat{0}|$ is the vacuum at boundary. Here $(z,x)$ are the radial coordinates in (\ref{radialcoord2}), and the prefactor already takes into account  the Weyl transformation from the sphere to the cylinder. $D$ is the dilatation operator which acts on the cylinder as the Hamiltonian. $|\mathcal{O}_{\Delta,J}|$ is a projector onto the conformal multiplet of which the primary is $\mathcal{O}_{\Delta,J}$. It can be written as a sum over descendants as
\begin{equation}
    |\mathcal{O}_{\Delta,J}|=\sum |m,j\rangle \langle m,j|\;,
\end{equation}
where each level-$m$ descendant $|m,j\rangle$ has dimension $\Delta+m$ and spin $j$ taking values
\begin{equation}
\label{rangej}
j \in\{J+m, J+m-2, \ldots, \max (J-m, J+m \bmod 2)\}\;.
\end{equation}
The contribution of the descendant $|m,j\rangle$ in the conformal block (\ref{radialquantization1}) is
\begin{equation}
z^{\Delta+m}\langle\widehat{0}|\widehat{\mathcal{O}}_{1,\text{cyl.}} (\mathbf{n}_1)| m,j \rangle_{\mu_1\cdots\mu_j}\langle m,j |^{\mu_1\cdots\mu_j} \mathcal{O}_{2,\text{cyl.}}(\mathbf{n}_2) \mathcal{O}_{3,\text{cyl.}}(-\mathbf{n}_2) | 0\rangle\;.
\end{equation}
Note that the two overlaps are determined by rotational symmetry, up to some constant coefficients
\begin{eqnarray}
&&\langle\widehat{0}|\widehat{\mathcal{O}}_{1,\text{cyl.}} (\mathbf{n}_1)| m,j \rangle_{\mu_1\cdots\mu_j} \propto \mathbf{n}_{1\mu_1} \cdots\mathbf{n}_{1\mu_j}-\text{traces}\;,\\
    && \langle m,j |^{\mu_1\cdots\mu_j} \mathcal{O}_{2,\text{cyl.}}(\mathbf{n}_2) \mathcal{O}_{3,\text{cyl.}}(-\mathbf{n}_2) | 0\rangle \propto \mathbf{n}_{2}^{\mu_1} \cdots\mathbf{n}_{2}^{\mu_j}-\text{traces}\;.
\end{eqnarray}
Therefore, we have the following ansatz for the radial expansion of the BB$\partial$ bulk channel conformal block\footnote{Here we have assumed that we are in the generic case where both $m$ and $J+k$ are high enough. Otherwise the terms lying outside the range of (\ref{rangej}) vanish.} 
\begin{equation}
\label{radial expansion}
g_{\Delta,J}^{\text{B}}(z,x) = (4z)^{\Delta-\Delta_2-\Delta_3} \sum_{m=0}^{\infty} z^m \sum_{\substack{k=-m \\ k+m=0,2,\cdots}}^{m} A_{m,k}\,\widehat{C}_{J+k}^{\nu} ( x )\;, \quad A_{0,k}=\delta_{0,k}\;,
\end{equation}
where $\nu=d/2-1$. We have also used the fact that the normalized Gegenbauer polynomials can be written in terms of the inner products of direction vectors
\begin{equation}
\begin{gathered}
|\mathbf{n}|^j|\mathbf{m}|^j \widehat{C}_{j}^{\nu}\left(\frac{\mathbf{n} \cdot \mathbf{m}}{|\mathbf{n}||\mathbf{m}|}\right)=(\mathbf{n}^{\mu_1} \cdots \mathbf{n}^{\mu_j}-\text { traces })(\mathbf{m}_{\mu_1} \cdots \mathbf{m}_{\mu_j}-\text { traces })\;.
\end{gathered}
\end{equation}
Finally, $A_{m,k}$ are unknown parameters and we have chosen the  normalization $A_{0,k}=\delta_{0,k}$ for the leading level coefficients.\footnote{This normalization is compatible with the asymptotic behavior (\ref{asybulkblock}) of $g_{\Delta,J}^{\text{B}}(z,x)$.}

\begin{figure}
    \centering
    \includegraphics[width=0.68\linewidth]{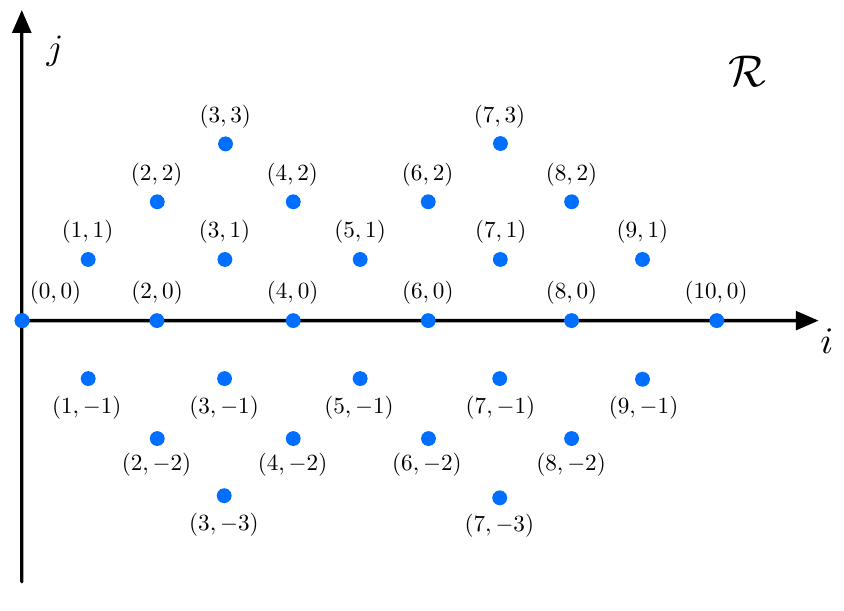}
    \caption{The points $\left(i,j\right)$ inside the region $\mathcal{R}$ which parameterize the coefficients $c_{i,j}$.}
    \label{recurregion}
\end{figure}

To determine the coefficients $A_{m,k}$,  we will use the bulk channel Casimir equation
\begin{equation}
\label{bulkcasimir}
\bigg(\frac{1}{2}(\mathbf{L}_2+\mathbf{L}_3)^2+\Delta(\Delta-d)+J(J+d-2) \bigg) G_{\Delta,J}^{\text{B}}=0\;,
\end{equation}
where $\mathbf{L}_2$ and $\mathbf{L}_3$ are generators of the full $SO(d+1,1)$ conformal group acting on point 2 and 3. It can be rewritten as a differential equation with respect to the $(z,x)$ cross ratios acting on the stripped conformal block, which we schematically denote as 
\begin{equation}
    \mathbf{Cas}^{\rm B} g_{\Delta,J}^{\text{B}}(z,x)=0\;.
\end{equation}
Substituting the ansatz (\ref{radial expansion}) into this equation, it is clear that we can solve the $A_{m,k}$ order by order starting from low $m$ by first Taylor expanding in $z$ and then expanding in $x$. However, a more systematic approach is to exploit the recursion relation \cite{Hogervorst:2013sma,Lauria:2017wav}. The idea is that the action of Casimir on the order $(m,k)$ term of the ansatz, {\it i.e.}, $\left(\ref{radial expansion}\right)$: $z^m A_{m,k}\widehat{C}_{J+k}^{\nu}(x)$, gives rise to a linear combination of such terms with shifted orders. Written schematically, we have
\begin{equation}
\mathbf{Cas^{\text{B}}}\left(z^m A_{m,k}\widehat{C}_{J+k}^{\nu}(x)\right)=-\frac{A_{m,k}}{(-1+z^4)^2(1-2xz+z^2)} {\rm Lin}\Big(\{z^{m+\delta}\widehat{C}_{J+k+\eta}^{\nu} \} \big|_{\delta\in \{0,1,\ldots,10\}}^{\eta\in\{-3,-2,\ldots,3\}}\Big)\;,
\end{equation}
where ${\rm Lin}\left(\cdots\right)$ denotes a particular linear combination of its arguments with specific coefficients. Note that the factor which we extracted in the front is important, otherwise the shift of $m$ will go to infinity. This shift relation implies a recursion relation for the coefficients $A_{m,k}$ with finitely many terms. It has the form 
\begin{equation}
\label{recurs}
\sum_{(i,j) \in \mathcal{R}} c_{i,j} \, A_{m-i,k+j}=0\;,
\end{equation}
where $\mathcal{R}$ denotes the region in Fig. \ref{recurregion}.  The $c_{i,j}$ coefficients are recorded in the $\mathtt{Mathematica}$ notebook, where we also record the results of $A_{m,k}$ up to $m=4$.\footnote{It is straightforward to solve the recursion relation (\ref{recurs})  up to a given order using the algorithm in the ancillary file of \cite{Lauria:2017wav}.} As we can see, $A_{m,k}$ becomes more and more complicated as  we increase $m$. Obtaining a closed form expression for the general expansion coefficients is difficult.  Here we only list the $A_{m,k}$ coefficients for the first two levels which are relatively simple
\begin{equation}
\begin{gathered}
A_{0,0}=1\;,\quad A_{1,1}=- \frac{2(\widehat{\Delta}_1(\Delta+\Delta_2-\Delta_3)+J(\widehat{\Delta}_1+\Delta_2-\Delta_3))}{\Delta+J}\;,\\
A_{1,-1}=-\frac{J(J+2\nu-1)(-\widehat{\Delta}_1(\Delta+\Delta_2-\Delta_3)+(J+2\nu)(\widehat{\Delta}_1+\Delta_2-\Delta_3))}{2(J+\nu-1)(J+\nu)(J-\Delta+2\nu)}\;.
\end{gathered}
\end{equation}

Note that the above method is particularly efficient for computing spinning conformal blocks because it does not depend much on the spin $J$. However, since obtaining the general expansion coefficients  is difficult, the result is not very convenient in certain analytic applications which require all order results. It would therefore be good to have other complementary methods. In Section \ref{Boundary channel CB}, we used an approach based on computing CPWs. This method can also be adapted here for the computation of the bulk channel 
 scalar conformal block. We will not give the explicit calculations but only present the final result for the  $J=0$ case, which reads\footnote{In \cite{Behan:2020nsf}, the authors also derived the BB$\partial$ bulk channel conformal block for two external bulk free fields $\phi$ and a boundary operator $\widehat{\mathcal{O}}$ with scalar exchange, which is a special case of our result (\ref{blockbulkscalar}). If two bulk operators have the same perpendicular distance ($\xi_2 = 1$), our result (\ref{blockbulkscalar}) reproduces their (B.13). However their conformal block for general configuration (B.12) seems to contain a typo and is slightly different from ours (\ref{blockbulkscalar}). More precisely, $\xi$ in their summand should be $\xi y_1/y_2$.} 
\begin{equation}
	\begin{aligned}
		\label{blockbulkscalar}
		&g_{\Delta,0}^{\text{B}}(\xi_1,\xi_2)=\xi_1^{\frac{\Delta-\Delta_{2}-\Delta_{3}}{2}} \times \\&\sum_{n=0}^{\infty} \bigg[\xi_{1}^{n}\frac{\pi(-\frac{1}{4})^{n} \Gamma(\Delta)  \csc (\pi(\Delta-\widehat{\Delta}_{1}+n)) \Gamma\big(\frac{2 n+\Delta+\Delta_{2}-\Delta_{3}}{2}\big) \Gamma\big(\frac{2 n+\Delta-\Delta_{2}+\Delta_{3}}{2}\big) \Gamma\big(\frac{d-2(n+\Delta)}{2}\big)}{n!\Gamma(\Delta-\widehat{\Delta}_{1}) \Gamma\big(\frac{\Delta+\Delta_{2}-\Delta_{3}}{2}\big) \Gamma\big(\frac{\Delta-\Delta_{2}+\Delta_{3}}{2}\big)  \Gamma(\frac{d}{2}-\Delta) \Gamma(2 n+\Delta) \Gamma(-2 n-\Delta+\widehat{\Delta}_{1}+1)}\\&\quad\times { }_{2} F_{1}\bigg(\widehat{\Delta}_{1}, \frac{2 n+\Delta+\Delta_{2}-\Delta_{3}}{2} ; 2 n+\Delta ; 1-\xi_{2}\bigg) \Bigg]\;.
	\end{aligned}
\end{equation}
One can check that the conformal block is symmetric under $2\leftrightarrow 3$ exchange and the bulk channel Casimir equation (\ref{bulkcasimir}) is satisfied order by order. Furthermore, taking the limit $\widehat{\Delta}_1 \to 0$, we find that $g_{\Delta,0}^{\text{B}}(\xi_1,\xi_2)$ reduces to the bulk channel conformal block for two-point functions (\ref{bulkCB2pt}). Unlike the expansion in the radial coordinates, the CPW method is sensitive to the spin. We will not further pursue the higher spin extension of this method in this paper and will leave it for future work.

\section{Witten diagrams: The B$\partial\partial$ case}
\label{Witten diagrams Bpp case}
Let us now use these results for conformal blocks to analyze the Witten diagrams which arise in a simple model for holographic boundary (or interface) CFTs. The setup is the same one as in \cite{DeWolfe:2001pq,Aharony:2003qf,Rastelli:2017ecj,Mazac:2018biw}, and is the simplest version of the Karch-Randall setup \cite{Karch:2000gx,Karch:2001cw}. In the Poincar\'e coordinates 
\begin{equation}
	\label{AdSmetric}
	d s^2=\frac{d z_0^2+d \vec{z}\;\!{}^2+d z_{\perp}^2}{z_0^2}\;,
\end{equation}
the holographic dual of the boundary or the interface is the $AdS_d$ subspace at $z_\perp=0$. On this subspace, there are localized degrees of freedom which can interact via local interactions with the fields in the bulk $AdS_{d+1}$. However, the presence of these localized degrees of freedom does not back-react to the geometry in the bulk. Note that when considering boundary CFTs, the space terminates at $z_\perp=0$ and we have only a half $AdS_{d+1}$ space. One imposes boundary conditions ({\it e.g.}, Dirichlet or Neumann) on the $z_\perp=0$ $AdS_d$ slice. However, one can use the method of images to double the half space to the full $AdS_{d+1}$ space (see, {\it e.g.}, \cite{Mazac:2018biw}). Therefore, it is sufficient to discuss the interface case where we have a probe $AdS_d$ probe brane inside $AdS_{d+1}$. As we mentioned in the introduction, these Witten diagrams are relevant for the functional approach for the BCFT two-point functions \cite{Kaviraj:2018tfd,Mazac:2018biw}. Recently, it was also realized certain Witten diagrams in this setup enjoy an infinite dimensional Yangian symmetry \cite{Rigatos:2022eos}. 

As in Section \ref{Kinematics of BCFT three-point functions}, the analysis of the Witten diagrams will also be facilitated by using the embedding space. The embedding vector associated to a point in $AdS_{d+1}$ is 
\begin{equation}
		Z^{A}=\frac{1}{z_{0}}\left(\frac{1+z_{0}^{2}+\vec{z}\;\!{}^{2}+z_{\perp}^2}{2}, \frac{1-z_{0}^{2}-\vec{z}\;\!{}^{2}-z_{\perp}^2}{2}, \vec{z},z_{\perp}\right)\;,
\end{equation}
satisfying $Z^2=-1$. In terms of the embedding vectors, the bulk-to-boundary propagator reads 
\begin{equation}
  G_{B\partial}^{\Delta_i}(P,Z)=\frac{1}{\left(-2P\cdot Z\right)^{\Delta_i}}=\bigg(\frac{z_0}{z_0^2+\left(\vec{z}-\vec{x}\right)^2+\left(z_\perp-x_\perp\right)^2}\bigg)^{\Delta_i}\;.  
\end{equation}
The (scalar) bulk-to-bulk propagator in $AdS_{d+1}$ is
\begin{equation}
    G_{BB,(d+1)}^{\Delta} (Z,W)= \frac{\Gamma(\Delta)}{2 \pi^{\frac{d}{2}} \Gamma\left(\Delta-\frac{d}{2}+1\right)} u^{-\Delta}{}_2 F_1\bigg(\Delta, \frac{2 \Delta-d+1}{2}, 2 \Delta-d+1,-\frac{4}{u}\bigg)\;,
\end{equation}
where $u$ is defined as
\begin{equation}
u=-2-2 Z \cdot W =\frac{(\vec{z}-\vec{w})^2+\left(z_{\perp}-w_{\perp}\right)^2+\left(z_{0}-w_{0}\right)^2}{z_{0} w_{0}}\;.
\end{equation}
The bulk-to-bulk propagator satisfies the equation of motion
\begin{equation}
\big(\square_{d+1}+\Delta(\Delta-d)\big) G_{B B,(d+1)}^{\Delta}(z, w)=\delta^{(d+1)}(z, w)\;,
\end{equation}
where $\square_{d+1}$ is the $AdS_{d+1}$ Laplacian. In the following, we will first consider the simpler case of B$\partial\partial$ three-point functions. Witten diagrams that correspond to the BB$\partial$ three-point functions will be considered in Section \ref{Witten diagrams: The BBp case}.

\subsection{Contact Witten diagram}
\begin{figure}
	\centering
	\includegraphics[width=0.71\linewidth]{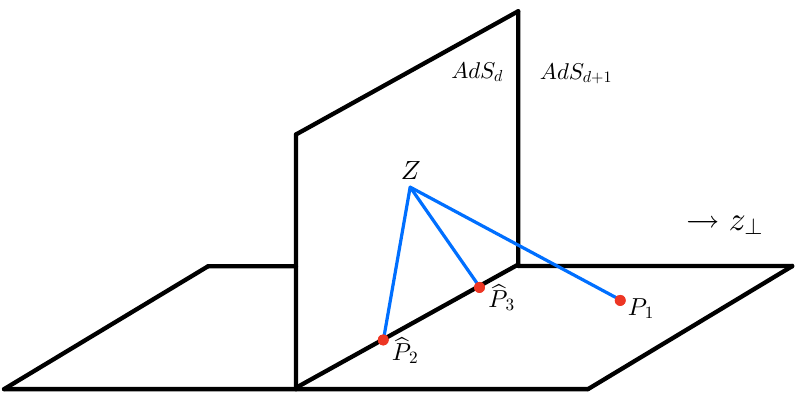}
	\caption{The contact Witten diagram for the B$\partial\partial$ three-point function, where the vertex is inserted at $Z$ and is integrated over the $AdS_{d}$ brane.}
	\label{Bppcontact}
\end{figure}

The simplest diagram is a contact Witten diagram with zero derivatives which arises from a cubic vertex on $AdS_d$ (Fig. \ref{Bppcontact}). It corresponds to the following integral 
\begin{equation}
\begin{aligned}
	\label{Bppcontactintegral}
W_{\Delta_1,\widehat{\Delta}_2,\widehat{\Delta}_3}^{\text{contact}}(P_1, \widehat{P}_{2}, \widehat{P}_{3}) 
=&\int_{A d S_{d}} dZ \,G_{B\partial}^{\Delta_1}(P_1,Z) G_{B\partial}^{\widehat{\Delta}_2}(\widehat{P}_2,Z) G_{B\partial}^{\widehat{\Delta}_3}(\widehat{P}_3,Z)\\\equiv&\int_{A d S_{d}} d Z\,(-2 P_1 \cdot Z)^{-\Delta_1}(-2 \widehat{P}_{2} \cdot Z)^{-\widehat{\Delta}_{2}}(-2 \widehat{P}_{3} \cdot Z)^{-\widehat{\Delta}_{3}},
\end{aligned}
\end{equation}
which is a generalization of $D$-functions in CFTs without boundaries. This diagram can be evaluated by first using the Schwinger parameterization and then integrating out the $AdS_d$ coordinates $Z$. We find\footnote{In the following, we use $\mathcal{W}$ to denote the Witten diagrams which have kinematic factors stripped off according to (\ref{2}) and (\ref{BBpartialstripped}).}
\begin{equation}
\label{contact}
\mathcal{W}_{\Delta_1,\widehat{\Delta}_2,\widehat{\Delta}_3}^{\text{contact}}(\xi) = \frac{\pi^{h^{\prime}}\Gamma\big(\frac{\sum_i\Delta_i-2h^{\prime}}{2}\big)\Gamma\big(\frac{\sum_i\Delta_i}{2}\big)}{2\,\Gamma(\Delta_1)\Gamma(\widehat{\Delta}_2)\Gamma(\widehat{\Delta}_3)}  \int_{0}^{\infty} \prod_{i} d t_{i} t_{i}^{\Delta_{i}-1} \frac{\delta(\sum_{i} t_{i}-1)}{(t_1^2+t_1t_2+t_1t_3+\xi t_2t_3)^{\frac{\sum_i\Delta_i}{2}}}\;,
\end{equation}
where $\sum_{i} \Delta_i=\Delta_1+\widehat{\Delta}_2+\widehat{\Delta}_3$. Using the MB representation (\ref{MBrep}) as before, we get 
\begin{eqnarray}
\nonumber \mathcal{W}_{\Delta_1,\widehat{\Delta}_2,\widehat{\Delta}_3}^{\text{contact}}(\xi)&=&\frac{\pi^{h^{\prime}}\Gamma\big(\frac{\sum_i\Delta_i-2h^{\prime}}{2}\big)}{2\prod_i \Gamma(\Delta_i)} 
\int \frac{d \tau}{2\pi i} \;\Gamma(-\tau)\Gamma(-\Delta_{23,1}-\tau)\Gamma( \Delta _ { 2 } + \tau ) \Gamma(\Delta_{3}+\tau) \xi^{\tau}\\
\label{Wc}&=&\frac{\pi^{h^{\prime}}}{2} \Gamma\bigg(\frac{\sum_i\Delta_i-2h^{\prime}}{2}\bigg)   \bigg[\frac{\Gamma(-\Delta_{23,1})}{\Gamma(\Delta_1)}{}_2 F_{1}(\widehat{\Delta}_2,\widehat{\Delta}_3;1+\Delta_{23,1};\xi)\\
\nonumber &&+\frac{\Gamma(\Delta_{12,3})\Gamma(\Delta_{13,2})\Gamma(\Delta_{23,1})}{\Gamma(\Delta_1)\Gamma(\widehat{\Delta}_2)\Gamma(\widehat{\Delta}_3)}\xi^{\frac{\Delta_1-\widehat{\Delta}_2-\widehat{\Delta}_3}{2}}{}_2 F_{1}(\Delta_{12,3},\Delta_{13,2};1-\Delta_{23,1};\xi)\bigg]\;,
	\end{eqnarray}
where $\Delta_{ij,k}=\frac{\Delta_i+\Delta_j-\Delta_k}{2}$. Note that this diagram has also been computed as a conformal Feynman integral in flat space in \cite{Loebbert:2020glj}, where the result is in complete agreement with ours.\footnote{General contact Witten diagrams were first calculated in Mellin space in \cite{Rastelli:2017ecj}. The equivalence of the conformal integrals with the AdS contact Witten diagrams was first pointed out in \cite{Rigatos:2022eos}.}

This explicit expression for the contact Witten diagram allows us to straightforwardly decompose it into conformal blocks. For generic external dimensions, we find the diagram decomposes into conformal blocks which are associated with $\partial_\perp^{2n}\mathcal{O}_1$ and $\widehat{\mathcal{O}}_2\widehat{\square}^n\widehat{\mathcal{O}}_3$ where $\widehat{\square}$ is the Laplacian on the boundary. Explicitly, the conformal block decomposition reads
\begin{equation}
\label{BppcontactCBdecom}
\mathcal{W}_{\Delta_1,\widehat{\Delta}_2,\widehat{\Delta}_3}^{\text{contact}}(\xi)=\sum_{n=0}^{\infty}A_n^{\text{B}\partial\partial}\,g_{\Delta_1+2n}(\xi)+\sum_{n=0}^{\infty}B_n^{\text{B}\partial\partial}\,g_{\widehat{\Delta}_2+\widehat{\Delta}_3+2n}(\xi)\;,
\end{equation}
where the OPE coefficients are
	\begin{equation}
	\begin{aligned}
	\label{opecontact}
&A_n^{\text{B}\partial\partial}\\&=\frac{(-1)^{n}\pi ^{h^{\prime}}\Gamma(-n+\Delta_{23,1} )\Gamma(n+\Delta_{12,3})\Gamma(n+\Delta_{13,2})\Gamma(-h^{\prime}+n+\Delta_1)\Gamma\big(\frac{\sum_i\Delta_i-2h^{\prime}}{2}+n\big)}{2n!\Gamma(\Delta_1)\Gamma(\widehat{\Delta}_2)\Gamma(\widehat{\Delta}_3)\Gamma(-h^{\prime}+2 n+\Delta _1)}\;,\\&B_{n}^{\text{B}\partial\partial}\\&=\frac{(-1)^{n}\pi ^{h^{\prime}}\Gamma(-n-\Delta_{23,1}) \Gamma (n+\widehat{\Delta} _2) \Gamma(n+\widehat{\Delta} _3) \Gamma (-h^{\prime}+n+\widehat{\Delta} _2+\widehat{\Delta}
		_3) \Gamma\big(\frac{\sum_i\Delta_i-2h^{\prime}}{2}+n\big)}{2n! \Gamma (\Delta _1) \Gamma (\widehat{\Delta} _2) \Gamma (\widehat{\Delta} _3) \Gamma (-h^{\prime}+2 n+\widehat{\Delta} _2+\widehat{\Delta}_3)}\;.
	\end{aligned}
	\end{equation}
Note that the two series of conformal blocks are related to each other by $\Delta_1 \leftrightarrow \widehat{\Delta}_2+\widehat{\Delta}_3$,  thanks to the symmetry in the decomposition coefficients
\begin{equation}
A_n^{\text{B}\partial\partial}\big|_{\Delta_1 \leftrightarrow\widehat{\Delta}_2+\widehat{\Delta}_3}=B_n^{\text{B}\partial\partial}\;.
\end{equation}
The form of this decomposition (as well as its symmetry) is reminiscent of the conformal block decomposition of a contact four-point function in pure $AdS_{d+1}$ (see, {\it e.g.}, (3.8) and (3.9) of \cite{Zhou:2018sfz}). Similar to $\mathcal{O}_2\widehat{\square}^n\mathcal{O}_3$, the  operators $\partial_\perp^{2n}\mathcal{O}_1$  can be viewed as a new type of ``double-trace'' operators for BCFTs.

For the special situation where the two towers of conformal dimensions overlap, {\it i.e.},  $\Delta_1=\widehat{\Delta}_2+\widehat{\Delta}_3+2m$,  with $m \in \mathbb{Z}$, the conformal blocks $g_{\Delta_1+2n}(\xi)$ and $g_{\widehat{\Delta}_2+\widehat{\Delta}_3+2n}(\xi)$ become degenerate and the coefficients become singular. This case can be obtained from the general case by taking the limit. In doing so, we find the derivative of conformal blocks appearing in the decomposition. Due to the symmetry in the two series of conformal blocks, let us assume without loss of generality that $m\geq0$. The decomposition then becomes
\begin{equation}
\mathcal{W}_{\Delta_1,\widehat{\Delta}_2,\widehat{\Delta}_3}^{\text{contact}}(\xi)=\sum_{n=0}^{\infty}C_n^{\text{B}\partial\partial}\,g_{\widehat{\Delta}_{2}+\widehat{\Delta}_3+2n}(\xi)+\sum_{n=0}^{\infty}D_n^{\text{B}\partial\partial}\,\partial_{\Delta}g_{\widehat{\Delta}_2+\widehat{\Delta}_3+2m+2n}(\xi)\;,
\end{equation}
where the new coefficients are
\begin{equation}
\begin{aligned}
&C_n^{\text{B}\partial\partial}\\&=\left\{\begin{array}{l}
\frac{(-1)^{n} \pi ^{h^{\prime}} \Gamma (m-n) \Gamma (n+\widehat{\Delta}_2) \Gamma(n+\widehat{\Delta} _3) \Gamma(-h^{\prime}+n+\widehat{\Delta} _2+\widehat{\Delta} _3) \Gamma (-h^{\prime}+m+n+\widehat{\Delta}
	_2+\widehat{\Delta} _3)}{2 n! \Gamma (\widehat{\Delta} _2) \Gamma (\widehat{\Delta} _3) \Gamma(2 m+\widehat{\Delta} _2+\widehat{\Delta} _3) \Gamma (-h^{\prime}+2 n+\widehat{\Delta} _2+\widehat{\Delta} _3)}\;, \quad n<m \\
\frac{(-1)^m \pi ^{h^{\prime}} \Gamma (n+\widehat{\Delta} _2) \Gamma (n+\Delta _3) \Gamma(-h^{\prime}+n+\widehat{\Delta} _2+\widehat{\Delta} _3) \Gamma (-h^{\prime}+m+n+\widehat{\Delta} _2+\widehat{\Delta} _3)}{2n!
	\Gamma(\widehat{\Delta} _2) \Gamma (\widehat{\Delta }_3)  \Gamma (2 m+\widehat{\Delta} _2+\widehat{\Delta} _3) \Gamma (-m+n+1) \Gamma (-h^{\prime}+2 n+\widehat{\Delta} _2+\widehat{\Delta} _3)} \\\times\left(-H_{\frac{-d-1}{2}+m+n+\widehat{\Delta} _2+\widehat{\Delta}
	_3} - H_{\frac{-d-1}{2}+n+\widehat{\Delta}_2+\widehat{\Delta}_3}+2 H_{\frac{-d-1}{2}+2 n+\widehat{\Delta}_2+\widehat{\Delta}_3}\right.\\\left.\quad+H_{n-m}-H_{n+\widehat{\Delta} _2-1}-H_{n+\widehat{\Delta}_3-1}+H_n\right)\;, \quad\quad\quad\quad\quad\quad\quad\quad\quad\quad\, n\geq m\\
\end{array}\right.
\end{aligned}
\end{equation}
	\begin{equation}
	\begin{aligned}
	&D_n^{\text{B}\partial\partial}=(-1)^m \pi ^{h^{\prime}}\\&\times\frac{ \Gamma (m+n+\widehat{\Delta} _2) \Gamma (m+n+\widehat{\Delta} _3) \Gamma(-h^{\prime}+m+n+\sum_i \widehat{\Delta}_i) \Gamma (-h^{\prime}+2 m+n+\sum_i \widehat{\Delta}_i)}{  n! \Gamma (\widehat{\Delta} _2) \Gamma(\widehat{\Delta} _3) \Gamma (m+n+1)\Gamma(2 m+\sum_i \widehat{\Delta}_i) \Gamma (-h^{\prime}+2 m+2 n+\sum_i \widehat{\Delta}_i)}\;,
	\end{aligned}
	\end{equation}
where $\sum_i\widehat{\Delta}_i\equiv\widehat{\Delta}_2+\widehat{\Delta}_3$. Here $H_n$ is the Harmonic number $H_n=\sum_{r=1}^n \frac{1}{r}$.

\subsection{Exchange Witten diagram}

\begin{figure}
	\centering
	\includegraphics[width=0.71\linewidth]{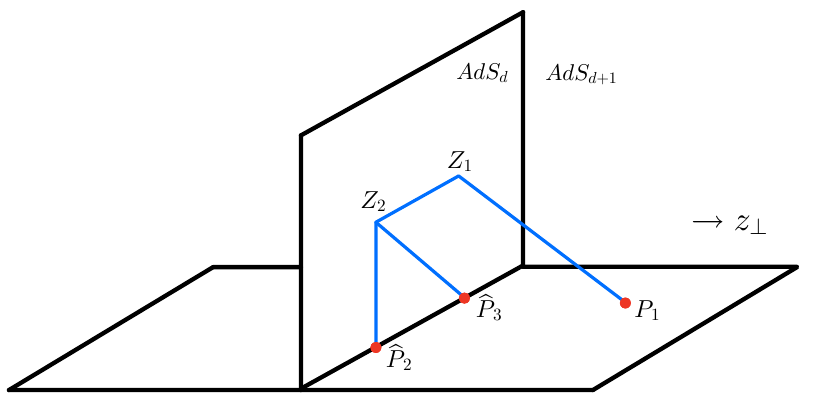}
	\caption{The exchange Witten diagram for the B$\partial\partial$ three-point function, where the vertex insertion points $Z_1$ and $Z_2$ in $AdS_{d}$ should be integrated over.}
	\label{Bppexchange}
\end{figure}

The next simplest diagram is the following exchange Witten diagram depicted in Fig. \ref{Bppexchange}
\begin{equation}
\begin{aligned}
\label{exch}
&W_{\widehat{\Delta}}^{\text{B}\partial\partial,\text{exchange}}(P_1, \widehat{P}_{2}, \widehat{P}_{3})\\&=\int_{A d S_{d}} d Z_{1} \int_{A d S_{d}} d Z_{2} \,G_{B \partial}^{\Delta_{1}}(P_{1}, Z_{1}) G_{B B, (d)}^{\widehat{\Delta}}(Z_{1}, Z_{2}) G_{B \partial}^{\widehat{\Delta}_{2}}(\widehat{P}_{2}, Z_{2})G_{B \partial}^{\widehat{\Delta}_{3}}(\widehat{P}_{3}, Z_{2})\;,
\end{aligned}
\end{equation}
which exchanges a scalar field living on the $AdS_d$ brane. This integral is difficult to compute in a closed form in position space due to the complexity of the bulk-to-bulk propagator. However, it is relatively simple to calculate its conformal block decomposition by exploiting the equation of motion (EOM) relation \cite{Zhou:2018sfz}. Let us define the following operator
\begin{equation}
\label{defofeom}
\mathbf{EOM}_{\widehat{\Delta}}^{\text{B}\partial\partial}=\frac{1}{2} (\widehat{\mathbf{L}}_{2}+\widehat{\mathbf{L}}_{3})^2+\widehat{\Delta}(\widehat{\Delta}-d+1)\;,
\end{equation}
which is the same as the boundary channel two-particle conformal Casimir up to a constant shift. In the bulk, it coincides with the EOM operator acting on the bulk-to-bulk propagator. As a result, the EOM operator turns the exchange Witten diagram into a contact Witten diagram
\begin{equation}
\label{eom}
\mathbf{EOM}_{\widehat{\Delta}}^{\text{B}\partial\partial}\left(W_{\widehat{\Delta}}^{\text{B}\partial\partial,\text{exchange}}(P_1, \widehat{P}_{2}, \widehat{P}_{3}) \right)=W_{\Delta_1,\widehat{\Delta}_2,\widehat{\Delta}_3}^{\text{contact}}(P_1, \widehat{P}_{2}, \widehat{P}_{3})\;.
\end{equation}
The derivation of this relation is almost identical to that in Section 2.2  of \cite{Zhou:2018sfz} so we will not repeat it here. Essentially, the conformal invariance of the $Z_2$ integral allows us to pass the action of the Casimir to the bulk-to-bulk propagagtor where it becomes just the AdS Laplacian. Making use of the EOM of the bulk-to-bulk propagator, this yields a delta function and we get a contact diagram. Written explicitly, the action of the operator on the Witten diagram reads
\begin{equation}
\begin{split}
&\mathbf{EOM}_{\widehat{\Delta}}^{\text{B}\partial\partial}\left(\mathcal{W}_{\widehat{\Delta}}(\xi)\right)\\&=4 (\xi -1) \xi^2  \mathcal{W}_{\widehat{\Delta}}''(\xi )+2 \xi\left(2(\widehat{\Delta}_2+\widehat{\Delta}_3+1)\xi+d-2\widehat{\Delta}_2-2\widehat{\Delta}_3-3\right) \mathcal{W}_{\widehat{\Delta}}'(\xi )\\&\quad+\left((\widehat{\Delta} -\widehat{\Delta}_2-\widehat{\Delta}_3) (-d+1+\widehat{\Delta}+\widehat{\Delta}_2+\widehat{\Delta}
_3)+4 \widehat{\Delta} _2 \widehat{\Delta} _3 \xi \right)\mathcal{W}_{\widehat{\Delta}}(\xi )\;,
\end{split}
\end{equation}
where the upper label in $\mathcal{W}_{\widehat{\Delta}}^{\text{B}\partial\partial,\text{exchange}}(\xi)$ has been omitted for notation simplicity.

 From the conformal block decomposition of the contact Witten diagram (\ref{BppcontactCBdecom}), the EOM relation gives the following decomposition of the exchange diagram
\begin{equation}
\label{expan}
\mathcal{W}_{\widehat{\Delta}}^{\text{B}\partial\partial,\text{exchange}}(\xi)=a^{\text{B}\partial\partial} \,g_{\widehat{\Delta}}(\xi)+\sum_{n=0}^{\infty}a_n^{\text{B}\partial\partial}\, g_{\Delta_1+2n}(\xi)+\sum_{n=0}^{\infty}b_n^{\text{B}\partial\partial}\, g_{\widehat{\Delta}_2+\widehat{\Delta}_3+2n}(\xi)\;.
\end{equation}
In addition to the double-trace operator conformal blocks there is also a single-trace conformal block which corresponds to the exchanged field in AdS and is annihilated by the EOM operator. 
The double-trace coefficients are related to those of the contact Witten diagram by
\begin{equation}
\begin{split}
\label{Bppexchdouble}
&a_n^{\text{B}\partial\partial}=\frac{A_n^{\text{B}\partial\partial}}{\widehat{\Delta}(\widehat{\Delta}
	-d+1)-(\Delta_1+2n)(\Delta_1+2n-d+1)}\;,\\ &b_n^{\text{B}\partial\partial}=\frac{B_n^{\text{B}\partial\partial}}{\widehat{\Delta}(\widehat{\Delta}-d+1)-(\widehat{\Delta}_2+\widehat{\Delta}_3+2n)(\widehat{\Delta}_2+\widehat{\Delta}_3+2n-d+1)}\;.
\end{split}
\end{equation}
Since the single-trace conformal block belongs to the kernel of the EOM operator, the associated coefficient $a^{\text{B}\partial\partial}$ is not determined by this relation. However, it can be computed by using an alternative method which is based on the split representation of the bulk-to-bulk propagator and the result is 
\begin{equation}
 \begin{split}
&a^{\text{B}\partial\partial}=\\&\frac{\pi^{h^{\prime}}\Gamma \big(\frac{-\widehat{\Delta}+\Delta _1}{2}\big)\Gamma \big(\frac{-\widehat{\Delta} +\widehat{\Delta}_2+\widehat{\Delta}_3}{2}\big) \Gamma \big(\frac{\widehat{\Delta} +\widehat{\Delta}_2-\widehat{\Delta}_3}{2}
		\big) \Gamma \big(\frac{\widehat{\Delta} -\widehat{\Delta}
			_2+\widehat{\Delta} _3}{2} \big) \Gamma \big(\frac{-2h^{\prime}+\widehat{\Delta} +\Delta _1}{2}\big) \Gamma \big(\frac{-2h^{\prime}+\widehat{\Delta}+\widehat{\Delta}_2+\widehat{\Delta} _3}{2}
		\big)}{8\,\Gamma(\Delta_{1}) \Gamma (\widehat{\Delta} _2) \Gamma(\widehat{\Delta}_3) \Gamma (1-h^{\prime}+\widehat{\Delta}
		)}\;.
  \end{split}
  \end{equation}
The calculation is presented in Appendix \ref{appendixspectral}, where we also reproduce the double-trace coefficients using this method.

\subsubsection{Dimensional reduction relations}
In \cite{Zhou:2020ptb} it was found that there is an intimate connection between the recursion relations of conformal blocks and those of exchange Witten diagrams. Essentially, every recursion for conformal blocks can induce a relation for exchange Witten diagrams which is obtained by replacing the conformal blocks with  the corresponding exchange Witten diagrams with the same exchanged dimension and spin. This correspondence was demonstrated in  \cite{Zhou:2020ptb} for four-point functions in CFTs without boundaries and BCFT two-point functions, and was further extended to two-point functions in $\mathbb{RP}^d$ CFTs in \cite{Giombi:2020xah}. As a byproduct of our analysis, here we will show that the same correspondence also holds for B$\partial\partial$ three-point functions in BCFTs. 

We will focus on a particular family of recursion relations, namely the dimensional reduction relations. Our starting point is the following relations for conformal blocks between $d$ and $d-1$ dimensions
\begin{equation}
\begin{split}
\label{recur1}
g_{\widehat{\Delta}}^{(d)}(\xi)={}&\sum_{j=0}^{\infty} \alpha_{j}^{(d)}(\widehat{\Delta}) \,g_{\widehat{\Delta}+2 j}^{(d-1)}(\xi)\;,\\\alpha_{j}^{(d)}(\widehat{\Delta})={}&\frac{\Gamma(j+\frac{1}{2})\left(\frac{\widehat{\Delta}+\widehat{\Delta}_{2}-\widehat{\Delta}_{3}}{2}\right)_{j}\left(\frac{\widehat{\Delta}-\widehat{\Delta}_{2}+\widehat{\Delta}_{3}}{2}\right)_{j}}{\sqrt{\pi} j !\left(\frac{-d+2 \widehat{\Delta}+3}{2}\right)_{j}\left(j+\frac{-d+2 \widehat{\Delta}+2}{2}\right)_{j}}\;,
\end{split}
\end{equation}
and between $d$ and $d-2$ dimensions\footnote{Similar relations for four-point function conformal blocks in CFTs without boundaries were first found in \cite{Hogervorst:2016hal,Kaviraj:2019tbg}.}
\begin{equation}
\begin{split}
\label{recur2}
g_{\widehat{\Delta}}^{(d-2)}(\xi)={}&g_{\widehat{\Delta}}^{(d)}(\xi)+\beta^{(d)}(\widehat{\Delta}) g_{\widehat{\Delta}+2}^{(d)}(\xi)\;,\\
\beta^{(d)}(\widehat{\Delta})={}&-\frac{(\widehat{\Delta}+\widehat{\Delta}_{2}-\widehat{\Delta}_{3})(\widehat{\Delta}-\widehat{\Delta}_{2}+\widehat{\Delta}_{3})}{(d-2\widehat{\Delta}-5)(d-2 \widehat{\Delta}-3)}\;.
\end{split}
\end{equation}
 Here we have added an additional upper index to distinguish different spacetime dimensions. These relations can be straightforwardly verified by using the explicit expression of the conformal block (\ref{block}). But they are also the same as the dimensional reduction formulas in  the $\mathbb{RP}^{d-1}$ case \cite{Giombi:2020xah}. This is because, as we pointed out at the end of Section \ref{Bpartialpartial case},  the conformal blocks coincide in the two cases.

Following the prescription of \cite{Zhou:2020ptb}, the recursions for Witten diagrams are obtained by replacing the conformal blocks by ``unit-normalized'' exchange Witten diagrams
\begin{equation}
\label{precur1}
\mathcal{P}_{\widehat{\Delta}}^{(d)}(\xi) =\sum_{j=0}^{\infty} \alpha_{j}^{(d)}(\widehat{\Delta}) \,\mathcal{P}_{\widehat{\Delta}+2 j}^{(d-1)}(\xi)\;,
\end{equation}
\begin{equation}
\label{precur2}
\mathcal{P}_{\widehat{\Delta}}^{(d-2)}(\xi) =\mathcal{P}_{\widehat{\Delta}}^{(d)}(\xi)+\beta^{(d)}(\widehat{\Delta}) \mathcal{P}_{\widehat{\Delta}+2}^{(d)}(\xi)\;.
\end{equation}
Here ``unit-normalized'' means that we have renormalized the exchange Witten diagrams  
\begin{equation}
\mathcal{P}_{\widehat{\Delta}}(\xi)=\frac{1}{a^{\text{B}\partial\partial}}\mathcal{W}_{\widehat{\Delta}}^{\text{B}\partial\partial,\text{exchange}}(\xi)\;,
\end{equation}
so that the single-trace conformal block $g^{(d)}_{\widehat{\Delta}}$ appears with the unit coefficient. We should stress that these Witten diagrams for BCFTs are not the same as those appearing in the $\mathbb{RP}^{d-1}$ setup, even though the conformal blocks are identical. These relations for exchange Witten diagrams can be proven by using the conformal block decomposition (\ref{expan})
\begin{equation}
\label{pexpan}
\mathcal{P}_{\widehat{\Delta}}^{(d)}(\xi)= g_{\widehat{\Delta}}^{(d)}(\xi)+\sum_{n=0}^{\infty}\mathbf{a}_n\, g_{\Delta_1+2n}^{(d)}(\xi)+\sum_{n=0}^{\infty}\mathbf{b}_n\, g_{\widehat{\Delta}_2+\widehat{\Delta}_3+2n}^{(d)}(\xi)\;,
\end{equation}
where
\begin{equation}
\mathbf{a}_{n}=\frac{a_n^{\text{B}\partial\partial}}{a^{\text{B}\partial\partial}}\;,\quad\mathbf{b}_{n}=\frac{b_n^{\text{B}\partial\partial}}{a^{\text{B}\partial\partial}}\;.
\end{equation}
Inserting this expansion into (\ref{precur2}), we find that the conformal block $g^{(d)}_{\widehat{\Delta}}$ cancels out by construction thanks to (\ref{recur2}). However, the cancellation of the remaining conformal blocks is nontrivial. Since they are linearly independent  their coefficients must vanish, which gives the following equations 
\begin{equation}
\begin{gathered}
\label{recureq1}
\mathbf{a}_n^{(d-2)}(\widehat{\Delta})+\mathbf{a}_{n-1}^{(d-2)}(\widehat{\Delta})\beta^{(d)}(\Delta_1+2n-2)=\mathbf{a}_n^{(d)}(\widehat{\Delta})+\beta^{(d)}(\Delta)\mathbf{a}_n^{(d)}(\widehat{\Delta}+2)\;,\\\mathbf{b}_n^{(d-2)}(\widehat{\Delta})+\mathbf{b}_{n-1}^{(d-2)}(\widehat{\Delta})\beta^{(d)}(\widehat{\Delta}_2+\widehat{\Delta}_3+2n-2)=\mathbf{b}_n^{(d)}(\widehat{\Delta})+\beta^{(d)}(\widehat{\Delta})\mathbf{b}_n^{(d)}(\widehat{\Delta}+2)\;.
\end{gathered}
\end{equation}
Using our explicit expression (\ref{block}) for the conformal blocks, these identities can be easily verified. Similarly, the recursion relation (\ref{precur1}) gives the following equations
\begin{equation}
\begin{gathered}
\label{recureq2}
\sum_{i=0}^{n}\alpha_{n-i}^{(d)}(\Delta_1+2i)\,\mathbf{a}_i^{(d)}(\widehat{\Delta})=\sum_{j=0}^{\infty}\alpha_{j}^{(d)}(\widehat{\Delta})\,\mathbf{a}_n^{(d-1)}(\widehat{\Delta}+2j)\;,\\\sum_{i=0}^{n}\alpha_{n-i}^{(d)}(\widehat{\Delta}_2+\widehat{\Delta}_3+2i)\,\mathbf{b}_i^{(d)}(\widehat{\Delta})=\sum_{j=0}^{\infty}\alpha_{j}^{(d)}(\widehat{\Delta})\,\mathbf{b}_n^{(d-1)}(\widehat{\Delta}+2j)\;.
\end{gathered}
\end{equation}
While it is difficult to prove these qualities analytically, we can verify them numerically to high accuracy.

\section{Witten diagrams: The BB$\partial$ case}
\label{Witten diagrams: The BBp case}
In this section, we consider tree-level Witten diagrams in the same setup for BB$\partial$ three-point functions.
\subsection{Contact Witten diagram}
\begin{figure}
	\centering
\includegraphics[width=0.7\linewidth]{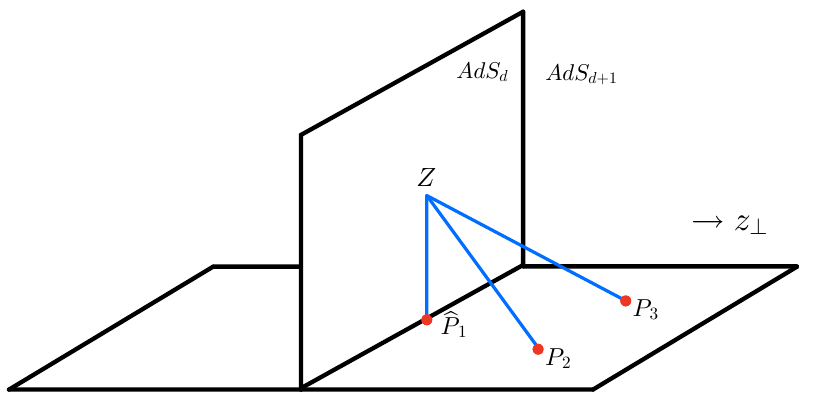}
	\caption{The contact Witten diagram for the BB$\partial$ three-point function, where the vertex insertion point $Z$ is integrated over the $AdS_d$ brane.}
	\label{BBpcontactfig}
\end{figure}

We begin by considering the simplest contact diagram (Fig. \ref{BBpcontactfig}) which arises from a zero-derivative contact vertex localized on the $AdS_d$ subspace
\begin{equation}
	\begin{aligned}
W_{\widehat{\Delta}_1,\Delta_2,\Delta_3}^{\text{contact}}(\widehat{P}_{1}, P_2, P_3)=\int_{A d S_{d}} d Z \,G_{B\partial}^{\widehat{\Delta}_1}(\widehat{P}_1,Z_1)G_{B\partial}^{\Delta_2}(P_2,Z_1) G_{B\partial}^{\Delta_3}(P_3,Z_1)\;.
	\end{aligned}
\end{equation}
This diagram has been studied in \cite{Rastelli:2017ecj} and has a simple expression in terms of the Mellin amplitude (as will be reviewed in Section \ref{Mellin space representation}). But in position space it depends non-trivially on the two cross ratios which makes computing it in a closed form difficult. Our goal here is to obtain the conformal block decomposition of this diagram.

\subsubsection{Boundary channel decomposition}
As was shown in \cite{Rastelli:2017ecj}, the BB$\partial$ contact Witten diagram is just a constant in a properly defined Mellin space. However, there are different choices for the independent variables which appear in the Mellin representation. For the boundary channel, a convenient choice is the following (see (\ref{3pointmellinrep})) 
    \begin{equation}
	\begin{aligned}
		\label{mellinCbdy}
		&\mathcal{W}_{\widehat{\Delta}_1,\Delta_2,\Delta_3}^{\text{contact}}(\zeta_1,\zeta_2)=\int \frac{d\alpha_2 d\alpha_3}{(2\pi i)^2}\bigg[ \zeta_1^{\alpha_2-\Delta_2} \zeta_2^{\alpha_3+\widehat{\Delta}_1-\Delta_3} \Gamma(\alpha_2)\Gamma(\alpha_3) \Gamma\Big(\frac{\alpha_2+\alpha_3}{2}\Big)  \\&\quad\times\frac{\Gamma\big(\frac{-\widehat{\Delta}_1+\Delta_2+\Delta_3-\alpha_2-\alpha_3}{2}\big)\Gamma\big(\frac{\widehat{\Delta}_1+\Delta_2-\Delta_3-\alpha_2+\alpha_3}{2}\big) \Gamma\big(\frac{\widehat{\Delta}_1-\Delta_2+\Delta_3+\alpha_2-\alpha_3}{2}\big)}{2\,\Gamma(\alpha_2+\alpha_3)} \mathcal{M}_{\widehat{\Delta}_1,\Delta_2,\Delta_3}^{\text{contact}}\Bigg]\;,
	\end{aligned}
\end{equation}
where 
\begin{equation}
\mathcal{M}_{\widehat{\Delta}_1,\Delta_2,\Delta_3}^{\text{contact}}=\frac{\pi^{h^{\prime}} \Gamma\big(\frac{\widehat{\Delta}_1+\Delta_2+\Delta_3-2h^{\prime}}{2}\big)}{2\,\Gamma(\widehat{\Delta}_1)\Gamma(\Delta_2)\Gamma(\Delta_3)}\;,
\end{equation}
is the contact Mellin amplitude. From (\ref{newcrossratiosbdy}), the boundary channel OPE limit corresponds to $\zeta_1 \to \infty$, $\zeta_2 \to \infty$. Therefore, we can deform the $\alpha_2$ and $\alpha_3$ contours to the left and obtain the following series expansion which is convergent in the boundary OPE limit
\begin{equation}
\begin{split}
\label{bdyseriescontact}
&\mathcal{W}_{\widehat{\Delta}_1,\Delta_2,\Delta_3}^{\text{contact}}(\zeta_1,\zeta_2)=\mathcal{M}_{\widehat{\Delta}_1,\Delta_2,\Delta_3}^{\text{contact}}\times\\&\sum_{n_2,n_3=0}^{\infty}\bigg[  \frac{(-1)^{n_2+n_3} \Gamma\big(\frac{-n_2-n_3}{2}\big) \Gamma \big(\frac{n_2-n_3+2\Delta_{12,3}}{2} \big) \Gamma\big(\frac{-n_2+n_3+2\Delta_{13,2}}{2}\big)
   \Gamma \big(\frac{n_2+n_3+2\Delta_{23,1}}{2} \big) }{2 \,
   n_2! n_3!\,\Gamma(-n_2-n_3)} \\& \times \zeta _1^{-\Delta _2-n_2} \zeta
   _2^{\Delta _1-\Delta _3-n_3} \\&+\frac{ (-1)^{n_2+n_3}  \Gamma(n_3+\widehat{\Delta}
   _1) \Gamma(n_2+n_3+\Delta _2) \Gamma(-2 n_2-2 n_3-2\Delta_{12,3}) \Gamma(-n_2-2 n_3-2\Delta_{12,3})}{n_2! n_3! \Gamma\big(\frac{-2 n_2-2 n_3-2\Delta_{12,3}}{2}\big)}\\& \times \zeta _1^{-\Delta _2-n_2} \zeta _2^{-\Delta _2-n_2-2 n_3}\\  
&+\frac{(-1)^{n_2+n_3} \Gamma
   (n_2+\widehat{\Delta}_1) \Gamma(n_2+n_3+\Delta _3)\Gamma (-2 n_2-2 n_3-2\Delta_{13,2}) \Gamma(-2
   n_2-n_3-2\Delta_{13,2}) }{n_2! n_3! \Gamma\big(\frac{-2 n_2-2 n_3-2\Delta_{13,2}}{2}\big)}\\&\times \zeta _1^{-\Delta _1-\Delta _3-2 n_2-n_3} \zeta _2^{\Delta _1-\Delta _3-n_3}\bigg]\;.
   \end{split}
\end{equation}
Using the asymptotic behavior (\ref{asybdyblock}) and the series expansion (\ref{blockboundary}) of the BB$\partial$ boundary conformal block, we can obtain the conformal block decomposition by matching (\ref{bdyseriescontact}) order by order. This gives
\begin{equation}
\begin{split}
\label{Boundarydecomofcontact}
&\mathcal{W}_{\widehat{\Delta}_1,\Delta_2,\Delta_3}^{\text{contact}}(\zeta_1,\zeta_2)=\sum_{m=0}^{\infty}\sum_{n=0}^{\infty} A_{m,n}^{\text{BB}\partial}\, g_{\Delta_2+2m,\Delta_3+2n}^{\partial}(\zeta_1,\zeta_2)\\&+ \sum_{m=0}^{\infty}\sum_{n=0}^{\infty} B_{m,n}^{\text{BB}\partial} \,g_{\Delta_2+2m,\widehat{\Delta}_1+\Delta_2+2m+2n}^{\partial}(\zeta_1,\zeta_2) + \sum_{m=0}^{\infty}\sum_{n=0}^{\infty}C_{m,n}^{\text{BB}\partial}  \,g_{\widehat{\Delta}_1+\Delta_3+2m+2n,\Delta_3+2n}^{\partial}(\zeta_1,\zeta_2) \;,
\end{split}
\end{equation}
with the OPE coefficients
\begin{equation}
	\begin{aligned}
			&A_{m,n}^{\text{BB}\partial} =(-1)^{m+n} \pi^{h^{\prime}}  \Gamma\bigg(\frac{-2h^{\prime}+2m+2n+\sum_i \Delta_i}{2}\bigg)\times\\&\frac{ \Gamma(-h^{\prime}+m+\Delta_2) \Gamma(-h^{\prime}+n+\Delta_3) \Gamma(m-n+\Delta_{12,3}) \Gamma(-m+n+\Delta_{13,2}) \Gamma(m+n+\Delta_{23,1}) }{ 2m!n!\Gamma(\widehat{\Delta}_1) \Gamma(\Delta_2)\Gamma(\Delta_3) \Gamma(-h^{\prime}+2m+\Delta_2) \Gamma(-h^{\prime}+2n+\Delta_3)}\;,
	\end{aligned}
\end{equation}
\begin{equation}
	\begin{aligned}
		&B_{m,n}^{\text{BB}\partial} = (-1)^{m+n} \pi^{h^{\prime}} \Gamma\bigg(\frac{-2h^{\prime}+2m+2n+\sum_i \Delta_i}{2}\bigg)\Gamma(n+\widehat{\Delta}_1)\\&\times\frac{\Gamma(-h^{\prime}+m+\Delta_2)\Gamma(2m+n+\Delta_2)\Gamma(-h^{\prime}+2m+n+\widehat{\Delta}_1+\Delta_2)\Gamma(-m-n-\Delta_{12,3}) }{2 m! n! \Gamma(\widehat{\Delta}_1)\Gamma(\Delta_2)\Gamma(\Delta_3)\Gamma(-h^{\prime}+2m+\Delta_2)\Gamma(-h^{\prime}+2m+2n+\widehat{\Delta}_1+\Delta_2)}\;,
	\end{aligned}
\end{equation}
\begin{equation}
		\begin{aligned}
			  & C_{m,n}^{\text{BB}\partial} =  (-1)^{m+n} \pi^{h^{\prime}} \Gamma\bigg(\frac{-2h^{\prime}+2m+2n+\sum_i \Delta_i}{2}\bigg)\Gamma(m+\widehat{\Delta}_1)\\&\times\frac{\Gamma(-h^{\prime}+n+\Delta_3)\Gamma(m+2n+\Delta_3)\Gamma(-h^{\prime}+m+2n+\widehat{\Delta}_1+\Delta_3)\Gamma(-m-n-\Delta_{13,2}) }{2 m! n! \Gamma(\widehat{\Delta}_1)\Gamma(\Delta_2)\Gamma(\Delta_3) \Gamma(-h^{\prime}+2n+\Delta_3) \Gamma(-h^{\prime}+2m+2n+\widehat{\Delta}_1+\Delta_3)}\;.
		\end{aligned}
	\end{equation}
 
\subsubsection{Bulk channel decomposition}
For the bulk channel conformal block decomposition, it is more convenient to make another choice of independent Mellin variables compared to (\ref{mellinCbdy}). Explicitly, we make the change of variables  
\begin{equation}
\delta_{23}=\frac{-\alpha_2-\alpha_3-\widehat{\Delta}_1+\Delta_2+\Delta_3}{2}\;,\quad \quad\gamma_{12}= \frac{-\alpha_2+\alpha_3+\widehat{\Delta}_1+\Delta_2-\Delta_3}{2}\;,
\end{equation}
and we get the following representation (see (\ref{3pointmellinrepbulk}))
\begin{equation}
	\begin{aligned}
 \label{mellinCbulk}
\mathcal{W}_{\widehat{\Delta}_1,\Delta_2,\Delta_3}^{\text{contact}}(\xi_1,\xi_2)&=\int \frac{d\delta_{23} d\gamma_{12}}{(2\pi i)^2} \bigg[\xi_1^{-\delta_{23}} \xi_2^{-\gamma_{12}}\Gamma(\delta_{23}) \Gamma(\gamma_{12}) \Gamma(\widehat{\Delta}_1-\gamma_{12})\Gamma(\Delta_2-\delta_{23}-\gamma_{12})\\\times&\;\!\Gamma(\Delta_3-\widehat{\Delta}_1-\delta_{23}+\gamma_{12})\frac{\Gamma\Big(\frac{-\widehat{\Delta}_1+\Delta_2+\Delta_3-2\delta_{23}}{2}\Big)}{\Gamma(-\widehat{\Delta}_1+\Delta_2+\Delta_3-2\delta_{23})} \mathcal{M}_{\widehat{\Delta}_1,\Delta_2,\Delta_3}^{\text{contact}}\Bigg]\;.
	\end{aligned}
\end{equation}
These different choices as well as the notation will become clear in Section \ref{MellinBBpartial}.  From (\ref{bulkOPElimit}) the bulk OPE limit corresponds to $\xi_1\to 0$ and $\xi_2 \to 1$, but directly taking residues in the representation (\ref{mellinCbulk}) leads to an expansion around $\xi_1=0$ and $\xi_2=0$. In order to obtain the series expansion in the bulk channel OPE limit, we first deform the $\gamma_{12}$ contour to the right real axis and sum over the residues. The expression can then be simplified by using the following hypergeometric identity 
\begin{eqnarray}
\label{identity3}
\nonumber&&{}_2F_1(a,b;c;z)=\frac{\Gamma (c)  \Gamma (a+b-c) }{\Gamma (a) \Gamma (b)}(1-z)^{-a-b+c}{}_2F_1(c-a,c-b;-a-b+c+1;1-z)\\\label{identity3}&&\quad+\frac{\Gamma (c) \Gamma (-a-b+c)  }{\Gamma (c-a) \Gamma (c-b)}{}_2F_1(a,b;a+b-c+1;1-z)\;,\quad\quad -a-b+c\notin
   \mathbb{Z}\;,
\end{eqnarray}
which leads to 
\begin{equation}
\begin{split}
\mathcal{W}_{\widehat{\Delta}_1,\Delta_2,\Delta_3}^{\text{contact}}(\xi_1,\xi_2)=  \int &\frac{d\delta_{23}}{2\pi i} \bigg[\frac{\xi _1^{-\delta _{23}} \Gamma (\delta _{23}) \Gamma (\widehat{\Delta}_1) \Gamma (\Delta _2-\delta _{23}) \Gamma (\Delta _3-\delta _{23}) \Gamma(\Delta_{23,1}-\delta_{23})}{\Gamma(\Delta _2+\Delta _3-2 \delta _{23})} \\&\times \;\! _2F_1\Big(\widehat{\Delta} _1,\Delta _2-\delta _{23};\Delta _2+\Delta _3-2 \delta _{23};1-\xi _2\Big)\bigg]\mathcal{M}_{\widehat{\Delta}_1,\Delta_2,\Delta_3}^{\text{contact}}\;.
   \end{split}
\end{equation}
We can further deform the $\delta_{23}$ contour to the left and obtain a convergent series expansion in bulk OPE limit
\begin{equation}
\begin{split}
\mathcal{W}_{\widehat{\Delta}_1,\Delta_2,\Delta_3}^{\text{contact}}(\xi_1,\xi_2)= \sum_{n=0}^{\infty} &\bigg[\frac{(-1)^n \xi _1^{n}  \Gamma(\widehat{\Delta}_1) \Gamma (\Delta _2+n) \Gamma(\Delta _3+n)\Gamma(\Delta_{23,1}+n)}{n!\Gamma(\Delta _2+\Delta _3+2n)} \\\times&  \;\!_2F_1\Big(\widehat{\Delta} _1,\Delta _2+n;\Delta _2+\Delta _3+2n;1-\xi _2\Big)\bigg]\mathcal{M}_{\widehat{\Delta}_1,\Delta_2,\Delta_3}^{\text{contact}}\;.
   \end{split}
\end{equation}
This expression can be easily compared with the bulk channel scalar conformal blocks (\ref{blockbulkscalar}). It is not difficult to find that only scalar operators with dimension $\Delta_2+\Delta_3+2n$ appear and the form of the $\xi_2$ dependence already automatically matches that in the conformal blocks. Therefore, we obtain the following decomposition of the BB$\partial$ contact diagram into bulk channel conformal blocks
\begin{equation}
	\begin{gathered}
		\label{expanbulkchannel}
\mathcal{W}_{\widehat{\Delta}_1,\Delta_2,\Delta_3}^{\text{contact}}(\xi_1,\xi_2)=\sum_{n=0}^{\infty} A_n^{\text{BB}\partial}\, g_{\Delta_2+\Delta_3+2n,0}^{\text{B}}(\xi_1,\xi_2)\;,
	\end{gathered}
\end{equation}
where the OPE coefficients are
\begin{equation}
\begin{aligned}
&A_n^{\text{BB}\partial}=\\&\frac{(-1)^{n} \pi^{h^{\prime}}\Gamma(n+\Delta_{2}) \Gamma(n+\Delta_{3}) \Gamma(n+\Delta_{23,1}) \Gamma(-\frac{1}{2}-h^{\prime}+n+\Delta_{2}+\Delta_{3}) \Gamma\big(\frac{-2h^{\prime}+2 n+\sum_i \Delta_i}{2}\big)}{2 n ! \Gamma(\Delta_{2}) \Gamma(\Delta_{3}) \Gamma(2 n+\Delta_{2}+\Delta_{3}) \Gamma(-\frac{1}{2}-h^{\prime}+2 n+\Delta_{2}+\Delta_{3})}\;.
\end{aligned}
\end{equation}

\subsection{Boundary exchange Witten diagram}
\label{Boundary exchange Witten diagram} 
\begin{figure}
    \centering
    \includegraphics[width=0.7\linewidth]{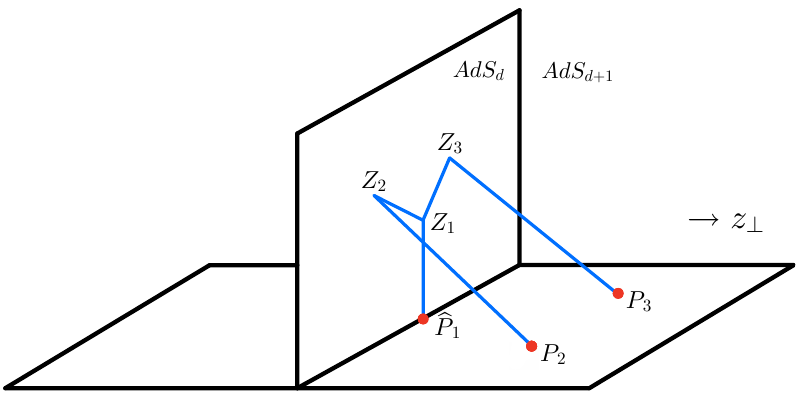}
    \caption{The boundary exchange Witten diagram for the BB$\partial$ three-point function, where the vertices are inserted at $Z_1$, $Z_2$ and $Z_3$ on the $AdS_{d}$ brane and are integrated over.}
    \label{BBpbdyexchangefig}
\end{figure}

Let us now turn to the BB$\partial$ boundary exchange diagram depicted in Fig. \ref{BBpbdyexchangefig}. This diagram is defined as
\begin{equation}
\label{defbdyexchdiag}
W^{\partial}_{\widehat{\Delta},\widehat{\Delta}^{\prime}}(\widehat{P}_1,P_2,P_3)=\int_{AdS_d} dZ_1\, G_{B \partial}^{\widehat{\Delta}_{1}}(\widehat{P}_{1}, Z_1) A_2(P_2,Z_1)A_3(P_3,Z_1)\;,
\end{equation}
where
\begin{equation}
	\begin{gathered}
		A_2(P_2,Z_1)=\int_{AdS_d} dZ_2\, G_{B \partial}^{\Delta_{2}}(P_2, Z_2) G_{B B,(d)}^{\widehat{\Delta}}(Z_2,Z_1)\;,\\ A_3(P_3,Z_1)=\int_{AdS_d} dZ_3 \, G_{B \partial}^{\Delta_{3}}(P_3, Z_3) G_{B B,(d)}^{\widehat{\Delta}^{\prime}}(Z_3,Z_1)\;.
	\end{gathered}
\end{equation}
We will compute this diagram in Mellin space in Section \ref{Mellin space representation}. For the moment, our focus is the conformal block decomposition. 

Let us first consider the conformal block decomposition in the boundary channel. We claim that the boundary exchange Witten diagram can be decomposed into boundary channel conformal blocks as follows
\begin{equation}
	\begin{aligned}
 \label{CBdecomofbdy}
&\mathcal{W}^{\partial}_{\widehat{\Delta},\widehat{\Delta}^{\prime}}= a^{\partial}\, g^{\partial}_{\widehat{\Delta},\widehat{\Delta}^{\prime}}+\sum_{m=0}^{\infty}b_{m}^{\partial}\, g^{\partial}_{\widehat{\Delta},\widehat{\Delta}+\widehat{\Delta}_1+2m} + \sum_{m=0}^{\infty} b^{\prime \partial}_m \,g_{\widehat{\Delta}^{\prime}+\widehat{\Delta}_1+2m,\widehat{\Delta}^{\prime}}^{\partial}+\sum_{m=0}^{\infty} c_m^{\partial}\, g_{\widehat{\Delta},\Delta_3+2m}^{\partial} \\&\quad+\sum_{m=0}^{\infty} c^{\prime \partial}_m \,g_{\Delta_2+2m,\widehat{\Delta}^{\prime}}^{\partial}+\sum_{m=0}^{\infty}\sum_{n=0}^{\infty}a_{m,n}^{\partial}\,g^{\partial}_{\Delta_2+2m,\Delta_3+2n}+\sum_{m=0}^{\infty}\sum_{n=0}^{\infty}b_{m,n}^{\partial}\,g^{\partial}_{\Delta_2+2m,\widehat{\Delta}_1+\Delta_2+2m+2n} \\&\quad+ \sum_{m=0}^{\infty}\sum_{n=0}^{\infty} c_{m,n}^{\partial}\,g^{\partial}_{\widehat{\Delta}_1+\Delta_3+2m+2n,\Delta_3+2n}\;.
	\end{aligned}
\end{equation}
The appearance of some of the conformal blocks is expected from the EOM relations. Defining the boundary EOM operators acting on $\mathcal{O}_2$ and $\mathcal{O}_3$ respectively
\begin{equation}
	\begin{gathered}
\mathbf{EOM}^{\partial,2}_{\widehat{\Delta}_0}=\frac{1}{2}\widehat{\mathbf{L}}_2^2+\widehat{\Delta}_0(\widehat{\Delta}_0-d+1)\;,\\ \mathbf{EOM}^{\partial,3}_{\widehat{\Delta}_0^{\prime}}=\frac{1}{2}\widehat{\mathbf{L}}_3^2+\widehat{\Delta}_0^{\prime}(\widehat{\Delta}_0^{\prime}-d+1)\;,
	\end{gathered}
\end{equation}
the exchange Witten diagram is related to the contact Witten diagram by 
\begin{equation}
	\label{eombdychannel}
\mathbf{EOM}^{\partial,3}_{\widehat{\Delta}^{\prime}}\left(\mathbf{EOM}^{\partial,2}_{\widehat{\Delta}}\big(W^{\partial}_{\widehat{\Delta},\widehat{\Delta}^{\prime}}(\widehat{P}_1,P_2,P_3)\big)\right)= W^{\text{contact}}_{\widehat{\Delta}_1,\Delta_2,\Delta_3}(\widehat{P}_1,P_2,P_3)\;.
\end{equation}
The derivation of this identity is again similar to the relation in \cite{Zhou:2018sfz}. The action of each EOM operator collapses the bulk-to-bulk propagator to which the corresponding bulk-to-boundary propagator is connected. Note that the boundary EOM operators are also the boundary channel Casimir operators with constant shifts and therefore act on the boundary channel conformal blocks diagonally. Using the boundary channel decomposition (\ref{Boundarydecomofcontact}) of the contact diagram, it follows that there must also be conformal blocks $g^{\partial}_{\Delta_2+2m,\Delta_3+2n}$, $g^{\partial}_{\Delta_2+2m,\widehat{\Delta}_1+\Delta_2+2m+2n}$, $g^{\partial}_{\widehat{\Delta}_1+\Delta_3+2m+2n,\Delta_3+2n}$ with coefficients
\begin{equation}
	\begin{aligned}
		&a_{m,n}^{\partial}=\frac{A_{m,n}^{\text{BB}\partial}}{\mathcal{C}_{\Delta_2+2m,\Delta_3+2n}}\;,\quad
		b_{m,n}^{\partial}=\frac{B_{m,n}^{\text{BB}\partial}}{\mathcal{C}_{\Delta_2+2m,\widehat{\Delta}_1+\Delta_2+2m+2n}}\;,\quad
		c_{m,n}^{\partial}=\frac{C_{m,n}^{\text{BB}\partial}}{\mathcal{C}_{\widehat{\Delta}_1+\Delta_3+2m+2n,\Delta_3+2n}}\;.
	\end{aligned}
\end{equation}
Here  $A_{m,n}^{\text{BB}\partial}$, $B_{m,n}^{\text{BB}\partial}$ and $B_{m,n}^{\text{BB}\partial}$ are the boundary OPE coefficients of the contact diagram (\ref{Boundarydecomofcontact}) and 
\begin{equation}
    \mathcal{C}_{\Delta_0,\Delta_0^{\prime}}=\left(\widehat{\Delta}(\widehat{\Delta}-d+1)-\Delta_0(\Delta_0-d+1)\right)\left(\widehat{\Delta}^{\prime}(\widehat{\Delta}^{\prime}-d+1)-\Delta_0^{\prime}(\Delta_0^{\prime}-d+1)\right)\;,
\end{equation}
is the product of the eigenvalues of the two EOM operators. On the other hand, we note that there can also be conformal blocks which are annihilated by the double actions of the EOM operators. These are boundary channel conformal blocks of which at least one of the exchange operators is $\widehat{\Delta}$ and  $\widehat{\Delta}'$. The EOM relation is agnostic about these operators but they turn out to be $g_{\widehat{\Delta},\widehat{\Delta}^{\prime}}^{\partial}$,  $g_{\widehat{\Delta},\widehat{\Delta}+\widehat{\Delta}_1+2m}^{\partial}$,  $g_{\widehat{\Delta},\Delta_3+2m}^{\partial}$, $g_{\widehat{\Delta}^{\prime}+\widehat{\Delta}_1+2m,\widehat{\Delta}^{\prime}}^{\partial}$ and $g_{\Delta_2+2m,\widehat{\Delta}^{\prime}}^{\partial}$. The appearance of these conformal blocks can be seen by using the split representation for the bulk-to-bulk propagators. This will be discussed in detail in Appendix \ref{appendixspectral1}, where the OPE coefficients $a^{\partial},b_m^{\partial},b^{\prime\partial}_m,c_m^{\partial},c_m^{\prime\partial}$ are also computed, see (\ref{ap}-\ref{01}).

Let us also briefly discuss the bulk channel decomposition of the boundary exchange Witten diagram. We will outline the main idea of the calculation although we will not follow it to the end and provide the explicit decomposition coefficients. Our strategy will be similar to that of \cite{Zhou:2018sfz} where four-point functions in CFTs without boundaries were considered. There the strategy relied on two facts. The first is the EOM identity which relates the exchange diagram to the contact diagram. The latter has a known conformal block decomposition in the crossed channel (with respect to the EOM operator). The second fact is the action of the EOM operator on a crossed channel conformal block  leads to a recursion relation which consists of finitely many crossed channel conformal blocks. Combining these two facts we can obtain recursion relations for the OPE decomposition coefficients which can be solved upon inputting a subset of them as the boundary condition. 

The situation here is similar. We already have the EOM relation (\ref{eombdychannel}) and we only need to find the action of the boundary channel EOM operator on bulk channel conformal blocks. This can be achieved by using the radial expansion (\ref{radial expansion}) of the conformal blocks $g_{\Delta,J}^{\text{B}}$. We find the following recursion relations with finitely many terms 
\begin{eqnarray}
\label{eomcrossing1}
&&\mathbf{EOM}_{\widehat{\Delta}_0}^{\partial,2} \left(g_{\Delta,J}^{\text{B}} \right)= \alpha_{-2,0}\, g_{\Delta-2,J}^{\text{B}}+ \alpha_{-1,1}\, g_{\Delta-1,J+1}^{\text{B}}+\alpha_{-1,-1}\, g_{\Delta-1,J-1}^{\text{B}}\\\nonumber&&\; +\alpha_{0,2}\, g_{\Delta,J+2}^{\text{B}}+ \alpha_{0,0}\, g_{\Delta,J}^{\text{B}}+\alpha_{0,-2}\, g_{\Delta,J-2}^{\text{B}}+\alpha_{1,1}\, g_{\Delta+1,J+1}^{\text{B}}+\alpha_{1,-1}\, g_{\Delta+1,J-1}^{\text{B}}+\alpha_{2,0}\, g_{\Delta+2,J}^{\text{B}}\;, 
\end{eqnarray}
\begin{eqnarray}
\label{eomcrossing2}
&&\mathbf{EOM}^{\partial,3}_{\widehat{\Delta}^{\prime}_0} \left(g_{\Delta,J}^{\text{B}} \right)= \beta_{-2,0}\, g_{\Delta-2,J}^{\text{B}}+ \beta_{-1,1}\, g_{\Delta-1,J+1}^{\text{B}}+\beta_{-1,-1}\, g_{\Delta-1,J-1}^{\text{B}}\\\nonumber&& \;+\beta_{0,2}\, g_{\Delta,J+2}^{\text{B}}+ \beta_{0,0}\, g_{\Delta,J}^{\text{B}}+\beta_{0,-2}\, g_{\Delta,J-2}^{\text{B}}+\beta_{1,1}\, g_{\Delta+1,J+1}^{\text{B}}+\beta_{1,-1}\, g_{\Delta+1,J-1}^{\text{B}}+\beta_{2,0}\, g_{\Delta+2,J}^{\text{B}}\;.
\end{eqnarray}
Here the coefficients $\alpha_{i,j}$ and $\beta_{i,j}$ are rational expressions which are independent of the cross ratios. As a result of the symmetry of the bulk channel conformal block
\begin{equation}
	G_{\Delta, J}^{\text{B}}(\widehat{P}_1, P_2, P_3)=(-1)^J G_{\Delta, J}^{\text{B}}(\widehat{P}_1, P_3, P_2)\;,
\end{equation}
the two sets of coefficients $\alpha_{i,j}$ and $\beta_{i,j}$ are related by
\begin{equation}
	\alpha_{i,j}=(-1)^j \beta_{i,j} \big|_{\Delta_2 \leftrightarrow \Delta_3}\;.
\end{equation}
Their explicit expressions are a bit too cumbersome to be recorded here.\footnote{They can be found in the ancillary $\mathtt{Mathematica}$ notebook.} Instead, let us mention a few properties. First, when the spins of the conformal blocks on the RHS of (\ref{eomcrossing1}) and (\ref{eomcrossing2}) become negative the associated coefficients vanish. This makes sure that only conformal blocks with physical spins can appear. Second, when acting on the spectrum $(\Delta,J)=(\Delta_2+\Delta_3+2n+J,J)$ with $n=0,1,2,\ldots$, the coefficients have correct zeros such that the spectrum is preserved. In other words, although (\ref{eomcrossing1}) and (\ref{eomcrossing2}) generate conformal blocks with lower conformal dimensions, the ones with $n<0$ have vanishing coefficients. Note this is only true when the spectrum assumes the special form $(\Delta,J)=(\Delta_2+\Delta_3+2n+J,J)$. Finally, when $\widehat{\Delta}_1 \to 0$, $J\to 0$ and $\widehat{\Delta}_0^{\prime} \to \widehat{\Delta}_0$, two relations become the same and both reduce to the two-point function relation (3.71) of \cite{Mazac:2018biw}.

In the crossed channel, we expect the conformal block decomposition of the boundary exchange diagram to take the following form
\begin{equation}
\mathcal{W}^{\partial}_{\widehat{\Delta},\widehat{\Delta}^{\prime}}(\xi_1,\xi_2)=\sum_{J=0}^{\infty} \sum_{n=0}^{\infty} a^{\partial,\text{cr}}_{n,J}g^{\text{B}}_{\Delta_2+\Delta_3+2n+J,J}(\xi_1,\xi_2)\;.
\end{equation}
Note it follows from the second property that we still have the same set of conformal blocks after the action of the EOM operators. Using the EOM identity (\ref{eombdychannel}) and the bulk channel conformal block decomposition of the contact Witten diagram (\ref{expanbulkchannel}), we have the following recursion relation for the OPE coefficients $a^{\partial,\text{cr}}_{n,J}$
\begin{equation}
\label{crossedrecur1}
    \sum_{i=-2}^{2} \sum_{j=-2}^{2} \mathcal{C}_{i,j} a^{\partial,\text{cr}}_{n-i,J+i+j}=\left\{\begin{array}{lc}
A_n^{\text{BB}\partial}\,, & \;J=0\\
0\,, &  J>0
\end{array}\right.\;.
\end{equation}
Here the coefficients $\mathcal{C}_{i,j}$ are combinations of products of the $\alpha$ and $\beta$ coefficients with $\Delta$ set to $\Delta_2+\Delta_3+2n+J$, which can be easily worked out from (\ref{eomcrossing1}) and (\ref{eomcrossing2}) and will not be spelled out here. The recursion relation can be solved upon inputting the seed data $a_{0,J}^{\partial,\text{cr}}$ and $a_{1,J}^{\partial,\text{cr}}$. However, we will not compute these seed coefficients here and leave them for future work.

\subsection{Bulk exchange Witten diagram}
\begin{figure}
    \centering
    \includegraphics[width=0.71\linewidth]{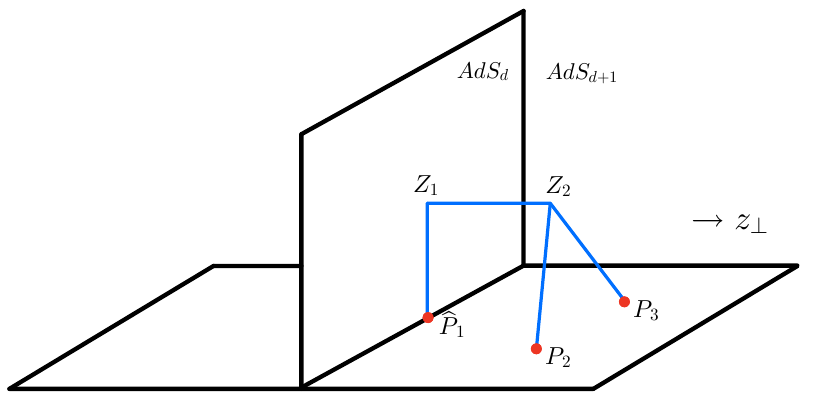}
    \caption{The bulk scalar exchange Witten diagram for the BB$\partial$ three-point function, where the vertex insertion point $Z_1$ is integrated over $AdS_{d}$ and $Z_2$ is integrated over $AdS_{d+1}$.}
    \label{BBpbulkexchangefig}
\end{figure}

Compared to the boundary exchange Witten diagrams, the bulk exchange Witten diagrams $W_{\Delta,J}^{\text{B}}$ (depicted in Fig. \ref{BBpbulkexchangefig}) have more possibilities because the bulk-to-bulk propagator can have arbitrary spins. For simplicity, in this subsection we will focus our position space discussion on the scalar case where $J=0$. The spinning case with $J=1$ will be considered in Mellin space in Section \ref{Mellin space representation}.

The scalar exchange diagram $W_{\Delta,0}^{\text{B}}$ is defined by the integral
\begin{equation}
	\begin{aligned}
 \label{defofbulkscalarexch}
		W_{\Delta,0}^{\text{B}}(\widehat{P}_1,P_2,P_3)=\int_{A d S_{d}} d Z_{1} \, G_{B \partial}^{\widehat{\Delta}_{1}}(\widehat{P}_{1}, Z_{1}) A(Z_1,P_2,P_3)\;,
	\end{aligned}
\end{equation}
where
\begin{equation}
\label{intA}
   A(Z_1,P_2,P_3)=\int_{A d S_{d+1}} d Z_{2} \, G_{B B,(d+1)}^{\Delta}(Z_{1}, Z_{2}) G_{B \partial}^{\Delta_{2}}(P_{2}, Z_{2})G_{B \partial}^{\Delta_{3}}(P_{3}, Z_{2})\;.
\end{equation}
We claim that this diagram should have the following form of conformal block decomposition
\begin{equation}
	\mathcal{W}_{\Delta,0}^{\text{B}}(\xi_1,\xi_2)=a^{\text{B}}\,g_{\Delta,0}^{\text{B}}(\xi_1,\xi_2)+\sum_n a_n^{\text{B}}\, g_{\Delta_2+\Delta_3+2 n,0}^{\text{B}}(\xi_1,\xi_2)\;,
\end{equation}
which follows from the EOM relation. Let us define the bulk channel EOM operator 
\begin{equation}
	\label{3pteom}
	\mathbf{EOM}^{\text{B}}_{\Delta_0,J} = \frac{1}{2}\left(\mathbf{L}_{2}+\mathbf{L}_{3}\right)^{2}+\Delta_0(\Delta_0-d)+J(J+d-2)\;,
\end{equation}
which is the bulk channel two-particle conformal Casimir operator with a constant shift. The action of this operator with $J=0$ turns the bulk scalar exchange Witten diagram into the contact Witten diagram 
\begin{equation}
\label{scalarbulkeom}
\mathbf{EOM}^{\text{B}}_{\Delta,0}\left(W_{\Delta,0}^{\text{B}}(\widehat{P}_1,P_2,P_3)\right)= W^{\text{contact}}_{\widehat{\Delta}_1,\Delta_2,\Delta_3}(\widehat{P}_1,P_2,P_3)\;.
\end{equation}
Since the EOM operator acts diagonally on the bulk channel conformal blocks, it follows that  $g_{\Delta_2+\Delta_3+2 n,0}^{\text{B}}(\xi_1,\xi_2)$, which are the only conformal blocks present in the decomposition of the contact diagram, should also be present in the decomposition of the exchange diagram. Their OPE coefficients are determined by the EOM relation to be
\begin{equation}
	a_n^{\text{B}}=\frac{A_n^{\text{BB}\partial}}{\Delta(\Delta-d)-(\Delta_2+\Delta_3+2 n)(\Delta_2+\Delta_3+2 n-d)}\;,
\end{equation}
where $A_n^{\text{BB}\partial}$ are bulk channel decomposition coefficients of the contact diagram (\ref{expanbulkchannel}). 

On the other hand, the conformal block $g_{\Delta,0}^{\text{B}}$ is annihilated by the EOM operator. To determine its associated coefficient $a^{\text{B}}$, we could use the split representation for the bulk-to-bulk propagator as the in boundary exchange diagram case (Appendix \ref{appendixspectral1}). Here instead we will demonstrate an alternative method which uses the integrated vertex identity \cite{DHoker:1999mqo,Zhou:2018sfz,Goncalves:2019znr}. The integrated vertex identity allows us to integrate out the bulk-to-bulk propagator in $A(Z_1,P_2,P_3)$ and replace it by a sum of products of bulk-to-boundary propagators
\begin{equation}
	\begin{aligned}\label{intvertexid}
		&A(Z_1,P_2, P_3)=\sum_{i=0}^{\infty}\left(-2 P_2 \cdot P_3\right)^i T_{\Delta_2,\Delta_3,i}\, G_{B \partial}^{\Delta_2+i}\left(P_2, Z_1\right) G_{B \partial}^{\Delta_3+i}\left(P_3, Z_1\right) \\
	&\quad+\sum_{i=0}^{\infty}\left(-2 P_2 \cdot P_3\right)^{\frac{\Delta-\Delta_2-\Delta_3+2 i}{2}} Q_{\Delta_2,\Delta_3,i} \,G_{B \partial}^{\frac{\Delta+\Delta_2-\Delta_3}{2}+i}\left(P_2, Z_1\right) G_{B \partial}^{\frac{\Delta-\Delta_2+\Delta_3}{2}+i}\left(P_3, Z_1\right)\;,
	\end{aligned}
\end{equation}
where
\begin{equation}
\label{defT}
T_{\Delta_2,\Delta_3,i}=\frac{\left(\Delta_2\right)_i\left(\Delta_3\right)_i}{\left(\Delta-\Delta_2-\Delta_3\right)\left(-d+\Delta+\Delta_2+\Delta_3\right)\left(\frac{-\Delta+\Delta_2+\Delta_3+2}{2}\right)_i\left(\frac{-d+\Delta+\Delta_2+\Delta_3+2}{2}\right)_i}\;,
\end{equation}
\begin{equation}
\begin{split}
\label{defQ}
		&Q_{\Delta_2,\Delta_3,i}= \frac{(-1)^i \Gamma\big(\frac{d-2 i-2 \Delta}{2}\big) \sin\big(\frac{\pi(d-2 \Delta)}{2}\big) \Gamma\big(\frac{-d+\Delta+\Delta_2+\Delta_3}{2}\big)}{4 \pi \Gamma(i+1) \Gamma(\Delta_2) \Gamma(\Delta_3)} \\ &\quad \times \frac{\Gamma\big(\frac{\Delta-\Delta_2+\Delta_3}{2}\big) \Gamma\big(\frac{\Delta+\Delta_2-\Delta_3}{2}\big) \Gamma\big(\frac{-\Delta+\Delta_2+\Delta_3}{2}\big) \Gamma\big(\frac{-\Delta+\Delta_2-\Delta_3+2}{2}\big) \Gamma\big(\frac{-\Delta-\Delta_2+\Delta_3+2}{2}\big)}{\Gamma\big(\frac{-\Delta+\Delta_2-\Delta_3-2 i+2}{2}\big) \Gamma\big(\frac{-\Delta-\Delta_2+\Delta_3-2 i+2}{2}\big)}\;.
\end{split}
\end{equation}
As a result, $W_{\Delta,0}^{\text{B}}$ can be written as a sum of contact diagrams with shifted external dimensions
\begin{equation}
	\begin{aligned}
		W_{\Delta,0}^{\text{B}}(P_i)=&\sum_{i=0}^{\infty}\left(-2 P_2 \cdot P_3\right)^i T_{\Delta_2,\Delta_3,i} W^{\text{contact}}_{\widehat{\Delta}_1,\Delta_2+i,\Delta_3+i} (P_i)\\&+\sum_{i=0}^{\infty}\left(-2 P_2 \cdot P_3\right)^{\frac{\Delta-\Delta_2-\Delta_3+2 i}{2}} Q_{\Delta_2,\Delta_3,i} W^{\text{contact}}_{\widehat{\Delta}_1,\frac{\Delta+\Delta_2-\Delta_3}{2}+i,\frac{\Delta-\Delta_2+\Delta_3}{2}+i}(P_i)\;,
	\end{aligned}
\end{equation}
written in terms of the cross ratios, we have
\begin{equation}
	\begin{aligned}
		\label{intervertex}
		\mathcal{W}_{\Delta,0}^{\text{B}}(\xi_1,\xi_2)=&\sum_{i=0}^{\infty}\xi_1^i T_{\Delta_2,\Delta_3,i} \mathcal{W}^{\text{contact}}_{\widehat{\Delta}_1,\Delta_2+i,\Delta_3+i}(\xi_1,\xi_2)\\& +\sum_{i=0}^{\infty}\xi_1^{\frac{\Delta-\Delta_2-\Delta_3}{2}+i} Q_{\Delta_2,\Delta_3,i} \mathcal{W}^{\text{contact}}_{\widehat{\Delta}_1,\frac{\Delta+\Delta_2-\Delta_3}{2}+i,\frac{\Delta-\Delta_2+\Delta_3}{2}+i}(\xi_1,\xi_2)\;.
	\end{aligned}
\end{equation}
It is not difficult to see that the leading contribution in the bulk OPE limit is  from the leading term in the second sum
\begin{equation}
	\begin{aligned}
 \label{asyscalarbulk}
\mathcal{W}_{\Delta,0}^{\text{B}}(\xi_1,\xi_2) \xrightarrow[]{\xi_1\to0,\;\xi_2\to 1} Q_{\Delta_2,\Delta_3,0} \left(A_0^{\text{BB}\partial}\Big|_{\Delta_2 \to \frac{\Delta+\Delta_2-\Delta_3}{2},\Delta_3 \to \frac{\Delta-\Delta_2+\Delta_3}{2}} \right)\xi_1^{\frac{\Delta-\Delta_2-\Delta_3}{2}}\;,
\end{aligned}
\end{equation}
after using the bulk channel decomposition of contact diagram (\ref{expanbulkchannel}). Since $g_{\Delta,0}^{\text{B}}$ is leading in the OPE limit, this gives 
\begin{equation}
\label{aB}
	a^{\text{B}}=\frac{\pi^{h^{\prime}}\Gamma(\frac{\Delta-\widehat{\Delta}_1}{2})\Gamma\big(\frac{1-d+\Delta+\widehat{\Delta}_1}{2}\big)\Gamma\big(\frac{\Delta+\Delta_2-\Delta_3}{2}\big)\Gamma\big(\frac{\Delta-\Delta_2+\Delta_3}{2}\big)\Gamma\big(\frac{-\Delta+\Delta_2+\Delta_3}{2}\big)\Gamma\big(\frac{-d+\Delta+\Delta_2+\Delta_3}{2}\big)}{8\,\Gamma(\Delta)\Gamma(1-\frac{d}{2}+\Delta)\Gamma(\Delta_2)\Gamma(\Delta_3)}\;.
\end{equation}

To decompose the bulk exchange Witten diagram into boundary channel conformal blocks, we can follow the same strategy as we did for the crossed channel decomposition of the boundary exchange Witten diagram. It requires us to compute the action of $\mathbf{EOM}^{\text{B}}_{\Delta_0,J}$ on boundary conformal block. On a generic function $g(\xi_1,\xi_2)$, the action reads
\begin{equation}
	\begin{aligned}
		\label{actionEOM}
		&\mathbf{EOM}^{\text{B}}_{\Delta,J} (g(\xi_1,\xi_2)) =\big(\Delta _2 (d-\Delta _3
		\left(\xi _1+2\right)+\widehat{\Delta}_1 (\xi _1-2 \xi _2+2))-d \Delta_0+\Delta _3 \left(d-\Delta _3\right)\\&+J (d+J-2)+\Delta_0^2-\Delta _2^2\big)g(\xi _1,\xi _2)+\xi _1 (2 d-\left(\Delta _2+\Delta _3+1\right) (\xi _1+4)+\widehat{\Delta}_1 (\xi _1-2 \xi _2+2))\\&g^{(1,0)}(\xi_1,\xi_2)+\big(-2 (\widehat{\Delta}_1+\Delta _2+1) \xi _2^2+\xi _2
		(\Delta _2 (\xi _1+2)-\Delta _3 (\xi _1+2)+\widehat{\Delta}_1 (\xi _1+4)+\xi _1+4)\\&-2 \widehat{\Delta}_1+2 \Delta _3-2\big)g^{(0,1)}(\xi _1,\xi _2)-(\xi _1+4) \xi _1^2
		g^{(2,0)}(\xi _1,\xi _2)-2 (\xi _2^2-1) \xi _1 g^{(1,1)}\left(\xi _1,\xi _2\right)\\&+\xi _2 \big((\xi _1-2 \xi _2+4) \xi _2-2\big) g^{(0,2)}(\xi _1,\xi _2)\;.
	\end{aligned}
\end{equation}
Using the explicit expression for boundary channel conformal blocks $\left(\ref{blockboundary0}\right)$, we find the following recursion relation
\begin{equation}
	\begin{aligned}
		\label{eomcrossing3}
\mathbf{EOM}^{\text{B}}_{\Delta_0,J}\left(g_{\widehat{\Delta},\widehat{\Delta}^{\prime}}^{\partial}\right)=\,& \alpha_{\widehat{\Delta},\widehat{\Delta}^{\prime}}\,g_{\widehat{\Delta}-1,\widehat{\Delta}^{\prime}-1}^{\partial}+\beta_{\widehat{\Delta},\widehat{\Delta}^{\prime}}\,g_{\widehat{\Delta},\widehat{\Delta}^{\prime}}^{\partial}+\gamma_{\widehat{\Delta},\widehat{\Delta}^{\prime}}\,g_{\widehat{\Delta}+1,\widehat{\Delta}^{\prime}+1}^{\partial}\\&+\lambda_{\widehat{\Delta},\widehat{\Delta}^{\prime}}\,g_{\widehat{\Delta}-1,\widehat{\Delta}^{\prime}+1}^{\partial}+\lambda^{\prime}_{\widehat{\Delta},\widehat{\Delta}^{\prime}}\,g_{\widehat{\Delta}+1,\widehat{\Delta}^{\prime}-1}^{\partial}\;,
	\end{aligned}
\end{equation}
where the coefficients are
\begin{eqnarray}
&&\alpha_{\widehat{\Delta},\widehat{\Delta}^{\prime}}=-(\widehat{\Delta}-\Delta_2)(\widehat{\Delta}^{\prime}-\Delta_3)\;,\\\nonumber&&
	\beta_{\widehat{\Delta},\widehat{\Delta}^{\prime}}=J(J+d-2)+\widehat{\Delta}(\widehat{\Delta}+1)+\widehat{\Delta}^{\prime}(\widehat{\Delta}^{\prime}+1)-\widehat{\Delta}_1(\widehat{\Delta}_1+1)+\Delta_0^2-\Delta_2^2-\Delta_3^2\\&&+d(-\widehat{\Delta}-\widehat{\Delta}^{\prime}-\Delta_0+\widehat{\Delta}_1+\Delta_2+\Delta_3)\;,\\&&
	\gamma_{\widehat{\Delta},\widehat{\Delta}^{\prime}}= \frac{\left[\parbox{11.00cm}{$(\widehat{\Delta}+\widehat{\Delta}^{\prime} -\widehat{\Delta}_1) (d-\widehat{\Delta} -\Delta _2-1) (d-\widehat{\Delta}-\widehat{\Delta}^{\prime}-\widehat{\Delta}_1-1) (2 d-\widehat{\Delta}
			-\widehat{\Delta}^{\prime}-\widehat{\Delta}_1-4) (d-\widehat{\Delta}-\widehat{\Delta}^{\prime} +\widehat{\Delta}_1-3)(d-\widehat{\Delta}^{\prime}-\Delta _3-1)$}\right]}{ (d-2 \widehat{\Delta} -3) (d-2 \widehat{\Delta}
		-1) (d-2 \widehat{\Delta}^{\prime}-3) (d-2 \widehat{\Delta}^{\prime}-1)},\\&&
\lambda_{\widehat{\Delta},\widehat{\Delta}^{\prime}}=\frac{(\widehat{\Delta}-\Delta_2)(\widehat{\Delta}-\widehat{\Delta}^{\prime}-\widehat{\Delta}_1)(d-3+\widehat{\Delta}-\widehat{\Delta}^{\prime}-\widehat{\Delta}_1)(1-d+\widehat{\Delta}^{\prime}+\Delta_3)}{(d-3-2\widehat{\Delta}^{\prime})(d-1-2\widehat{\Delta}^{\prime})}\;,\\&&
\lambda^{\prime}_{\widehat{\Delta},\widehat{\Delta}^{\prime}}=\lambda_{\widehat{\Delta},\widehat{\Delta}^{\prime}}\Big|_{\widehat{\Delta} \leftrightarrow \widehat{\Delta}^{\prime},\Delta_2\leftrightarrow \Delta_3} \;.
\end{eqnarray}
When we set $\widehat{\Delta}_1,J \to 0$ and $\Delta_2\to \widehat{\Delta}_1,\, \Delta_3 \to \Delta_2$, and further $\widehat{\Delta}^{\prime}\to \widehat{\Delta}=\widehat{\Delta}_1+n$, the above identity reduces to the two-point counterpart which was given in (3.65) of \cite{Mazac:2018biw}. 

In the crossed channel, we expect the bulk exchange Witten diagram to have the following form of conformal block decomposition
\begin{equation}
    \begin{aligned}
\mathcal{W}_{\Delta,0}^{\text{B}}(\xi_1,\xi_2)=&\sum_{m=0}^{\infty}\sum_{n=0}^{\infty} a_{m,n}^{\text{B},\text{cr}}\, g_{\Delta_2+m,\Delta_3+n}^{\partial}(\xi_1,\xi_2)+ \sum_{m=0}^{\infty}\sum_{n=0}^{\infty} b_{m,n}^{\text{B},\text{cr}} \,g_{\Delta_2+m,\widehat{\Delta}_1+\Delta_2+m+n}^{\partial}(\xi_1,\xi_2) \\&+ \sum_{m=0}^{\infty}\sum_{n=0}^{\infty}c_{m,n}^{\text{B},\text{cr}}  \,g_{\widehat{\Delta}_1+\Delta_3+m+n,\Delta_3+n}^{\partial}(\xi_1,\xi_2) \;.
\end{aligned}
\end{equation}
Note that the spectra of operators are also preserved by the recursion relation (\ref{eomcrossing3}) after the action of the EOM operator. Using the boundary channel decomposition of contact diagram (\ref{Boundarydecomofcontact}), we arrive at the following recursion relations for the decomposition coefficients
\begin{equation}
\begin{split}
\label{bulkcrossrecur1}
 & \alpha_{\Delta_2+m+1,\Delta_3+n+1}  a_{m+1,n+1}^{\text{B},\text{cr}}+\beta_{\Delta_2+m,\Delta_3+n} a_{m,n}^{\text{B},\text{cr}}+\gamma_{\Delta_2+m-1,\Delta_3+n-1} a_{m-1,n-1}^{\text{B},\text{cr}}+\\&\lambda_{\Delta_2+m+1,\Delta_3+n-1} a_{m+1,n-1}^{\text{B},\text{cr}}+\lambda^{\prime}_{\Delta_2+m-1,\Delta_3+n+1} a_{m-1,n+1}^{\text{B},\text{cr}} =\left\{\begin{array}{lc}
A_{\frac{m}{2},\frac{n}{2}}^{\text{BB}\partial}\,, & \;m,n \text { even } \\
0\,, &  \text { others }
\end{array}\right.\;,
  \end{split}
\end{equation}
\begin{equation}
    \begin{aligned}
    \label{bulkcrossrecur2}
       & \alpha_{\Delta_2+m+1,\widehat{\Delta}_1+\Delta_2+m+n+1} b_{m+1,n}^{\text{B},\text{cr}}+\beta_{\Delta_2+m,\Delta_1+\Delta_2+m+n} b_{m,n}^{\text{B},\text{cr}}+\gamma_{\Delta_2+m-1,\Delta_1+\Delta_2+m+n-1} b_{m-1,n}^{\text{B},\text{cr}}+\\&\lambda_{\Delta_2+m+1,\widehat{\Delta}_1+\Delta_2+m+n-1} b_{m+1,n-2}^{\text{B},\text{cr}}+\lambda^{\prime}_{\Delta_2+m-1,\widehat{\Delta}_1+\Delta_2+m+n+1} b_{m-1,n+2}^{\text{B},\text{cr}} =\left\{\begin{array}{lc}
B_{\frac{m}{2},\frac{n}{2}}^{\text{BB}\partial}\,, & \;m,n \text { even } \\
0\,, &  \text { others }
\end{array}\right.\;,
    \end{aligned}
\end{equation}
and similarly for $c_{m,n}^{\text{B},\text{cr}}$ with $\mathcal{O}_2 \leftrightarrow \mathcal{O}_3$ in (\ref{bulkcrossrecur2}). The recursion relation (\ref{bulkcrossrecur1}) can be solved after inputting the seed coefficients $a_{0,n}^{\text{B},\text{cr}}$ and $a_{m,0}^{\text{B},\text{cr}}$, and the relation (\ref{bulkcrossrecur2}) for $b_{m,n}^{\text{B},\text{cr}}$ (and the relation for $c_{m,n}^{\text{B},\text{cr}}$) can be solved with the seed coefficients $b_{0,n}^{\text{B},\text{cr}}$\,($c_{m,0}^{\text{B},\text{cr}}$).

\section{Mellin space representation}
\label{Mellin space representation}
\subsection{A brief review of the formalism}
A natural language to discuss holographic correlators is the Mellin space  \cite{Mack:2009mi,Penedones:2010ue}. In this formalism, holographic correlators are described by Mellin amplitudes which have simple analytic structure resembling that of the flat-space amplitudes. This formalism has also been extended to BCFTs \cite{Rastelli:2017ecj} and CFTs with defects  \cite{Goncalves:2018fwx}, where the Mellin amplitudes have the interpretation of form factors with a fixed source corresponding to the boundary or the defect. Here we will study the Witten diagrams considered in previous sections in Mellin space and obtain complementary results. In particular, these Witten diagrams which are difficult to evaluate in position space admit simple expressions in Mellin space. Moreover, as we will see, the Mellin space results also allow us to cross check our results for the conformal blocks.

We begin with a brief review of the BCFT Mellin formalism. We consider the correlation function of $n$ bulk and $m$ boundary operators
\begin{equation}
	\mathcal{C}_{n, m} \equiv\langle \mathcal{O}_{\Delta_1}(P_1) \cdots \mathcal{O}_{\Delta_n}(P_n) \widehat{\mathcal{O}}_{\widehat{\Delta}_1}(\widehat{P}_1) \cdots \widehat{\mathcal{O}}_{\widehat{\Delta}_m}(\widehat{P}_m)\rangle\;.
\end{equation}
We can represent it as a multi-dimensional inverse Mellin transformation \cite{Rastelli:2017ecj}
\begin{equation}
\begin{split}
		\label{BCFTMellintotal}
		\mathcal{C}_{n, m}=\int&\prod_{i<j} [d\delta_{i j}]\,(-2 P_{i} \cdot P_{j})^{-\delta_{i j}} \prod_{i, I}[d\gamma_{i I}]\,(-2 P_{i} \cdot \widehat{P}_{I})^{-\gamma_{i I}} \prod_{I<J} [d\beta_{I J}]\,(-2 \widehat{P}_{I} \cdot \widehat{P}_{J})^{-\beta_{I J}} \\
		& \times \prod_{i} [d\alpha_{i}]\,(P_{i} \cdot N_b)^{-\alpha_{i}}\, M(\delta_{i j}, \gamma_{i I}, \beta_{I J}, \alpha_{i})\;, 
\end{split}
\end{equation}
where $\delta_{ij}$, $\gamma_{iI}$, $\beta_{IJ}$, $\alpha_i$ are the BCFT Mellin variables. The rescaling relation (\ref{homon}) requires them to satisfy the following relations 
\begin{equation}
	\sum_{j} \delta_{i j}+\sum_{I} \gamma_{i I}+\alpha_{i}=\Delta_{i}\;, \quad
	\sum_{i} \gamma_{i I}+\sum_{J} \beta_{I J}=\widehat{\Delta}_{I}\;, 
\end{equation}
and $[d f]$ denotes the measure of the independent integration variables.\footnote{We have also absorbed into definition a factor of $\frac{1}{2\pi i}$ for each independent variable.} We further extract a factor of Gamma functions from 
$M(\delta_{i j}, \gamma_{i I}, \beta_{I J}, \alpha_{i})$ 
\begin{equation}
		\label{BCFTMellin}
		M(\delta_{i j}, \gamma_{i J}, \beta_{I J}, \alpha_{i})=	 \prod_{i<j} \Gamma(\delta_{i j}) \prod_{i, J} \Gamma(\gamma_{i J}) \prod_{I<J} \Gamma(\beta_{I J}) \prod_{i} \Gamma(\alpha_{i}) \frac{ \Gamma\left(-\frac{\mathcal{P}^{2}}{2}\right)}{\Gamma\left(-\mathcal{P}^{2}\right) }\, \mathcal{M}(\delta_{i j}, \gamma_{i J}, \beta_{I J}, \alpha_{i})\;,
\end{equation}
where 
\begin{equation}
    \mathcal{P}^{2}=-\sum_i \alpha_i\;,
\end{equation}
and $\mathcal{M}(\delta_{i j}, \gamma_{i J}, \beta_{I J}, \alpha_{i})$ is the BCFT Mellin amplitude. Here all the Gamma factors except for the last one have poles corresponding to multi-trace operators which are universal in holographic theories with large central charges. Factoring out these Gamma functions separates the multi-trace operator contribution from the single-trace one in the Mellin amplitude. This is in close analogy with the case without boundaries \cite{Mack:2009mi,Penedones:2010ue}.  The additional $\Gamma(-\mathcal{P}^2/2)/\Gamma(-\mathcal{P}^2)$ factor in the definition is motivated by the fact the Mellin amplitudes of contact Witten diagrams become a constant after extracting this factor. More precisely, \cite{Rastelli:2017ecj} showed that contact Mellin amplitudes are
\begin{equation}
	\label{BCFTcontactdiagram}
\mathcal{M}^{\text{contact}}_{\Delta_1,\ldots,\Delta_n, \widehat{\Delta}_{n+1},\ldots,\widehat{\Delta}_{n+m}}=\frac{\pi^{h^{\prime}} \Gamma\Big(\frac{\sum_{i=1}^n \Delta_i+\sum_{j=1}^m \widehat{\Delta}_{n+j}-2h^{\prime}}{2}\Big)}{2\prod_{i=1}^n \Gamma(\Delta_i) \prod_{j=1}^{m}\Gamma(\widehat{\Delta}_{n+j})}\;.
\end{equation}
However, the zeroes in $\Gamma(-\mathcal{P}^2/2)/\Gamma(-\mathcal{P}^2)$ in principle also allow the Mellin amplitude to have poles at these locations. As we will see later, this is a source of subtlety when we consider fields exchanged in the boundary channel. 
Since the Mellin representation of contact Witten diagrams has been analyzed in full generality, in this section we will focus on obtaining the Mellin amplitudes for exchange Witten diagrams.

\subsection{Witten diagrams of two-point function}
\label{Warm up: Witten diagrams of 2-point function}
As a warm-up, let us first consider the Witten diagrams for two-point functions. We demonstrate two methods, namely the EOM relation in Mellin space and integrated vertex identities, which will also be used in the study of three-point functions in the ensuing subsections. Compared to \cite{Rastelli:2017ecj}, in which the two-point Witten diagrams were also studied, the improvement here is the explicit expressions of two-point exchange Mellin amplitudes for generic conformal dimensions. We will also point out some interesting subtleties which were not noticed in \cite{Rastelli:2017ecj}. These subtleties will also be present in the three-point functions. Moreover, we will demonstrate in the simplest case how to use the Mellin amplitudes to cross check the position space expressions for conformal blocks. This gives us an alternative way to compute the conformal blocks.

For the two-point function $\left(\ref{2ptfunction}\right)$, the general Mellin representation becomes
\begin{equation}
	\begin{aligned}
		\label{mellin2pt}
		\langle \mathcal{O}_{\Delta_1}(P_1) \mathcal{O}_{\Delta_2}(P_2)\rangle= \int& [d \delta_{12} d \alpha_{1} d \alpha_{2} ]\Big[(-2 P_1 \cdot P_2)^{-\delta_{12}}  (P_{1} \cdot N_b)^{-\alpha_{1}}(P_{2} \cdot N_b)^{-\alpha_{2}} \\&\times \Gamma(\delta_{12})\Gamma(\alpha_{1}) \Gamma(\alpha_{2}) \frac{\Gamma\left(-\frac{\mathcal{P}^{2}}{2}\right)}{\Gamma(-\mathcal{P}^{2})} \mathcal{M}(\alpha_1,\alpha_2, \delta_{12})\bigg]\;.
	\end{aligned}
\end{equation}
Here $\delta_{12}+\alpha_{1}=\Delta_{1},  \delta_{12}+\alpha_{2}=\Delta_{2}$ and we can choose $s\equiv\Delta_1+\Delta_2-2\delta_{12}$ as the independent Mellin variable. Stripping off the kinematic factor (\ref{2ptfunction}), we get
\begin{equation}
	\begin{aligned}
		\label{mellin2pt1}
		&\mathcal{G}(\zeta)= \int \frac{ds}{4 \pi i}2^s \zeta^{-\frac{\Delta_1+\Delta_2-s}{2}} \frac{\Gamma\big(\frac{\Delta_1+\Delta_2-s}{2}\big)\Gamma\big(\frac{\Delta_1-\Delta_2+s}{2}\big) \Gamma\big(\frac{\Delta_2-\Delta_1+s}{2}\big) \Gamma(\frac{s}{2})}{\Gamma(s)}  \mathcal{M}(s)\;.
	\end{aligned}
\end{equation}

\vspace{0.3cm}
\noindent{\bf Using integrated vertex identities}
\vspace{0.2cm}

\noindent Let us first compute the Mellin amplitudes by using the integrated vertex identities. The bulk exchange Witten diagram
is defined to be
\begin{equation}
   W^{\text{2pt,B}}_\Delta=\int_{AdS_d}dZ_1\int_{AdS_{d+1}}dZ_2 \,G^\Delta_{BB,(d+1)}(Z_1,Z_2)G^{\Delta_1}_{B\partial}(P_1,Z_2)G^{\Delta_2}_{B\partial}(P_2,Z_2)\;. 
\end{equation}
It follows from the identity (\ref{intvertexid}) that the diagram can be expressed as the following infinite sums of contact Witten diagrams
\begin{equation}
\begin{aligned}
&\mathcal{W}_{\Delta}^{\text {2pt,B }}(\zeta)\\&=\sum_{i=0}^{\infty}T_{\Delta_1,\Delta_2,i} \,\zeta^i \mathcal{W}_{\Delta_1+i, \Delta_2+i}^{\text {contact }}(\zeta)+\sum_{i=0}^{\infty} Q_{\Delta_1,\Delta_2,i} \,\zeta^{\frac{\Delta-\Delta_1-\Delta_2}{2}+i} \mathcal{W}_{\frac{\Delta+\Delta_1-\Delta_2}{2}+i, \frac{\Delta-\Delta_1+\Delta_2}{2}+i}^{\text{contact}}(\zeta)\;,
\end{aligned}
\end{equation}
where the coefficients $T$, $Q$ were defined in (\ref{defT}), (\ref{defQ}). Using (\ref{BCFTcontactdiagram}), this gives the following expression for the exchange Mellin amplitude
\begin{equation}
\begin{split}
\label{mellinbulk2point}
\mathcal{M}^{\text{2pt,B}}_{\Delta}(s)=&\sum_{i=0}^{\infty} T_{\Delta_1,\Delta_2,i} \,\frac{\pi^{h^{\prime}} \Gamma\big(\frac{\Delta_1+\Delta_2+2i-2h^{\prime}}{2}\big)\Gamma\big(\frac{\Delta_1+\Delta_2+2i-s}{2}\big)}{2\,\Gamma(\Delta_1+i)\Gamma(\Delta_2+i)\Gamma\big(\frac{\Delta_1+\Delta_2-s}{2}\big)}\\	&+\sum_{i=0}^{\infty} Q_{\Delta_1,\Delta_2,i} \,\frac{\pi^{h^{\prime}}\Gamma\big(\frac{\Delta-s+2i}{2}\big)\Gamma\big(\frac{\Delta+2i-2h^{\prime}}{2}\big)}{2\, \Gamma\big(\frac{\Delta+\Delta_1-\Delta_2+2i}{2}\big)\Gamma\big(\frac{\Delta-\Delta_1+\Delta_2+2i}{2}\big)\Gamma\big(\frac{\Delta_1+\Delta_2-s}{2}\big)}\;.
\end{split}
\end{equation}
Both infinite sums can be performed and we find
\begin{equation}
\begin{split}\label{M2ptBresummed}
   &\mathcal{M}^{\text{2pt,B}}_{\Delta}(s)\\&= -\frac{\pi ^{h^{\prime}} \Gamma \big(\frac{-\Delta +\Delta _1+\Delta _2}{2} \big) \Gamma
   \big(\frac{-2 h^{\prime}+\Delta +\Delta _1+\Delta _2-1}{2}\big) }{8 \,\Gamma (\Delta _1) \Gamma (\Delta
   _2) \Gamma \big(\frac{\Delta _1+\Delta _2-s}{2}\big)}\Bigg[\Gamma \bigg(\frac{\Delta
   _1+\Delta _2-2h^{\prime}}{2}\bigg) \Gamma \bigg(\frac{\Delta _1+\Delta _2-s}{2}\bigg)\\&\times
   {}_3\tilde{F}_2\bigg(1,\frac{\Delta_1+\Delta _2-2 h^{\prime}}{2} ,\frac{\Delta
   _1+\Delta _2-s}{2};\frac{-\Delta +\Delta _1+\Delta _2+2}{2},\frac{-2 h^{\prime}+\Delta
   +\Delta _1+\Delta _2+1}{2};1\bigg)\\&-\Gamma \bigg(\frac{\Delta
   }{2}-h^{\prime}\bigg) \Gamma \bigg(\frac{\Delta -s}{2}\bigg) \,{}_2\tilde{F}_1\bigg(\frac{\Delta}{2}-h^{\prime},\frac{\Delta-s}{2};\Delta-h^{\prime} +\frac{1}{2};1\bigg)\Bigg]\;,
   \end{split}
\end{equation}
where ${}_p\tilde{F}_q$ is the regularized hypergeometric function. From these expressions, it seems that the Mellin amplitude has two series of poles at $s=\Delta+2m$ and $s=\Delta_1+\Delta_2+2m$, with $m=0,1,2,\ldots$. However, we expect the latter series to be fictitious and is actually not present. This is because we already have poles at $s=\Delta_1+\Delta_2+2m$ from the Gamma function $\Gamma(\frac{\Delta_1+\Delta_2-s}{2})$ in the definition (\ref{mellin2pt1}). The presence of the second series of poles in the Mellin amplitude leads to double poles in the integrand which should not be present at tree level.\footnote{Upon taking residues, these double poles give rise to logarithmic singularities which are associated with anomalous dimensions. However, anomalous dimensions should not appear at tree level in the bulk channel of two-point functions \cite{Mazac:2018biw}.} To check this, let us only take the poles at $s=\Delta+2m$ and try to write the Mellin amplitude as a sum over these poles
\begin{equation}
	\label{ansatz2ptex}
	\mathcal{M}^{\text{2pt,B}}_{\Delta}(s)=\sum_{m=0}^{\infty} \frac{\mathcal{R}_m^{\text{2pt},\text{B}}}{s-\Delta-2m}\;,
\end{equation}
where the residues can be easily extracted from (\ref{mellinbulk2point}) to be 
\begin{equation}
	\label{twofunctionQn}
	\mathcal{R}_m^{\text{2pt},\text{B}}=-\frac{\pi^{h^{\prime}}\Gamma\big(\frac{1-d+\Delta}{2}\big)\Gamma\big(\frac{-d+\Delta+\Delta_1+\Delta_2}{2}\big)\left(\frac{1+\Delta}{2}\right)_m \left(\frac{2+\Delta-\Delta_1-\Delta_2}{2}\right)_m}{4m! \Gamma(\Delta_1)\Gamma(\Delta_2)\Gamma(1-\frac{d}{2}+m+\Delta)}\;.
\end{equation}
We can perform this sum which gives 
\begin{equation}\label{M2ptBalternative}
    \mathcal{M}^{\text{2pt,B}}_{\Delta}(s)=\frac{\pi^{h^{\prime}}\Gamma\big(\frac{\Delta-2h^{\prime}}{2}\big)\Gamma\big(\frac{\Delta-s}{2}\big)\Gamma\big(\frac{\Delta+\Delta_1+\Delta_2-2h^{\prime}-1}{2}\big)}{8\,\Gamma(\Delta_1) \Gamma(\Delta_2)} { }_3 \tilde{F}_2\Bigg(\begin{array}{c|c}
\frac{\Delta+1}{2},\frac{\Delta-s}{2},\frac{\Delta-\Delta_1-\Delta_2+2}{2}\\
\frac{\Delta-s+2}{2},\frac{1}{2}-h^{\prime}+\Delta
\end{array}\, 1\Bigg)\;.
\end{equation}
One can verify numerically that (\ref{M2ptBresummed}) coincides with (\ref{M2ptBalternative}), which verifies the second series of poles is in fact absent. 

Let us also mention that the Mellin amplitude (\ref{ansatz2ptex}) also gives us an alternative way to compute the bulk channel conformal block. This is because the poles at $s=\Delta+2m$ corresponds to the exchange of an operator with dimension $\Delta$ in the bulk channel. In other words, the conformal block has the same poles and residues as the exchange Witten diagram in Mellin space. Using (\ref{ansatz2ptex}) in (\ref{mellin2pt1}) and taking residues at these poles, one obtains bulk channel conformal block (\ref{bulkCB2pt}).

We now turn to the boundary channel exchange diagram which is defined by
\begin{equation}
  W^{\text{2pt},\partial}_{\widehat{\Delta}}=\int_{AdS_d} dZ_1 dZ_2 \,G^{\Delta_1}_{B\partial}(P_1,Z_1) G^{\widehat{\Delta}}_{BB,(d)}(Z_1,Z_2) G^{\Delta_2}_{B\partial}(P_2,Z_2)\;. 
\end{equation}
For the $Z_1$ integral, we have the following integrated vertex identity \cite{Mazac:2018biw}
\begin{equation}
\begin{split}\label{boundaryvertexid}
\int_{AdS_d} dZ_1 \,G^{\Delta_1}_{B\partial}(P_1,Z_1) G^{\widehat{\Delta}}_{BB,(d)}(Z_1,Z_2)={}& \sum_{i=0}^{\infty} (x_{1,\perp})^{2 i} R_{\widehat{\Delta},\Delta_1,i}  G^{\Delta_1+2i}_{B\partial}(P_1,Z_2)\\
{}&+\sum_{i=0}^{\infty} (x_{1,\perp})^{\widehat{\Delta}-\Delta_1+2 i} S_{\widehat{\Delta},\Delta_1,i} G^{\widehat{\Delta}+2i}_{B\partial}(P_1,Z_2)\;,
\end{split}
\end{equation}
where 
\begin{equation}
	\label{R}
R_{\widehat{\Delta},\Delta_k,i}=\frac{(\Delta_k)_{2 i}}{2(\widehat{\Delta}-\Delta_k)\left(\frac{-\widehat{\Delta}+\Delta_k+2}{2}\right)_i\left(\frac{-d+\widehat{\Delta}+\Delta_k+1}{2}\right)_{i+1}}\;,
\end{equation}
\begin{equation}
	\label{S}
S_{\widehat{\Delta},\Delta_k,i}=\frac{(-1)^{i+1} \sin (\pi \Delta_k) \Gamma(1-\Delta_k) \cos\big(\frac{\pi(d-2 \widehat{\Delta})}{2}\big) \Gamma\big(\frac{\Delta_k-\widehat{\Delta}}{2}\big) \Gamma\big(\frac{d-2 i-2 \widehat{\Delta}-1}{2}\big)}{4 i ! \sin (\pi \widehat{\Delta}) \sin \big(\frac{\pi(-d+\widehat{\Delta}+\Delta_k+1)}{2}\big) \Gamma(-2 i-\widehat{\Delta}+1) \Gamma\big(\frac{d-\widehat{\Delta}-\Delta_k+1}{2}\big)}\;.
\end{equation}
This gives us the following representation of the boundary channel exchange Witten diagram as sums of contact Witten diagrams \cite{Mazac:2018biw}
\begin{equation}
\mathcal{W}^{\text{2pt},\partial}_{\widehat{\Delta}}(\zeta)=\sum_{i=0}^{\infty} 2^{-2 i} R_{\widehat{\Delta},\Delta_1,i}  \mathcal{W}_{\Delta_1+2 i, \Delta_2}^{\text {contact }}(\zeta)+\sum_{i=0}^{\infty} 2^{-\widehat{\Delta}+\Delta_1-2 i} S_{\widehat{\Delta},\Delta_1,i} \mathcal{W}_{\widehat{\Delta}+2 i, \Delta_2}^{\text {contact }}(\zeta)\;.
\end{equation}
In Mellin space, we then have
\begin{equation}
\begin{split}
\label{mellinbdy2point}
&\mathcal{M}^{\text{2pt},\partial}_{\widehat{\Delta}}(\delta_{12})=\sum_{i=0}^{\infty}  R_{\widehat{\Delta},\Delta_1,i} \frac{2^{-2 i}\pi^{h^{\prime}}\Gamma\big(\frac{\Delta_1+\Delta_2+2i-2h^{\prime}}{2}\big) (\Delta_1-\delta_{12})_{2i}}{2\,\Gamma(\Delta_1+2i) \Gamma(\Delta_2)\left(\frac{1+\Delta_1+\Delta_2-2\delta_{12}}{2}\right)_i}\\	&\quad+\sum_{i=0}^{\infty}  S_{\widehat{\Delta},\Delta_1,i} \frac{2^{-\widehat{\Delta}+\Delta_1-2i}\pi^{h^{\prime}}\Gamma(\widehat{\Delta}+2i-\delta_{12})\Gamma\big(\frac{\widehat{\Delta}+2i+\Delta_2-2h^{\prime}}{2}\big)\Gamma\big(\frac{1+\Delta_1+\Delta_2-2\delta_{12}}{2}\big)}{2\,\Gamma(\widehat{\Delta}+2i)\Gamma(\Delta_1-\delta_{12})\Gamma(\Delta_2)\Gamma\big(\frac{1+2i+\widehat{\Delta}+\Delta_2-2\delta_{12}}{2}\big)}\;,
\end{split}
\end{equation}
where it is more convenient to choose $\delta_{12}$ as the Mellin variable. Again this expression can be resummed and we get the following boundary exchange Mellin amplitude 
\begin{equation}
\begin{aligned}
\label{bdymellinbdy2point}
&\mathcal{M}^{\text{2pt},\partial}_{\widehat{\Delta}}(\delta_{12})= \frac{\pi ^{h^{\prime}+1} \Gamma \big(\frac{\Delta _1-\widehat{\Delta}}{2}\big) \Gamma
   \big(\frac{-2 \delta _{12}+\Delta _1+\Delta _2+1}{2}\big) \sec \big(\frac{\pi 
   (\widehat{\Delta} +\Delta _1-2 h^{\prime}+1)}{2} \big) }{8\,\Gamma
   (\Delta _1) \Gamma (\Delta _2) \Gamma (\Delta _1-\delta _{12}) \Gamma
   (h^{\prime}-\frac{\widehat{\Delta} }{2}-\frac{\Delta _1}{2}+1)} \times\\&\quad  \Bigg[\Gamma(\Delta _1-\delta _{12})
   \Gamma \bigg(\frac{\Delta _1+\Delta _2-2 h^{\prime}}{2} \bigg) \,{ }_4 \tilde{F}_3\Bigg(\begin{array}{c|c}
1,\frac{\Delta _1-\delta_{12}}{2},\frac{\Delta _1-\delta_{12}+1}{2},\frac{\Delta _1+\Delta_2-2h^{\prime}}{2}\\
\frac{\Delta _1-\widehat{\Delta}+2}{2},\frac{\widehat{\Delta}+\Delta_1+2-2h^{\prime} }{2},\frac{\Delta _1+\Delta_2+1-2\delta_{12}}{2}
\end{array}\, 1\Bigg)  \\&\quad-2^{\Delta _1-\widehat{\Delta}}\Gamma (\widehat{\Delta} -\delta _{12})\Gamma\Big(\frac{-2 h^{\prime}+\widehat{\Delta} +\Delta _2}{2}\Big) \, { }_3 \tilde{F}_2\Bigg(\begin{array}{c|c}
\frac{\widehat{\Delta} -\delta
   _{12}}{2},\frac{\widehat{\Delta} -\delta _{12}+1}{2},\frac{\widehat{\Delta}+\Delta
   _2-2 h^{\prime}}{2}\\
1-h^{\prime}+\widehat{\Delta},\frac{\widehat{\Delta}+\Delta _2 -2 \delta _{12}+1}{2}
\end{array}\, 1\Bigg)\Bigg]\;.
    \end{aligned}
\end{equation}
Note that in addition to the poles at $\delta_{12}=\widehat{\Delta}+2m$ which correspond to the exchanged boundary operator with dimension $\widehat{\Delta}$, there appears to be another series of poles at $\delta_{12}=\frac{\Delta_1+\Delta_2+1}{2}+m$. However, unlike the bulk exchange diagram case the latter series is not fictitious and must be present in the Mellin amplitude. This can be seen when we consider the EOM relation in Mellin space later in this subsection. We can also see the presence of these poles explicitly by considering a special case with $\Delta_1=\Delta_2=2$, $\widehat{\Delta}=1$ and $d=2$. We find the amplitude simplifies into
\begin{equation}
\mathcal{M}^{\text{2pt},\partial}_{\widehat{\Delta}=1}(\delta_{12})=\frac{\pi(3-2\delta_{12})\Gamma(2-\delta_{12})-4 \sqrt{\pi} \;\!\Gamma(\frac{5}{2}-\delta_{12})}{4\delta_{12}(\delta_{12}-1)\Gamma(2-\delta_{12})}\;.
\end{equation}
Clearly the Mellin amplitude contains these additional poles. One may be concerned that the presence of these poles leads to unexpected operator exchange in the boundary OPE channel. However, we note that the $\Gamma(-\frac{\mathcal{P}^2}{2})/\Gamma(-\mathcal{P}^2)$ in the definition (\ref{mellin2pt}) contains zeroes precisely at $\delta_{12}=\frac{\Delta_1+\Delta_2+1}{2}+m$. Therefore, the full integrand of the Mellin representation contains no unphysical poles. We should also comment that when the conformal dimensions satisfy the truncation conditions, {\it i.e.}, $\Delta_1-\widehat{\Delta}=2\mathbb{Z}_+$ or $\Delta_2-\widehat{\Delta}=2\mathbb{Z}_+$, the Mellin amplitude is a rational function with only finitely many poles in the series $\delta_{12}=\widehat{\Delta}+2m$, as was already pointed out in \cite{Rastelli:2017ecj}.

Now that we have clarified the pole structure, let us try to extract the conformal block from the exchange Mellin amplitude as in the bulk channel case. We only need to focus on the poles at $\delta_{12}=\widehat{\Delta}+2m$ which correspond to the exchange of the boundary operator with dimension $\widehat{\Delta}$
 \begin{equation}
    \label{boundary2ptMellin}
\mathcal{M}^{\text{2pt},\partial}_{\widehat{\Delta}}(\delta_{12})\supset \sum_{m=0}^{\infty} \frac{\mathcal{R}_m^{\text{2pt},\partial}}{\delta_{12}-\widehat{\Delta}-m}\;.
\end{equation}
The residues can be easily obtained from (\ref{mellinbdy2point}) and read
\begin{equation}
 \begin{split}
 \label{resibdy}
  \mathcal{R}_m^{\text{2pt},\partial}= & \frac{\pi ^{d/2} \Gamma \big(\frac{-2 \widehat{\Delta} +\Delta _1+\Delta _2-1}{2} \big) \Gamma \big(\frac{\widehat{\Delta} +\Delta _1-2h^{\prime}}{2}\big) \Gamma \big(\frac{\widehat{\Delta}+\Delta _2-2h^{\prime}}{2}\big) }{2 m!\Gamma(\Delta _1) \Gamma (\Delta _2)\Gamma \big(\frac{-\widehat{\Delta} +\Delta _1-1}{2}\big)  \Gamma \big(\frac{-\widehat{\Delta}+\Delta _2-1}{2} \big) \Gamma (1-h^{\prime}+\widehat{\Delta}) }\\&\times \frac{\left(\widehat{\Delta} -\Delta _1+2\right)_{m-1} \left(\widehat{\Delta} -\Delta _2+2\right)_{m-1} \left(-h^{\prime}+\widehat{\Delta}
   +\frac{3}{2}\right)_{m-1}}{\left(\widehat{\Delta}-\frac{\Delta _1+\Delta_2-3}{2}\right)_{m-1} (-2h^{\prime}+2 \widehat{\Delta} +2)_{m-1}}\;.
   \end{split}
\end{equation}
Inserting (\ref{boundary2ptMellin}) into (\ref{mellin2pt1}) and taking the residues at these poles one recovers the boundary channel conformal block (\ref{twopointblockboundry}) up to an overall factor.

\vspace{0.3cm}
\noindent{\bf Using EOM relations}
\vspace{0.2cm}

\noindent Another efficient way to compute Mellin amplitudes is to use the EOM relations. The bulk channel EOM operator for the two-point function is given by
  \begin{equation}
  	\label{2pteom}
  	\mathbf{EOM}_{\Delta}=\frac{1}{2}(\mathbf{L}_1+\mathbf{L}_2)^2+\Delta(\Delta-d)\;.
  \end{equation}
This differential operator translates into a difference operator in Mellin space
\begin{equation}
	\begin{gathered}
		\label{mellin2ptaction}
		\mathbf{EOM}_{\Delta} (\mathcal{M}(s))=(s(d-s)+\Delta(\Delta-d)) \mathcal{M}(s) -(\Delta_1+\Delta_2-s)(s-1) \mathcal{M}(s-2)\;.
	\end{gathered}
\end{equation}
For the bulk channel exchange Witten diagram, it gives the following EOM relation for the Mellin amplitudes
\begin{equation}
	\label{2pteomidentity}
\mathbf{EOM}_{\Delta}\left(\mathcal{M}^{\text{2pt,B}}_{\Delta}(s)\right)=\mathcal{M}_{\Delta_1,\Delta_2}^{\text{contact}}\;.
\end{equation}
The poles in the exchange Mellin amplitude (\ref{ansatz2ptex}) are shifted under the action of the difference operator. Requiring the absence of poles in the EOM identity implies the following recursion relation for the residues 
\begin{equation}
	\begin{aligned}
		\label{recursion2pt}
	((\Delta+2m)(d-\Delta-2m)+\Delta(\Delta-d)) \mathcal{R}_m^{\text{2pt},\text{B}}  =(1-2m-\Delta)(2m+\Delta-\Delta_1-\Delta_2) \mathcal{R}_{m-1}^{\text{2pt},\text{B}}\;,
	\end{aligned}
\end{equation}
which fully solves $\mathcal{R}_m^{\text{2pt},\text{B}}$ once the boundary value with $m=0$ is specified
\begin{equation}
\mathcal{R}_m^{\text{2pt},\text{B}}=\mathcal{R}_0^{\text{2pt},\text{B}}\frac{(\Delta +1) \left(\Delta -\Delta _1-\Delta _2+2\right) \Gamma \big(-\frac{d}{2}+\Delta +1\big) \left(\frac{\Delta +3}{2}\right)_{m-1} \left(\frac{\Delta -\Delta _1-\Delta _2+4}{2}\right)_{m-1}}{4 m! \Gamma \big(-\frac{d}{2}+m+\Delta +1\big)}\;.
\end{equation}
To obtain $\mathcal{R}_0^{\text{2pt},\text{B}}$ one can consider the bulk channel OPE limit $\zeta\to 0$ which is dominated by the pole at $s=\Delta$. In the bulk channel conformal block decomposition, this is given by the leading term in the conformal block $g^{\text{2pt,B}}(\zeta)$ and the OPE coefficient was given in \cite{Mazac:2018biw}. 
Comparing the two sides allows us to reproduce (\ref{twofunctionQn}). Alternatively, we can also fix $\mathcal{R}_0^{\text{2pt},\text{B}}$ by requiring that the regular terms in the EOM identity (\ref{2pteomidentity}) are also matched. This leads to the following nontrivial identity
\begin{equation}
\label{regur1}
    \sum_{m=0}^{\infty} (-\Delta_1-\Delta_2+d+1) \mathcal{R}_m^{\text{2pt,B}}=\mathcal{M}^{\text{contact}}_{\Delta_1,\Delta_2}\;.
\end{equation}
Note that the LHS contains $\Delta$ as a parameter while the contact Mellin amplitude on the RHS does not. Using the explicit expression (\ref{twofunctionQn}),  we have numerically checked that the relation (\ref{regur1}) is satisfied.

Similarly, we can consider the boundary exchange diagrams. The boundary channel EOM operator is defined by
\begin{equation}
\mathbf{EOM}_{\widehat{\Delta}}=\frac{1}{2} \widehat{\mathbf{L}}_1^2+\widehat{\Delta}(\widehat{\Delta}-d+1)\;,
\end{equation}
and has the following action on the Mellin amplitude
\begin{equation}
\begin{split}
\label{EOMAactonmellinbdy1}
\mathbf{EOM}_{\widehat{\Delta}}\left(\mathcal{M}(\delta_{12})\right)=&\left(-\delta_{12}(1-d+\delta_{12})+\widehat{\Delta}(\widehat{\Delta}-d+1)\right) \mathcal{M}(\delta_{12})\\&+\frac{(d-2\delta_{12})(\Delta_1-\delta_{12})(\Delta_2-\delta_{12})}{1+\Delta_1+\Delta_2-2\delta_{12}}\mathcal{M}(\delta_{12}-1)\;.
\end{split}
\end{equation}
The EOM relation for the boundary channel exchange Mellin amplitude reads
\begin{equation}
\label{eombdy2pt}
\mathbf{EOM}_{\widehat{\Delta}}\left(\mathcal{M}_{\widehat{\Delta}}^{\text{2pt},\partial}(\delta_{12})\right)=\mathcal{M}^{\text{contact}}_{\Delta_1,\Delta_2}\;.
\end{equation}
Note that unlike the bulk channel case, (\ref{EOMAactonmellinbdy1}) contains a pole at $\delta_{12}=\frac{\Delta_1+\Delta_2+1}{2}$ after the action. If the Mellin amplitude did not contain any poles in the series $\delta_{12}=\frac{\Delta_1+\Delta_2+1}{2}+m$, it would be impossible to cancel such poles so that (\ref{eombdy2pt}) is satisfied. On the other hand, as a necessary condition the poles in the series $\delta_{12}=\widehat{\Delta}+2m$ must also vanish in (\ref{eombdy2pt}). This gives us the following recursion relations for the residues 
\begin{equation}
    \frac{(d-2\widehat{\Delta}-2m)(\Delta_1-\widehat{\Delta}-m)(\Delta_2-\widehat{\Delta}-m)}{1+\Delta_1+\Delta_2-2\widehat{\Delta}-2m} \mathcal{R}_{m-1}^{\text{2pt},\partial}-m(1-d+m+2\widehat{\Delta}) \mathcal{R}_{m}^{\text{2pt},\partial}=0\;.
\end{equation}
This relation determines $\mathcal{R}_{m}^{\text{2pt},\partial}$ up to the $m=0$ boundary value, which can be fixed by using the conformal block decomposition of the Witten diagram. Finally, using (\ref{bdymellinbdy2point}) we have also checked numerically that (\ref{eombdy2pt}) is satisfied.

\subsection{Witten diagrams of B$\partial\partial$ three-point function}
In Section \ref{Witten diagrams Bpp case}, we discussed Witten diagrams for the B$\partial\partial$ three-point function, and obtained explicit expressions for contact and exchange diagram in position space. Here we consider the same diagrams in Mellin space. For B$\partial\partial$ three-point functions, 
 the Mellin representation (\ref{BCFTMellintotal}) becomes
\begin{equation}
	\begin{aligned}
		\label{mellintype1_1}
		\langle \widehat{\mathcal{O}}_{\widehat{\Delta}_1}(\widehat{P}_1) \mathcal{O}_{\Delta_2}(P_2) \mathcal{O}_{\Delta_3}(P_3)\rangle=\int& [d \gamma_{12} d \gamma_{13} d \beta_{23} d \alpha_{1}]\left[(-2 P_1 \cdot \widehat{P}_2)^{-\gamma_{12}}(-2 P_{1}\cdot \widehat{P}_{3} )^{-\gamma_{13}}\right.\\&\left.\times(-2\widehat{P}_2\cdot \widehat{P}_3)^{-\beta_{23}}\left(P_{1} \cdot N_b\right)^{-\alpha_{1}}   M(\gamma_{12},\gamma_{13},\beta_{23},\alpha_1)\right]\;,
	\end{aligned}
\end{equation}
with
\begin{equation}
	\begin{gathered}
	M(\gamma_{12},\gamma_{13},\beta_{23},\alpha_1) =  \Gamma(\gamma_{12}) \Gamma(\gamma_{13}) \Gamma(\beta_{23}) \Gamma(\alpha_1)  \frac{\Gamma(\frac{- \mathcal{P}^2}{2})}{\Gamma(- \mathcal{P}^2)}\mathcal{M} (\gamma_{12},\gamma_{13},\beta_{23},\alpha_1)\;, \\ \gamma_{12}+\gamma_{13}+\alpha_1=\Delta_1\;, \quad \gamma_{12}+\beta_{23}=\widehat{\Delta}_2\;,\quad \gamma_{13}+\beta_{23}=\widehat{\Delta}_3\;,\\ -\mathcal{P}^2=\Delta_1-\gamma_{12}-\gamma_{13}\;.
	\end{gathered}
\end{equation}
Choosing $s=\widehat{\Delta}_2+\widehat{\Delta}_3-2\beta_{23}$ to be independent variable, we get 
\begin{equation}
\begin{split}
	\label{mellintype1_2}
	&\mathcal{G}^{\text{B}\partial\partial}(\xi)= \\&\int  \frac{d s}{4\pi i}\, \xi^{\frac{s-\widehat{\Delta}_2-\widehat{\Delta}_3}{2}} \Gamma\bigg(\frac{\widehat{\Delta}_2-\widehat{\Delta}_3+s}{2}\bigg)\Gamma\bigg(\frac{\widehat{\Delta}_3-\widehat{\Delta}_2+s}{2}\bigg)\Gamma\bigg(\frac{\widehat{\Delta}_2+\widehat{\Delta}_3-s}{2}\bigg)\Gamma\bigg(\frac{\Delta_1-s}{2}\bigg) \mathcal{M}(s)\;.
 \end{split}
\end{equation}

 We focus on the exchange diagram (\ref{exch}). Using the integrated vertex identity (\ref{boundaryvertexid}) in the definition, we can write the exchange diagram as the following infinite sums of contact diagrams 
\begin{equation}
\mathcal{W}_{\widehat{\Delta}}^{\text{B}\partial\partial,\text{exchange}}(\xi)=\sum_{i=0}^{\infty}  R_{\widehat{\Delta},\Delta_1,i} \mathcal{W}^{\text{contact}}_{\Delta_1+2i,\widehat{\Delta}_2,\widehat{\Delta}_3}(\xi)+ \sum_{i=0}^{\infty} S_{\widehat{\Delta},\Delta_1,i} \mathcal{W}^{\text{contact}}_{\widehat{\Delta}+2i,\widehat{\Delta}_2,\widehat{\Delta}_3}(\xi)\;,
\end{equation}
where the coefficients were given in (\ref{R}) and (\ref{S}). In Mellin space, this translates to
\begin{equation}
	\begin{aligned}
		\label{10}
		\mathcal{M}_{\widehat{\Delta}}^{\text{B}\partial\partial,\text{exchange}}(s)=& \sum_{i=0}^{\infty}  R_{\widehat{\Delta},\Delta_1,i} \frac{\pi^{h^{\prime}}\Gamma(\frac{\Delta_1-s+2i}{2})\Gamma(\frac{\Delta_1+\widehat{\Delta}_2+\widehat{\Delta}_3+2i-2h^{\prime}}{2}) }{2\,\Gamma(\Delta_1+2i)\Gamma(\widehat{\Delta}_2)\Gamma(\widehat{\Delta}_3) \Gamma(\frac{\Delta_1-s}{2})}\\&+ \sum_{i=0}^{\infty} S_{\widehat{\Delta},\Delta_1,i} \frac{\pi^{h^{\prime}}\Gamma(\frac{\widehat{\Delta}-s+2i}{2})\Gamma(\frac{\widehat{\Delta}+\widehat{\Delta}_2+\widehat{\Delta}_3+2i-2h^{\prime}}{2})}{2\,\Gamma(\widehat{\Delta}+2i)\Gamma(\widehat{\Delta}_2)\Gamma(\widehat{\Delta}_3)\Gamma(\frac{\Delta_1-s}{2})}\;.
	\end{aligned}
\end{equation}
Similar to the bulk channel exchange diagram in the two-point function case, the Mellin amplitudes have physical poles at $s=\widehat{\Delta}+2m$ which correspond to the exchange of the operator with dimension $\widehat{\Delta}$. By contrast, the poles at $s=\Delta_1+2m$ coming from the first line of (\ref{10}) are in fact spurious. To see this, we can perform the sum in (\ref{10}) and obtain
\begin{equation}
\begin{split}
\label{MBppsum1}
&\mathcal{M}_{\widehat{\Delta}}^{\text{B}\partial\partial,\text{exchange}}(s)\\&=\frac{\pi ^{h^{\prime}+1} \Gamma \big(\frac{\Delta _1-\widehat{\Delta} }{2}\big) \sec \big(\frac{\pi (\widehat{\Delta} +\Delta _1-2 h^{\prime}+1)}{2}\big) }{8 \,\Gamma (\Delta _1) \Gamma (\widehat{\Delta} _2) \Gamma (\widehat{\Delta} _3) \Gamma \big(\frac{2h^{\prime}-\widehat{\Delta}-\Delta_1+2 }{2}\big) \Gamma \big(\frac{\Delta _1-s}{2}\big)} \Bigg[\Gamma \bigg(\frac{-2 h^{\prime}+\Delta _1+\widehat{\Delta} _2+\widehat{\Delta} _3}{2}\bigg)
   \Gamma\bigg(\frac{\Delta_1-s}{2}\bigg) \\&\times{}_3\tilde{F}_2\bigg(1,\frac{\Delta _1-s}{2},\frac{-2 h^{\prime}+\Delta _1+\widehat{\Delta} _2+\widehat{\Delta} _3}{2};\frac{-\widehat{\Delta} +\Delta _1+2}{2},\frac{-2 h^{\prime}+\widehat{\Delta} +\Delta _1+2}{2};1\bigg)\,-\\&\Gamma \bigg(\frac{-2 h^{\prime}+\widehat{\Delta} +\widehat{\Delta} _2+\widehat{\Delta} _3}{2} \bigg) \Gamma \bigg(\frac{\widehat{\Delta} -s}{2}\bigg) {}_2\tilde{F}_1\Bigg(\frac{\widehat{\Delta} -s}{2},\frac{-2 h^{\prime}+\widehat{\Delta}
   +\widehat{\Delta}_2+\widehat{\Delta}_3}{2};1-h^{\prime}+\widehat{\Delta} ;1\Bigg)\Bigg]\;.
   \end{split}
\end{equation}
On the other hand, the physical poles
\begin{equation}
\label{MellinBpp}
\mathcal{M}_{\widehat{\Delta}}^{\text{B}\partial\partial,\text{exchange}}(s)=\sum_{m=0}^{\infty}\frac{\mathcal{R}^{\text{B}\partial\partial}_m}{s-\widehat{\Delta}-2m}\;,
\end{equation}
with residues
\begin{equation}
\label{RBpp}
	\mathcal{R}^{\text{B}\partial\partial}_m=\frac{(-1)^{m+1} \pi ^{h^{\prime}}\Gamma\big(\frac{-2 h^{\prime}+\widehat{\Delta} +\Delta _1}{2}\big) \Gamma\big(\frac{-2 h^{\prime}+\widehat{\Delta} +\widehat{\Delta} _2+\widehat{\Delta} _3}{2}\big) \left(\frac{-2 m-\widehat{\Delta} +\Delta _1}{2}\right)_m \left(\frac{-2 m-\widehat{\Delta}
			+\widehat{\Delta} _2+\widehat{\Delta} _3}{2}\right)_m}{4 m! \Gamma (\Delta _1) \Gamma (\widehat{\Delta} _2) \Gamma (\widehat{\Delta} _3) \Gamma (-h^{\prime}+\widehat{\Delta} +1) \left(h^{\prime}-m-\widehat{\Delta} \right)_m}\;,
\end{equation}
give upon resummations
\begin{equation}\label{MBppalternative}
\begin{split}
&\mathcal{M}_{\widehat{\Delta}}^{\text{B}\partial\partial,\text{exchange}}(s)=\\&\frac{\pi^{h^{\prime}}\Gamma\big(\frac{\Delta-s}{2}\big)\Gamma\big(\frac{-2h^{\prime}+\widehat{\Delta}+\Delta_1}{2}\big)\Gamma\big(\frac{-2h^{\prime}+\widehat{\Delta}+\widehat{\Delta}_2+\widehat{\Delta}_3}{2}\big)}{8\,\Gamma(\Delta_1) \Gamma(\widehat{\Delta}_2)\Gamma(\widehat{\Delta}_3)} { }_3 \tilde{F}_2\Bigg(\begin{array}{c|c}
\frac{\widehat{\Delta}-s}{2},\frac{\widehat{\Delta}-\Delta_1+2}{2},\frac{\widehat{\Delta}-\widehat{\Delta}_2-\widehat{\Delta}_3+2}{2}\\
\frac{\widehat{\Delta}-s+2}{2},1-h^{\prime}+\widehat{\Delta}
\end{array}\, 1\Bigg)\;.
\end{split}
\end{equation}
One can check numerically that (\ref{MBppalternative}) is the same as (\ref{MBppsum1}), which is only possible if the spurious poles are absent. 

One can also go to Mellin space and use the EOM relation to check that the unphysical poles are absent. The EOM relation (\ref{eom}) becomes 
\begin{equation}
\mathbf{EOM}_{\widehat{\Delta}}^{\text{B}\partial\partial}\left(\mathcal{M}_{\widehat{\Delta}}^{\text{B}\partial\partial,\text{exchange}}(s)\right)=\mathcal{M}^{\rm contact}_{\Delta_1,\widehat{\Delta}_2,    \widehat{\Delta}_3}\;,    
\end{equation}
where the operator acts as 
\begin{equation}
\mathbf{EOM}_{\widehat{\Delta}}^{\text{B}\partial\partial} \left(\mathcal{M}(s)\right)=(\widehat{\Delta}-s)(\widehat{\Delta}+s-d+1)\mathcal{M}(s)+(\widehat{\Delta}_1-s)(\widehat{\Delta}_2+\widehat{\Delta}_3-s)\mathcal{M}(s-2)\;.
\end{equation}
One can check that the full Mellin amplitude (\ref{MBppalternative}) satisfies this relation. Focusing only on the singularities in (\ref{MellinBpp}) instead, one gets the following recursion relation for the residues 
\begin{equation}
\label{recurbpp}
  (-2m-\widehat{\Delta}+\Delta_1)(-2m-\widehat{\Delta}+\widehat{\Delta}_2+\widehat{\Delta}_3)\mathcal{R}^{\text{B}\partial\partial}_{m-1}+2 m (-1 + d - 2 m - 2 \widehat{\Delta})\mathcal{R}^{\text{B}\partial\partial}_m=0\;,
\end{equation}
which allows us to solve $\mathcal{R}^{\text{B}\partial\partial}_m$ up to an overall factor. As a consistency check, let us plug (\ref{MellinBpp}) in the Mellin representation (\ref{mellintype1_2}) and take the residues at $s=\widehat{\Delta}+2m$. This reproduces precisely the B$\partial\partial$ conformal block  (\ref{block}).

\subsection{Witten diagrams of BB$\partial$ three-point function}\label{MellinBBpartial}
For BB$\partial$ three-point functions, the Mellin representation $\left(\ref{BCFTMellintotal}\right)$ is
\begin{equation}
	\begin{aligned}
		\label{mellin3pt}
		&\langle \widehat{\mathcal{O}}_{\widehat{\Delta}_1}(\widehat{P}_1) \mathcal{O}_{\Delta_2}(P_2) \mathcal{O}_{\Delta_3}(P_3)\rangle= \int [d \gamma_{12} d \gamma_{13} d \delta_{23}d \alpha_{2}d \alpha_{3} ] \left[ (-2 P_2 \cdot P_3)^{-\delta_{23}}\right.\\&\left.\;\times(-2 \widehat{P}_{1}\cdot P_{2} )^{-\gamma_{12}} (-2 \widehat{P}_{1}\cdot P_{3} )^{-\gamma_{13}} 
		 \left(P_{2} \cdot N_b\right)^{-\alpha_{2}}  \left(P_{3} \cdot N_b\right)^{-\alpha_{3}} M(\gamma_{12},\gamma_{13},\delta_{23} ,\alpha_2 ,\alpha_3) \right]\;,
	\end{aligned}
\end{equation}
with
\begin{equation}
	\begin{gathered}
		M(\gamma_{12},\gamma_{13},\delta_{23} ,\alpha_2 ,\alpha_3)= \Gamma(\delta_{23}) \Gamma(\gamma_{12}) \Gamma(\gamma_{13}) \Gamma(\alpha_2) \Gamma(\alpha_3) \frac{\Gamma(\frac{- \mathcal{P}^2}{2})}{\Gamma(- \mathcal{P}^2)}\mathcal{M}(\gamma_{12},\gamma_{13},\delta_{23} ,\alpha_2 ,\alpha_3)\;,\\ \gamma_{12}+\gamma_{13}=\widehat{\Delta}_1\;,\quad\delta_{23}+\gamma_{12}+\alpha_2=\Delta_2\;, \quad \delta_{23}+\gamma_{13}+\alpha_3=\Delta_3 \;,
		\\-\mathcal{P}^2=\widehat{\Delta}_1+\Delta_2+\Delta_3-2\gamma_{12}-2\gamma_{13}-2\delta_{23}\;.
	\end{gathered}
\end{equation}
A convenient choice of the independent Mellin variables is $\alpha_2$ and $\alpha_3$ which selects the cross ratios $\zeta_1$, $\zeta_2$ defined in (\ref{newcrossratiosbdy}) as the dual variables. The corresponding Mellin representation for the stripped correlator is
\begin{equation}
	\begin{aligned}
		\label{3pointmellinrep}
		&\mathcal{G}^{\text{BB}\partial}(\zeta_1,\zeta_2)=\int \frac{d\alpha_2 d\alpha_3}{(2\pi i)^2}\bigg[ \zeta_1^{\alpha_2-\Delta_2} \zeta_2^{\alpha_3+\widehat{\Delta}_1-\Delta_3} \Gamma(\alpha_2)\Gamma(\alpha_3) \Gamma\Big(\frac{\alpha_2+\alpha_3}{2}\Big) \\&\quad\times\frac{\Gamma\big(\frac{-\widehat{\Delta}_1+\Delta_2+\Delta_3-\alpha_2-\alpha_3}{2}\big)\Gamma\big(\frac{\widehat{\Delta}_1+\Delta_2-\Delta_3-\alpha_2+\alpha_3}{2}\big) \Gamma\big(\frac{\widehat{\Delta}_1-\Delta_2+\Delta_3+\alpha_2-\alpha_3}{2}\big)}{2\,\Gamma(\alpha_2+\alpha_3)}\mathcal{M}(\alpha_2,\alpha_3)\Bigg]\;.
	\end{aligned}
\end{equation}
This parameterization turns out to be useful for considering exchange diagrams in the boundary channel. We can also choose $\delta_{23}$ and $\gamma_{12}$ as the independent variables. This gives instead
\begin{equation}
	\begin{aligned}
		\label{3pointmellinrepbulk}
	\mathcal{G}^{\text{BB}\partial}(\xi_1,\xi_2)=\int& \frac{d\delta_{23} d\gamma_{12}}{(2\pi i)^2} \bigg[\xi_1^{-\delta_{23}} \xi_2^{-\gamma_{12}}\Gamma(\delta_{23}) \Gamma(\gamma_{12}) \Gamma(\widehat{\Delta}_1-\gamma_{12})\Gamma(\Delta_2-\delta_{23}-\gamma_{12})\\&\times\Gamma(\Delta_3-\widehat{\Delta}_1-\delta_{23}+\gamma_{12})\frac{\Gamma\big(\frac{-\widehat{\Delta}_1+\Delta_2+\Delta_3-2\delta_{23}}{2}\big)}{\Gamma(-\widehat{\Delta}_1+\Delta_2+\Delta_3-2\delta_{23})} \mathcal{M}(\delta_{23},\gamma_{12})\Bigg]\;,
	\end{aligned}
\end{equation}
where the dual variables are $\xi_1$, $\xi_2$ defined in (\ref{BBpcrossratio}). This choice of variables is more convenient for considering exchange diagrams in the bulk channel. For later convenience, let us also define $s=\Delta_2+\Delta_3-2\delta_{23}$.

As in the previous subsections, we consider exchange Witten diagrams in Mellin space. In the bulk channel, the exchanged field can have arbitrary integer valued spin $J$. We will consider explicitly the cases with $J=0$ and $J=1$ in Section \ref{BBpartialMellinbulkJeq0} and Section \ref{BBpartialMellinbulkJeq1} respectively. While higher $J$ is technically more complicated, the two cases considered here are sufficient to demonstrate the main features. In Section \ref{BBpartialMellinboundary} we consider the boundary channel exchange Witten diagram.

\subsubsection{Bulk exchange diagram with $J=0$}\label{BBpartialMellinbulkJeq0}

 The bulk channel scalar exchange Witten diagram $W_{\Delta,0}^{\text{B}}$ was already discussed in Section \ref{Witten diagrams: The BBp case} where it was expressed in (\ref{intervertex}) as infinite sums of contact diagrams. In Mellin space, this translates to
\begin{equation}
	\begin{aligned}
		\label{scalarcontactrep}
		\mathcal{M}_{\Delta,0}^{\text{B}}(s)=& \sum_{i=0}^{\infty}T_{\Delta_2,\Delta_3,i} \frac{\pi^{h^{\prime}}\Gamma\big(\frac{\Delta_2+\Delta_3+2i-s}{2}\big)\Gamma\big(\frac{\widehat{\Delta}_1+\Delta_2+\Delta_3+2i-2h^{\prime}}{2}\big)}{2\,\Gamma(\widehat{\Delta}_1)\Gamma(\Delta_2+i)\Gamma(\Delta_3+i)\Gamma\big(\frac{\Delta_2+\Delta_3-s}{2}\big)}\\&+\sum_{i=0}^{\infty} Q_{\Delta_2,\Delta_3,i} \frac{\pi^{h^{\prime}}\Gamma\big(\frac{\Delta-s+2i}{2}\big)\Gamma\big(\frac{\Delta+\widehat{\Delta}_1+2i-2h^{\prime}}{2}\big)}{2\,\Gamma(\widehat{\Delta}_1)\Gamma\big(\frac{\Delta+\Delta_2-\Delta_3+2i}{2}\big) \Gamma\big(\frac{\Delta-\Delta_2+\Delta_3+2i}{2}\big)\Gamma\big(\frac{\Delta_2+\Delta_3-s}{2}\big)}\;,
	\end{aligned}
\end{equation}
which can be resummed and gives
\begin{equation}
\begin{split}
\label{MBBpB0sum}
    &\mathcal{M}_{\Delta,0}^{\text{B}}(s)=\\&-\frac{\pi ^{h^{\prime}} \Gamma \big(\frac{-\Delta +\Delta _2+\Delta _3}{2}\big) \Gamma
   \big(\frac{-2 h^{\prime}+\Delta +\Delta _2+\Delta _3-1}{2}\big)}{8 \,\Gamma (\widehat{\Delta}_1) \Gamma (\Delta _2)
   \Gamma(\Delta _3) \Gamma \big(\frac{-s+\Delta _2+\Delta _3}{2}\big)}\bigg[\Gamma \bigg(\frac{-2 h^{\prime}+\sum_i \Delta_i}{2}\bigg)
   \Gamma \bigg(\frac{-s+\Delta _2+\Delta _3}{2}\bigg)\\&\times{}_3\tilde{F}_2\bigg(1,\frac{-s+\Delta _2+\Delta _3}{2},\frac{-2 h^{\prime}+\sum_{i} \Delta_i}{2};\frac{-\Delta +\Delta _2+\Delta _3+2}{2},\frac{-2 h+\Delta +\Delta
   _2+\Delta _3+1}{2};1\bigg)\\&-\Gamma \bigg(\frac{-2 h^{\prime}+\Delta +\widehat{\Delta} _1}{2}\bigg) \Gamma \bigg(\frac{\Delta -s}{2}\bigg) \,
   {}_2\tilde{F}_1\bigg(\frac{\Delta -s}{2},\frac{-2 h^{\prime}+\Delta +\widehat{\Delta}_1}{2};-h^{\prime}+\Delta
   +\frac{1}{2};1\bigg)\Bigg]\;.
   \end{split}
\end{equation}
This bulk exchange amplitude is similar to the one in B$\partial\partial$ three-point functions and two-point functions. Only the poles at $s=\Delta+2m$ are physical while the poles at $s=\Delta_2+\Delta_3+2m$ are spurious. As a result, we can rewrite the Mellin amplitude as 
\begin{equation}
\label{scalarexchmellin}
	\mathcal{M}_{\Delta,0}^{\text{B}}(s)=\sum_{m=0}^{\infty} \frac{\mathcal{R}^{\text{B},0}_{m}}{s-\Delta-2m}\;,
\end{equation}
where the residues are constants
\begin{equation}
\label{residueBBpscalar}
	\mathcal{R}^{\text{B},0}_{m}=-\frac{\pi^{h^{\prime}}\Gamma \big(\frac{-2h^{\prime}+\Delta +\widehat{\Delta} _1}{2} \big) \Gamma\big(\frac{-2h^{\prime}-1+\Delta +\Delta _2+\Delta _3}{2}\big) \left(\frac{-2 m-\Delta +\widehat{\Delta} _1+1}{2}\right)_m \left(\frac{\Delta -\Delta
			_2-\Delta _3+2}{2} \right)_m}{4 m!\Gamma(\widehat{\Delta}_1)\Gamma(\Delta_2)\Gamma(\Delta_3) \Gamma \big(\frac{-2h^{\prime}+1+2\Delta}{2}\big)  \left(\frac{2h^{\prime}+1-2m-2\Delta}{2}\right)_m}\;.
\end{equation}
This expression can be resummed and we find 
\begin{equation}\label{MBBpB0alternative}
\begin{split}
&\mathcal{M}_{\Delta,0}^{\text{B}}(s)\\&=\frac{\pi^{h^{\prime}}\Gamma\big(\frac{\Delta-s}{2}\big)\Gamma\big(\frac{-2h^{\prime}+\Delta+\widehat{\Delta}_1}{2}\big)\Gamma\big(\frac{-1-2h^{\prime}+\Delta+\Delta_2+\Delta_3}{2}\big)}{8\,\Gamma(\widehat{\Delta}_1) \Gamma(\Delta_2)\Gamma(\Delta_3)} { }_3 \tilde{F}_2\bigg(\begin{array}{c|c}
\frac{\Delta-s}{2},\frac{\Delta-\widehat{\Delta}_1+1}{2},\frac{\Delta-\Delta_2-\Delta_3+2}{2}\\
\frac{\Delta-s+2}{2},\frac{1}{2}-h^{\prime}+\Delta
\end{array}\, 1\bigg)\;.
\end{split}
\end{equation}
We checked that (\ref{MBBpB0alternative}) is the same as (\ref{MBBpB0sum}) numerically. As we mentioned before, an interesting limit is  $\widehat{\Delta}_1 \to 0$ (and $\Delta_2 \to \Delta_1$, $\Delta_3 \to \Delta_2$) which reduces the three-point function to a two-point function. Note that the Mellin amplitudes already have the same form even before taking the limit. It is easy to check that $\mathcal{R}^{\text{B},0}_{m}$ nicely reduces to residues $\mathcal{R}^{\text{2pt,B}}_m$ in (\ref{twofunctionQn}) for two-point functions.

Let us also examine the EOM relation (\ref{scalarbulkeom}) in Mellin space which now translates to 
\begin{equation}\label{MellinEOMbulkJeq0}    \mathbf{EOM}^{\text{B}}_{\Delta,0}\left(\mathcal{M}^{\text{B}}_{\Delta,0}(s)\right)=\mathcal{M}^{\rm contact}_{\widehat{\Delta}_1,\Delta_2,\Delta_3}\;.
\end{equation}
The spin-0 EOM operator acts on the Mellin amplitude as
\begin{eqnarray}
    \nonumber&&\mathbf{EOM}^{\text{B}}_{\Delta,0}\left(\mathcal{M}(\delta_{23},\gamma_{12})\right)=-2 \gamma_{12}(\gamma_{12}-\delta_{23}-\widehat{\Delta}_1+\Delta_3) \mathcal{M}(\delta_{23},\gamma_{12}+1)-2(\widehat{\Delta}_1-\gamma_{12})\\\nonumber&&\times(\Delta_2-\gamma_{12}-\delta_{23}) \mathcal{M}(\delta_{23},\gamma_{12}-1)-2\delta_{23}(-\widehat{\Delta}_1+\Delta_2+\Delta_3-2\delta_{23}-1) \mathcal{M}(\delta_{23}+1,\gamma_{12})\\\nonumber&&+\left(-2 \gamma _{12} (2 \widehat{\Delta} _1+\Delta _2-\Delta _3)+4 \gamma _{12}^2+d (-2 \delta _{23}-\Delta +\Delta _2+\Delta _3)-(-2 \delta _{23}+\Delta _2+\Delta _3)^2\right.\\&&\left.+2 \widehat{\Delta} _1
   (\Delta _2-\delta _{23})+\Delta ^2\right) \mathcal{M}(\delta_{23},\gamma_{12})\;.
\end{eqnarray}
Focusing only on the singular terms from the poles in (\ref{scalarexchmellin}), we obtain a simple recursion relation for the residues
\begin{equation}
    2m\left(d-2(m+\Delta)\right) \mathcal{R}^{\text{B},0}_{m}+(2m+\Delta-\widehat{\Delta}_1-1)(2m+\Delta-\Delta_2-\Delta_3) \mathcal{R}^{\text{B},0}_{m-1}=0\;,
\end{equation}
which determines all these coefficients up to the initial condition $\mathcal{R}^{\text{B},0}_{0}$. Moreover, one can check that with the correct normalization as in (\ref{residueBBpscalar}) the EOM identity (\ref{MellinEOMbulkJeq0}) holds also for the regular terms. 

Finally, if we substitute the exchange amplitude (\ref{scalarexchmellin}) into the Mellin representation (\ref{3pointmellinrepbulk}), we expect that the Mellin integral reproduces the BB$\partial$ bulk scalar conformal block (\ref{blockbulkscalar}). To check this, we deform the $s$ contour to sum over the residues from $s=\Delta+2m$ poles where each summand is a complex integral of $\gamma_{12}$. We deform the $\gamma_{12}$ contour to the left to sum over the residues from $\gamma_{12}=-p$ and $\gamma_{12}=-m-q+(-\Delta+2\widehat{\Delta}_1+\Delta_2-\Delta_3)/2$ with $p,q=0,1,2,\ldots$, and obtain ${}_2 F_1$ hypergeometric functions. After using the hypergeometric identity (\ref{identity3}) we reproduce (\ref{blockbulkscalar}). 

\subsubsection{Bulk exchange diagram with $J=1$}\label{BBpartialMellinbulkJeq1}

Let us now consider the simplest spinning exchange Witten diagram which is defined by\footnote{Here we work with the Poincar\'e coordinates instead of the embedding space.}
\begin{equation}
	\begin{aligned}
		\label{spin1exch}
		&W_{\Delta,1}^{\text{B}}(\widehat{x}_1,x_2,x_3)\\&=\int_{A d S_{d}} \frac{d^d w}{w_0^d} \int_{A d S_{d+1}} \frac{d^{d+1} z}{z_0^{d+1}} \nabla_{\widehat{\mu}} G_{B \partial}^{\widehat{\Delta}_{1}}(\widehat{x}_{1}, w) G^{\Delta,\widehat{\mu} \nu}_{BB,(d+1)}(w, z)\left( G_{B \partial}^{\Delta_{\mathcal{O}}}(x_{2}, z) \stackrel{\leftrightarrow}{\nabla}_{\nu} G_{B \partial}^{\Delta_{\mathcal{O}}}(x_{3},z)\right).
	\end{aligned}
\end{equation}
For simplicity we will consider the case where the conformal dimensions of two bulk operators are the same $\Delta_2=\Delta_3=\Delta_{\mathcal{O}}$. Moreover, we assume that the spin 1 vector field $V_\mu$ is minimally coupled to the current made out of $\phi_2$, $\phi_3$ (dual to $\mathcal{O}_2$, $\mathcal{O}_3$ respectively)
\begin{equation}
V^\mu J_\mu=V^\mu \phi_2\stackrel{\leftrightarrow}{\nabla}_{\nu}\phi_3\;,
\end{equation}
and couples to the defect field $\widehat{\phi}_1$ (dual to $\widehat{\mathcal{O}}_1$) via
\begin{equation}
	\nabla ^{\widehat{\mu}} \widehat{\phi}_1 V_{\widehat{\mu}}\;.
\end{equation}
Here the hatted index $\widehat{\mu}$ is only along  the $AdS_{d}$ subspace while $\mu$ is along all directions of $AdS_{d+1}$. Let us denote the $z$-integral in $\left(\ref{spin1exch}\right)$ as
\begin{equation}
	A^{\widehat{\mu}}(w,x_2,x_3)=\int_{A d S_{d+1}} \frac{d^{d+1} z}{z_0^{d+1}} G^{\Delta,\widehat{\mu} \nu}_{BB,(d+1)}(w, z)\left( G_{B \partial}^{\Delta_{\mathcal{O}}}(x_{2}, z) \stackrel{\leftrightarrow}{\nabla}_{\nu} G_{B \partial}^{\Delta_{\mathcal{O}}}(x_{3}, z)\right).
\end{equation}
This integral has been computed in \cite{DHoker:1999mqo} in a special truncating case and an integrated vertex identity was derived (see also \cite{Goncalves:2019znr,Bissi:2022mrs}). In Appendix \ref{Integrated vertex identity for vector exchange}, we generalize the integrated vertex identity to general $\Delta_{\mathcal{O}}$ and $\Delta$. Using this identity we can express the spin-1 exchange diagram as the following infinite sums of contact diagrams
\begin{equation}
	\begin{aligned}
 \label{spin1exchangeascontact}
		&\mathcal{W}^{\text{B}}_{\Delta,1}(\xi_1,\xi_2)=-2\widehat{\Delta}_1\sum_{k=0}^{\infty} P_k (\Delta_{\mathcal{O}}+k) \xi_1^k \xi_2 \mathcal{W}^{\text{contact}}_{\widehat{\Delta}_1+1,\Delta_{\mathcal{O}}+k+1,\Delta_{\mathcal{O}}+k}(\xi_1,\xi_2)+2\widehat{\Delta}_1 \\&\times\sum_{k=0}^{\infty} P_k (\Delta_{\mathcal{O}}+k)\xi_1^k \mathcal{W}^{\text{contact}}_{\widehat{\Delta}_1+1,\Delta_{\mathcal{O}}+k,\Delta_{\mathcal{O}}+k+1}(\xi_1,\xi_2)-2\widehat{\Delta}_1 \sum_{k=0}^{\infty} Q_k \left(\frac{\Delta-1}{2}+k\right) \xi_1^{\frac{\Delta-1}{2}-\Delta_{\mathcal{O}}+k} \\&  \times\xi_2\mathcal{W}^{\text{contact}}_{\widehat{\Delta}_1+1,\frac{\Delta+1}{2}+k,\frac{\Delta-1}{2}+k}(\xi_1,\xi_2)+ 2\widehat{\Delta}_1 \sum_{k=0}^{\infty}\bigg[ Q_k \left(\frac{\Delta-1}{2}+k\right) \xi_1^{\frac{\Delta-1}{2}-\Delta_{\mathcal{O}}+k} \\&\times\mathcal{W}^{\text{contact}}_{\widehat{\Delta}_1+1,\frac{\Delta-1}{2}+k,\frac{\Delta+1}{2}+k}(\xi_1,\xi_2)\bigg]\;,
	\end{aligned}
\end{equation}
where the expressions for the coefficients $P_k$ and $Q_k$ can be found in (\ref{Pk}), (\ref{Qk}). Using the Mellin representation $\left(\ref{3pointmellinrep}\right)$, we can translate this position space expression into Mellin space and get
\begin{equation}
    \begin{split}
    \label{spin1contactrep}
        &\mathcal{M}_{\Delta,1}^{\text{B}}(s,\gamma_{12})=\Big(\gamma_{12}-\frac{\widehat{\Delta}_1}{2}\Big) \times\\&\Bigg[\sum_{k=0}^{\infty}\frac{2\pi ^{h^{\prime}}  \Gamma (-h^{\prime}+k+\Delta_{\mathcal{O}}+\frac{\Delta _1}{2}+1) \Gamma (k-\frac{s}{2}+\Delta_{\mathcal{O}})\left[\left(\frac{3-\Delta }{2}+\Delta_{\mathcal{O}}\right)_k\left(\frac{2-2h^{\prime}+\Delta
   +2\Delta_{\mathcal{O}}}{2} \right)_k\right]^{-1}}{ (\Delta -2 \Delta_{\mathcal{O}}-1)
   (\Delta +2 \Delta_{\mathcal{O}}-2h^{\prime}) \Gamma(\Delta _1) \Gamma (\Delta_{\mathcal{O}})^2  \Gamma(\Delta_{\mathcal{O}}-\frac{s}{2}) }\\& +\sum_{k=0}^{\infty}\frac{\pi^{h^{\prime}}\Gamma\big(\frac{1-\Delta}{2}+\Delta_{\mathcal{O}}\big)\Gamma\big(\frac{\Delta}{2}-h^{\prime}+\Delta_{\mathcal{O}}\big)\Gamma\big(\frac{1-2h^{\prime}+2k+\Delta+\widehat{\Delta}_1}{2}\big)\Gamma\big(\frac{-1+2k-s+\Delta}{2}\big)}{2k!\;\!\Gamma(\widehat{\Delta}_1)\Gamma(\Delta_{\mathcal{O}})^2 \Gamma(\Delta_{\mathcal{O}}-\frac{s}{2})\Gamma(1-\frac{d}{2}+\Delta+k)}\Bigg] \;.
\end{split}
\end{equation}
The summations in (\ref{spin1contactrep}) can be performed, which leads to
\begin{equation}
\begin{split}
\label{MBppB1sum1}
  &\mathcal{M}_{\Delta,1}^{\text{B}}(s,\gamma_{12})\\&= -\frac{\pi ^{h^{\prime}} \big(\gamma_{12}-\frac{\widehat{\Delta}_1}{2}\big)\Gamma(\frac{1-\Delta}{2}+\Delta _{\mathcal{O}}) \Gamma (-h^{\prime}+\frac{\Delta
   }{2}+\Delta _{\mathcal{O}}) }{2\, \Gamma(\widehat{\Delta}_1)\Gamma (\Delta _{\mathcal{O}}){}^2  \Gamma
   (\Delta _{\mathcal{O}}-\frac{s}{2})} \Bigg[\Gamma \bigg(1-h^{\prime}+\Delta
   _{\mathcal{O}}+\frac{\widehat{\Delta} _1}{2}\bigg) \Gamma \Big(\Delta _{\mathcal{O}}-\frac{s}{2}\Big)\\& \times {}_3\tilde{F}_2\bigg(1,\Delta
   _{\mathcal{O}}-\frac{s}{2},-h^{\prime}+\Delta _{\mathcal{O}}+\frac{\widehat{\Delta} _1}{2}+1;-\frac{\Delta }{2}+\Delta _{\mathcal{O}}+\frac{3}{2},-h^{\prime}+\frac{\Delta
   }{2}+\Delta _{\mathcal{O}}+1;1\bigg)-\\&\Gamma\bigg(\frac{1-2 h^{\prime}+\Delta +\widehat{\Delta}_1}{2}\bigg) \Gamma
   \bigg(\frac{\Delta-s-1}{2} \bigg) {}_2\tilde{F}_1\bigg(\frac{\Delta-s -1}{2},\frac{1-2 h^{\prime}+\Delta +\widehat{\Delta}_1}{2};\Delta-h^{\prime} +\frac{1}{2};1\bigg)\Bigg]\;.
   \end{split}
\end{equation}
As in the scalar case, we expect that only the poles at $s=\Delta-1+2m$ are physical while the poles at $s=2\Delta_{\mathcal{O}}+2m$ are spurious. This gives the following tentative expression for the spin-1 exchange Mellin amplitudes as a sum over simple poles
\begin{equation}
	\label{ansatzJ1}
	\mathcal{M}_{\Delta,1}^{\text{B}}(s,\gamma_{12})=\sum_{m=0}^{\infty} \frac{\mathcal{R}^{\text{B},1}_{m}(\gamma_{12})}{s-(\Delta-1)-2m}\;.
\end{equation}
However, the numerators are now linear in the other independent Mellin variable $\gamma_{12}$
\begin{equation}
	\begin{aligned}
 \label{spin1residue}
	&\mathcal{R}^{\text{B},1}_{m}(\gamma_{12})=\\&-\frac{\pi^{h^{\prime}}\Gamma\big(\frac{1-2h^{\prime}+\Delta+\widehat{\Delta}_1}{2}\big)  \Gamma\big(\frac{-2h^{\prime}+\Delta}{2}+\Delta_{\mathcal{O}}\big) \left(\frac{1-\Delta}{2}-m+\Delta_{\mathcal{O}}\right)_m\left(\frac{2-2m-\Delta+\widehat{\Delta}_1}{2}\right)_{m} }{m!\Gamma(\widehat{\Delta}_1)\Gamma(\Delta_{\mathcal{O}})^2 \Gamma(-h^{\prime}+\frac{1}{2}+m+\Delta)} \bigg(\gamma_{12}-\frac{\widehat{\Delta}_1}{2}\bigg)\;.
	\end{aligned}
\end{equation}
Resumming (\ref{ansatzJ1}) we get
\begin{equation}\label{MBBpB1alternative}
\begin{split}
&\mathcal{M}_{\Delta,1}^{\text{B}}(s,\gamma_{12})=\bigg(\gamma_{12}-\frac{\widehat{\Delta}_1}{2}\bigg)\\&\times\frac{\pi^{h^{\prime}}\Gamma\big(\frac{\Delta-s-1}{2}\big)\Gamma\big(\frac{1-2h^{\prime}+\Delta+\widehat{\Delta}_1}{2}\big)\Gamma\big(\frac{-2h^{\prime}+\Delta}{2}+\Delta_{\mathcal{O}}\big)}{2\,\Gamma(\widehat{\Delta}_1) \Gamma(\Delta_{\mathcal{O}})^2} { }_3 \tilde{F}_2\Bigg(\begin{array}{c|c}
\frac{\Delta-s-1}{2},\frac{\Delta-2\Delta_{\mathcal{O}}+1}{2},\frac{\Delta-\widehat{\Delta}_1}{2}\\
\frac{\Delta-s+1}{2},\frac{1}{2}-h^{\prime}+\Delta
\end{array}\, 1\Bigg) \;.
\end{split}
\end{equation}
We checked numerically that (\ref{MBBpB1alternative}) is equal to (\ref{MBppB1sum1}), which confirms the absence of the spurious poles.

The spin-1 exchange Mellin amplitude also satisfies the following EOM relation in Mellin space
\begin{equation}
\label{eomspin1}
    \mathbf{EOM}^{\text{B}}_{\Delta,1}\left( \mathcal{M}_{\Delta,1}^{\text{B}}(s,\gamma_{12})\right) =\mathcal{M}^{\text{contact,2-der}}_{\widehat{\Delta}_1,\Delta_{\mathcal{O}},\Delta_{\mathcal{O}}}(\gamma_{12})\;.
\end{equation}
Here $\mathcal{M}^{\text{contact,2-der}}_{\widehat{\Delta}_1,\Delta_{\mathcal{O}},\Delta_{\mathcal{O}}}(\gamma_{12})$ is the Mellin amplitude of contact diagram with a pair of derivatives in the vertex defined as below
\begin{equation}
	\begin{aligned}
 \label{contact1derivative}
W^{\text{contact,2-der}}_{\widehat{\Delta}_1,\Delta_{\mathcal{O}},\Delta_{\mathcal{O}}}(x_i)=\int_{A d S_{d}} \frac{d^d w}{w_0^d}  \nabla_{\widehat{\mu}} G_{B \partial,(d)}^{\widehat{\Delta}_{1}}(\widehat{x}_{1}, w) g^{\widehat{\mu}\nu}(w)\left( G_{B \partial}^{\Delta_{\mathcal{O}}}(x_{2}, w) \stackrel{\leftrightarrow}{\nabla}_{\nu} G_{B \partial}^{\Delta_{\mathcal{O}}}(x_{3},w)\right)\;.
\end{aligned}
\end{equation}
and is linear in $\gamma_{12}$ 
\begin{equation}
\label{spin1contactmellin}
\mathcal{M}^{\text{contact,2-der}}_{\widehat{\Delta}_1,\Delta_{\mathcal{O}},\Delta_{\mathcal{O}}}(\gamma_{12})=\frac{2 \pi^{h^{\prime}}\Gamma \big(\frac{\widehat{\Delta}_1+2\Delta_{\mathcal{O}}+2-2h^{\prime}}{2}\big)\big(\gamma_{12}-\frac{\widehat{\Delta}_1}{2}\big)}{\Gamma(\widehat{\Delta}_1)\Gamma(\Delta_{\mathcal{O}})^2}\;.
\end{equation}
This Mellin amplitude can be obtained from the position space relation (\ref{spin1contactasDfunction}). The spin-1 EOM operator has the following action
\begin{eqnarray}
    \nonumber&&\mathbf{EOM}^{\text{B}}_{\Delta,1}\left(\mathcal{M}(\delta_{23},\gamma_{12})\right)=-2 \gamma_{12}(\gamma_{12}-\delta_{23}-\widehat{\Delta}_1+\Delta_{\mathcal{O}}) \mathcal{M}(\delta_{23},\gamma_{12}+1)-2(\widehat{\Delta}_1-\gamma_{12})\\\nonumber&&\times(\Delta_{\mathcal{O}}-\gamma_{12}-\delta_{23}) \mathcal{M}(\delta_{23},\gamma_{12}-1)-2\delta_{23}(-\widehat{\Delta}_1+2\Delta_{\mathcal{O}}-2\delta_{23}-1) \mathcal{M}(\delta_{23}+1,\gamma_{12})+\\\nonumber&&\left(-4\widehat{\Delta}_1 \gamma _{12}+4 \gamma _{12}^2+2 (\widehat{\Delta} _1-2 \Delta_{\mathcal{O}}+2\delta_{23}+d)
   (\Delta_{\mathcal{O}}-\delta _{23})+\Delta (\Delta-d)+d-1\right) \mathcal{M}(\delta_{23},\gamma_{12})\;.
\end{eqnarray}
We will use the ansatz (\ref{ansatzJ1}) for the exchange Mellin amplitude and we write the residue $\mathcal{R}^{\text{B},1}_{m}(\gamma_{12})$ as
\begin{equation}
    \mathcal{R}^{\text{B},1}_{m}(\gamma_{12})=\mathcal{A}^{\text{B},1}_{m} \gamma_{12}+ \mathcal{B}^{\text{B},1}_{m}\;,
\end{equation}
where $\mathcal{A}^{\text{B},1}_{m}$ and $\mathcal{B}^{\text{B},1}_{m}$ are independent of $s$. Requiring the poles to cancel we get two recursion relations
\begin{eqnarray}
   \label{spin1recur1} &&2m(d-2(m+\Delta)) \mathcal{A}^{\text{B},1}_{m}+(-2+2m+\Delta-\widehat{\Delta}_1)(-1+2m+\Delta-2\Delta_{\mathcal{O}}) \mathcal{A}^{\text{B},1}_{m-1}=0\\
   \nonumber&&
   2\left(-1+\Delta+m\left(d-2(-1+m+\Delta)\right)\right) \mathcal{B}^{\text{B},1}_{m}+(-2+2m+\Delta-\widehat{\Delta}_1)(-1+2m+\Delta-2 \Delta_{\mathcal{O}})\\\label{spin1recur2}&&\times\mathcal{B}^{\text{B},1}_{m-1}+\widehat{\Delta}_1(-1+2m+\Delta) \mathcal{A}^{\text{B},1}_{m}=0\;.
\end{eqnarray}
The recursion relation (\ref{spin1recur1}) can be first solved and determines $\mathcal{A}^{\text{B},1}_{m}$ up to an overall constant $C_A$
\begin{equation}
\label{solutionAm}
    \mathcal{A}^{\text{B},1}_{m}=C_A\frac{\left(\frac{\Delta-\widehat{\Delta}_1+2}{2}\right)_{m-1} \left(\frac{\Delta-2\Delta_{\mathcal{O}}+3}{2}\right)_{m-1}}{m!\left(\Delta-\frac{d}{2}+2\right)_{m-1}}\;.
\end{equation}
Inserting it into (\ref{spin1recur2}) leads to an inhomogeneous recurrence equation for $\mathcal{B}^{\text{B},1}_{m}$. The boundary value $\mathcal{B}^{\text{B},1}_{0}$ can then be determined by setting $m=0$ in (\ref{spin1recur2}) and requiring $\mathcal{B}^{\text{B},1}_{-1}=0$. This allows us to solve (\ref{spin1recur2}) and gives
\begin{equation}
    \mathcal{B}^{\text{B},1}_{m}=-\frac{\widehat{\Delta}_1}{2} \mathcal{A}^{\text{B},1}_{m}\;.
\end{equation}
Therefore, we have solved $\mathcal{R}_m^{\text{B},1}$ from the recursion relations up to an undetermined overall constant $C_A$. This constant can be fixed by requiring the EOM identity (\ref{eomspin1}) to hold also for the regular terms. The result is just (\ref{spin1residue}).

Let us conclude the discussion of bulk channel exchange Mellin amplitudes by making some comments for the cases with spins $J\geq 2$. For $J=2$, the integrated vertex identity has been worked out for gravitons in \cite{DHoker:1999mqo} and for massive symmetric tensors in \cite{Arutyunov:2002fh,Rastelli:2017udc}. For $J>2$ such identities are currently unknown. Extrapolating the $J=0$ and $J=1$ cases studied here, we expect the spin-$J$ exchange Mellin amplitude to have the same form 
\begin{equation}
	\mathcal{M}_{\Delta,J}^{\text{B}}(s,\gamma_{12})=\sum_{m=0}^{\infty} \frac{\mathcal{R}^{\text{B},J}_{m}(\gamma_{12})}{s-(\Delta-J)-2m}\;,
\end{equation}
where $\mathcal{R}^{\text{B},J}_{m}(\gamma_{12})$ is a degree-$J$ polynomial in $\gamma_{12}$. The EOM relations, which now have $2J$-derivative contact terms on the RHS, should still allow us to fix the numerators $\mathcal{R}^{\text{B},J}_{m}(\gamma_{12})$. But the explicit analysis will become increasingly difficult with higher values of $J$. We leave a systematic discussion of the Mellin amplitudes of these higher-spin exchange Witten diagrams for the future.

\subsubsection{Boundary exchange diagram}\label{BBpartialMellinboundary}

Let us now consider the boundary exchange diagram in Mellin space. Using the integrated vertex identity (\ref{boundaryvertexid}) twice in its definition (\ref{defbdyexchdiag}), we obtain the following representation in position space which expresses the exchange diagram as double infinite sums of contact diagrams
\begin{equation}
	\begin{aligned}
		\label{14}
		&\mathcal{W}^{\partial}_{\widehat{\Delta},\widehat{\Delta}^{\prime}}(\xi_1,\xi_2)=\\&\quad\sum_{i=0}^{\infty}\sum_{j=0}^{\infty} R_{\widehat{\Delta},\Delta_2,i} R_{\widehat{\Delta}^{\prime},\Delta_3,j} \mathcal{W}^{\text{contact}}_{\widehat{\Delta}_1,\Delta_2+2i,\Delta_3+2j}+\sum_{i=0}^{\infty}\sum_{j=0}^{\infty} R_{\widehat{\Delta},\Delta_2,i} S_{\widehat{\Delta}^{\prime},\Delta_3,j} \mathcal{W}^{\text{contact}}_{\widehat{\Delta}_1,\Delta_2+2i,\widehat{\Delta}^{\prime}+2j}
   \\&\quad+\sum_{i=0}^{\infty}\sum_{j=0}^{\infty} S_{\widehat{\Delta},\Delta_2,i}R_{\widehat{\Delta}^{\prime},\Delta_3,j}\mathcal{W}^{\text{contact}}_{\widehat{\Delta}_1,\widehat{\Delta}+2i,\Delta_3+2j}
  +\sum_{i=0}^{\infty}\sum_{j=0}^{\infty} S_{\widehat{\Delta},\Delta_2,i}S_{\widehat{\Delta}^{\prime},\Delta_3,j} \mathcal{W}^{\text{contact}}_{\widehat{\Delta}_1,\widehat{\Delta}+2i,\widehat{\Delta}^{\prime}+2j}\;.
	\end{aligned}
\end{equation}
Using the Mellin representation (\ref{3pointmellinrepbulk}), we obtain the following representation of the Mellin amplitude
\begin{equation}
	\begin{aligned}
 \label{bdymellinvertexidentity}
		&\mathcal{M}^{\partial}_{\widehat{\Delta},\widehat{\Delta}^{\prime}}(\alpha_2,\alpha_3)=\\&\sum_{i=0}^{\infty}\sum_{j=0}^{\infty} R_{\widehat{\Delta},\Delta_2,i} R_{\widehat{\Delta}^{\prime},\Delta_3,j}\,\mathcal{F}_{\Delta_2+2i,\Delta_3+2j}(\alpha_2,\alpha_3)+\sum_{i=0}^{\infty}\sum_{j=0}^{\infty} R_{\widehat{\Delta},\Delta_2,i} S_{\widehat{\Delta}^{\prime},\Delta_3,j}\, \mathcal{F}_{\Delta_2+2i,\widehat{\Delta}^{\prime}+2j}(\alpha_2,\alpha_3)\\&+\sum_{i=0}^{\infty}\sum_{j=0}^{\infty} S_{\widehat{\Delta},\Delta_2,i}R_{\widehat{\Delta}^{\prime},\Delta_3,j} \,\mathcal{F}_{\widehat{\Delta}+2i,\Delta_3+2j}(\alpha_2,\alpha_3)+\sum_{i=0}^{\infty}\sum_{j=0}^{\infty}   S_{\widehat{\Delta},\Delta_2,i}S_{\widehat{\Delta}^{\prime},\Delta_3,j} \,\mathcal{F}_{\widehat{\Delta}+2i,\widehat{\Delta}^{\prime}+2j}(\alpha_2,\alpha_3)\;,
	\end{aligned}
\end{equation}
where we have defined the function
\begin{equation}
\begin{split}
\label{funf}
    &\mathcal{F}_{\Delta_0,\Delta_0^{\prime}}(\alpha_2,\alpha_3)\\&=\frac{2^{\Delta_2+\Delta_3-\Delta_0-\Delta_0^{\prime}}\pi^{h^{\prime}}\Gamma\big(\frac{-2h^{\prime}+\Delta_0+\Delta_0^{\prime}+\widehat{\Delta}_1}{2}\big)\Gamma(\alpha_2+\Delta_0-\Delta_2)\Gamma(\alpha_3+\Delta_0^{\prime}-\Delta_3)\Gamma\big(\frac{1+\alpha_2+\alpha_3}{2}\big)}{2\,\Gamma(\widehat{\Delta}_1)\Gamma(\Delta_0)\Gamma(\Delta_0^{\prime})\Gamma\big(\frac{1+\alpha_2+\alpha_3+\Delta_0+\Delta_0^{\prime}-\Delta_2-\Delta_3}{2}\big)\Gamma(\alpha_2)\Gamma(\alpha_3)}\;.
    \end{split}
\end{equation}
The analytic structure of this Mellin amplitude is more difficult to analyze due to the presence of the double infinite sums. In addition to the poles at $\alpha_2=-\widehat{\Delta}+\Delta_2-m$, $\alpha_3=-\widehat{\Delta}^{\prime}+\Delta_3-n$, which we expect to be associated with the exchange of the operators with dimensions $\widehat{\Delta}$ and $\widehat{\Delta}'$, there are many other poles in (\ref{bdymellinvertexidentity}). For simplicity we will restrict ourselves to the truncating case where the conformal dimensions satisfy $\Delta_2-\widehat{\Delta}=2p\in \mathbb{Z}^{>0}_{\rm even}$, $\Delta_3-\widehat{\Delta}^{\prime}=2q\in \mathbb{Z}^{>0}_{\rm even}$. Remarkable simplifications happen in this case and we find the Mellin amplitude only contains finitely many poles in these two series. More precisely, we find that the Mellin amplitude consists of simultaneous poles as well as single poles
\begin{equation}
\label{BBpbdymellinansatz}
\begin{split}
\mathcal{M}_{\widehat{\Delta}, \widehat{\Delta}^{\prime}}^{\partial}\left(\alpha_2, \alpha_3\right)={}&\sum_{m=0}^{2p-1} \sum_{n=0}^{2q-1} \frac{\mathcal{R}_{m,n}^{\partial,\widehat{\Delta},\widehat{\Delta}^{\prime}}}{\big(\alpha_2+\widehat{\Delta}-\Delta_2+m\big)\big(\alpha_3+\widehat{\Delta}^{\prime}-\Delta_3+n\big)}\\
{}&+ \sum_{m=0}^{2p-1} \frac{\mathcal{P}^{(p-1)}_m(\alpha_3)}{\alpha_2+\widehat{\Delta}-\Delta_2+m}+ \sum_{n=0}^{2q-1} \frac{\mathcal{P}^{(q-1)}_n(\alpha_2)}{\alpha_3+\widehat{\Delta}^{\prime}-\Delta_3+n}\;,
\end{split}
\end{equation}
where the simultaneous-pole coefficients $\mathcal{R}_{m,n}^{\partial,\widehat{\Delta},\widehat{\Delta}^{\prime}}$ are independent of the Mellin variables while the single-pole coefficients $\mathcal{P}^{(p-1)}_m(\alpha_3)$, $\mathcal{P}^{(q-1)}_n(\alpha_2)$ are degree $p-1$ and degree $q-1$ polynomials of $\alpha_3$ and $\alpha_2$ respectively. The general form of the single-pole coefficients is more difficult to write down, but the simultaneous-pole coefficients can be easily written down in a closed form
\begin{equation}
\begin{split}
\label{residueBBpbdy}
    \mathcal{R}_{m,n}^{\partial,\widehat{\Delta},\widehat{\Delta}^{\prime}}&=\sum_{i=0}^{\left\lfloor\frac{m}{2}\right\rfloor} \sum_{j=0}^{\left\lfloor\frac{n}{2}\right\rfloor}\Bigg[ S_{\widehat{\Delta},\Delta_2,i}S_{\widehat{\Delta}^{\prime},\Delta_3,j} \frac{(-1)^{m+n} \pi ^{h^{\prime}} 2^{-\widehat{\Delta} -\widehat{\Delta}^{\prime}+\Delta _2+\Delta _3-2 i-2 j-1}}{ (m-2 i)! (n-2 j)! } \times\\& \frac{ \Gamma \big(\frac{-m-n-\widehat{\Delta}-\widehat{\Delta}^{\prime}+\Delta _2+\Delta _3+1}{2}\big) \Gamma \big(\frac{-2 h^{\prime}+2 i+2
   j+\widehat{\Delta} +\widehat{\Delta}^{\prime}+\widehat{\Delta}_1}{2}\big)}{\Gamma (\widehat{\Delta}_1)\Gamma (2
   i+\widehat{\Delta}) \Gamma(2 j+\widehat{\Delta}^{\prime}) \Gamma (-m-\widehat{\Delta} +\Delta _2) \Gamma (-n-\widehat{\Delta}^{\prime}+\Delta _3) \Gamma \big(\frac{2i+2j-m-n+1}{2}\big)} \Bigg]\;.
   \end{split}
\end{equation}
 Note that in particular the Mellin amplitude has no poles from the $\Gamma(\frac{1+\alpha_2+\alpha_3}{2})$ factor which comes from $\mathcal{F}_{\widehat{\Delta}+2i,\widehat{\Delta}^{\prime}+2j}(\alpha_2,\alpha_3)$. However, in the general case we do not expect such poles to be absent as they can be cancelled by the zeros at these locations provided by the Gamma function in the denominator of the Mellin definition (\ref{3pointmellinrep}). This is similar to the situation of the boundary exchange diagram of the two-point function. Let us also note that the residues (\ref{residueBBpbdy}) are extracted from the last sum of (\ref{bdymellinvertexidentity}) and the truncation conditions were not used in obtaining them. Therefore, they also apply to the general case except that the upper limits of the summations in (\ref{BBpbdymellinansatz}) are extended to infinity. Inserting (\ref{BBpbdymellinansatz}) into the Mellin representation (\ref{3pointmellinrep}) and taking residues at these poles, we expect to reproduce the expansion for BB$\partial$ boundary conformal block (\ref{blockboundary}) up to an overall normalization. We checked this numerically for the first few orders.

\section{Discussions and outlook}\label{outlook}
In this paper, we performed a study of three-point functions in boundary CFTs where we either have two boundary and one bulk operators, or one boundary and two bulk operators. We computed the conformal blocks of them in different OPE channels, extending previous results in the literature. We also studied the Witten diagrams of such three-point functions in a simple holographic setup. In position space, we discussed in detail how to decompose them into conformal blocks. In Mellin space, we obtained simple expressions of these diagrams and used them to cross check our results for the conformal blocks. In the following we outline a few future directions. 
\begin{itemize}
    \item One of our future goals is to establish a functional approach to BCFT three-point functions that extends the two-point function case \cite{Paulos:2019gtx,Mazac:2019shk}. In the latter case, it was found that the double-trace conformal blocks in two different OPE channels give rise to a new basis for decomposing correlators. Roughly, this can be argued from the form of the conformal block decomposition of exchange Witten diagram, which allows us to express a generic conformal block as a linear combination of double-trace conformal blocks in both channels. The dual of these basis vectors is a basis of analytic functionals. Applying them to the exchange Witten diagrams, we find that their actions are simply related to the conformal block decomposition coefficients. We expect that a similar logic should apply to the BB$\partial$ three-point function. Moreover, the conformal block decomposition of the Witten diagrams studied in this paper will also be useful in developing the functional approach.   
\item Relatedly, it would be very interesting to explore the Lorentzian inversion formula and dispersion relation at the three-point function level which will be similar to the ones for four-point functions in CFTs without boundaries \cite{Caron-Huot:2017vep,Carmi:2019cub} and two-point functions in CFTs with co-dimension $q\geq 2$ defects \cite{Lemos:2017vnx,Liendo:2019jpu,Barrat:2022psm,Bianchi:2022ppi}. 
    \item In this paper, we have restricted ourselves to the simplest higher-point functions in BCFT. We have left out the more complicated case with three bulk operators which has three conformal cross ratios. It would be interesting to analyze this case in the future as it provides a stepping stone between four-point functions and five-point functions in CFTs without boundaries, where the number of cross ratios jumps from two to five. 
    We have also limited ourselves to correlators of scalar operators. The case with spinning correlators will have much richer kinematics. It would be useful to adapt the weight-shifting technology \cite{Karateev:2017jgd,Costa:2018mcg} to facilitate the analysis of conformal blocks and Witten diagrams.  
    \item Finally, it would be interesting to extend our results to include supersymmetry and consider top-down models constructed using probe D-branes ({\it e.g.}, in the setup of \cite{DeWolfe:2001pq}). One can then try to bootstrap correlators in these theories by following the strategy of \cite{Rastelli:2016nze,Rastelli:2017udc} using the various Witten diagrams studied in this paper as ingredients.  
\end{itemize}

\acknowledgments

We thank Ilija Buric for useful discussions. This work is supported by funds from University of Chinese Academy of Sciences (UCAS), funds from the Kavli Institute for Theoretical Sciences (KITS), the Fundamental Research Funds for the Central Universities, and the NSFC Grant No. 12275273. X.Z. also wishes to thank the Kavli Institute for Theoretical Physics (KITP) for hospitality during the workshop ``Bootstrapping Quantum Gravity'' where part of the work was done. The work in KITP was supported in part by the National Science Foundation under Grant No. NSF PHY-1748958.

\appendix

\section{BB$\partial$ boundary conformal partial wave}
\label{BBpartial boundary conformal partial wave}
The conformal partial wave (CPW) was first introduced for four-point function $\langle \phi_1 \phi_2 \phi_3 \phi_4 \rangle$ in CFTs without boundaries \cite{Dolan:2000ut,Simmons-Duffin:2012juh}. In $d$ dimensions it is defined as an integral gluing together two three-point structures\footnote{In this appendix we work in ordinary position space instead of the embedding space.}
\begin{equation}
	\label{4ptCPW}
	\Psi_{\Delta}^{\text{4pt}}(x_i)= \frac{1}{\mathcal{N}_{\Delta}} \langle\langle\phi_1(x_1) \phi_2(x_2) |\phi| \phi_3(x_3) \phi_4(x_4)\rangle\rangle\;,
\end{equation}
with a projector $|\phi|$ defined by
\begin{equation}
	|\phi| \equiv \int_{\mathbb{R}^d} d^d x d^d y|\phi(x)\rangle \frac{1}{(x-y)^{2(d-\Delta)}}\langle\phi(y)|\;.
\end{equation}
It is clear that (\ref{4ptCPW}) is conformal invariant. The prefactor $1/\mathcal{N}_{\Delta}$ is independent of spacetime coordinates and fixes the normalization of the CPW. As in  \cite{Simmons-Duffin:2012juh}, we can choose it in such a way that inserting the projector $|\phi|$ into the two-point function $\langle\phi \phi\rangle$ gives the identity transformation
\begin{equation}
	\label{4ptCPWnorm}
	\mathcal{N}_{\Delta}=\frac{\pi^d \Gamma(\Delta-\frac{d}{2}) \Gamma(\frac{d}{2}-\Delta)}{\Gamma(\Delta) \Gamma(d-\Delta)} \Rightarrow \langle \phi(x_1) |\phi| \phi(x_2) \rangle= \langle \phi(x_1)\phi(x_2) \rangle\;.
\end{equation}
We can further introduce the shadow operator $\widetilde{\phi}(x)$ of the operator $\phi(x)$ as
\begin{equation}
	\widetilde{\phi}(x)\equiv \int_{\mathbb{R}^d} d^d y \frac{1}{(x-y)^{2(d-\Delta)}} \phi(y)\;,
\end{equation}
and the operator $\widetilde{\phi}$ has conformal dimension $d-\Delta$.\footnote{We should distinguish between the shadow operator $\widetilde{\phi}$ and the local operator$\phi_{d-\Delta}$ with the shadow dimension  $d-\Delta$. As was emphasized in \cite{Simmons-Duffin:2012juh}, unlike $\phi_{d-\Delta}$ the operator $\widetilde{\phi}$ is nonlocal and therefore does not violate unitarity. However, $\phi_{d-\Delta}$ only appears in conformal invariant structures $\langle\langle \dots \rangle \rangle$ and will not appear in any physical theory.}  In terms of the shadow operator the projector $|\phi|$ can be rewritten as
\begin{equation}
	\label{shadowprojector}
	|\phi|\equiv\int_{\mathbb{R}^d} d^d x |\phi(x)\rangle \langle\widetilde{\phi}(x)|=\int_{\mathbb{R}^d} d^d x |\widetilde{\phi}(x)\rangle \langle\phi(x)|\;.
\end{equation}
The last identity indicates an exchange symmetry of the projector, which is clear from the definition. A crucial property following from the definition of CPW (\ref{4ptCPW}) is that it is an eigenfunction of the two-particle quadratic conformal Casimir of the conformal group $SO(d+1,1)$ \cite{Simmons-Duffin:2012juh}
\begin{equation}
	\label{Casimireqfor4ptCPW}
	\bigg(\frac{1}{2}(\mathbf{L}_1+\mathbf{L}_2)^2+\Delta(\Delta-d) \bigg) \Psi^{\text{4pt}}_{\Delta}(x_i)=0\;.
\end{equation}
Note that both the scalar conformal block $G_{\Delta,0}^{\text{4pt}}(x_i)$ and its shadow $G_{d-\Delta,0}^{\text{4pt}}(x_i)$  satisfy the same Casimir equation. The symmetry of the conformal integral (\ref{4ptCPW}) under $\Delta\leftrightarrow d-\Delta$ requires that the CPW is a linear combination of the conformal block and its shadow
\begin{equation}
	\Psi_{\Delta}^{\text{4pt}}(x_i)=\mathcal{K}\, G_{\Delta,0}^{\text{4pt}}(x_i)+\widetilde{\mathcal{K}}\, G_{d-\Delta,0}^{\text{4pt}}(x_i)\;,
\end{equation}
where the coefficients $\mathcal{K}$ and $\widetilde{\mathcal{K}}$ can be found in \cite{Simmons-Duffin:2017nub}. Note the two terms have different singular behaviors in the OPE limit $x_{12}^2\to 0$. This can be manifested by using the two orderings in (\ref{shadowprojector}). For example, if we choose the first ordering and take $\phi_1$ to approach $\phi_2$, the three-point function $\langle\langle\phi_1\phi_2\phi(x)\rangle\rangle$ in (\ref{4ptCPW}) selects only $\phi$ in the OPE. The gives rise to the singular behavior $|x_{12}|^{\Delta-\Delta_1-\Delta_2}$  in $\Psi_{\Delta}^{\text{4pt}}(x_i)$ which  corresponds to the conformal block $G_{\Delta,0}^{\text{4pt}}(x_i)$. We should emphasize that we see only one behavior but not the other is because of the subtlety of the integral. The OPE limit was taken only at the level of the integrand. In the integral the integrated point can probe regions close to $x_1$ and $x_2$, which will introduce new singularities. 

The same idea can be applied to correlation functions in BCFTs. We define the boundary channel CPW for BB$\partial$ three-point function $\langle\widehat{\mathcal{O}}_1 \mathcal{O}_2 \mathcal{O}_3\rangle$ as 
\begin{equation}
	\label{bdyCPW}
\Psi^{\partial}_{\widehat{\Delta},\widehat{\Delta}^{\prime}}(x_i)=	\frac{1}{\mathcal{N}_{\widehat{\mathcal{O}}}} \int_{\mathbb{R}^{d-1}} d^{d-1}\widehat{x}\,d^{d-1}\widehat{y} \,\langle\langle \widehat{\mathcal{O}}_1 \widehat{\mathcal{O}}(\widehat{x}) \widehat{\mathcal{O}}'(\widehat{y}) \rangle \rangle \langle\langle \widetilde{\widehat{\mathcal{O}}}(\widehat{x}) \mathcal{O}_2\rangle\rangle \langle\langle\widetilde{\widehat{\mathcal{O}}'}(\widehat{y}) \mathcal{O}_3\rangle\rangle\;,
\end{equation}
where $\widetilde{\widehat{\mathcal{O}}}(\widehat{x})$ and $\widetilde{\widehat{\mathcal{O}}'}(\widehat{y})$ are the shadow operators of $\widehat{\mathcal{O}}(\widehat{x})$ and $\widehat{\mathcal{O}}'(\widehat{y})$ respectively
\begin{equation}
	\label{bdyshadowoperator}
\widetilde{\widehat{\mathcal{O}}}(\widehat{x})=\int_{\mathbb{R}^{d-1}} d^{d-1} \widehat{z}\frac{1}{(\widehat{x}-\widehat{z})^{2\widetilde{\widehat{\Delta}}}}\widehat{\mathcal{O}}(\widehat{z})\;,\quad \widetilde{\widehat{\mathcal{O}}'}(\widehat{y})=\int_{\mathbb{R}^{d-1}} d^{d-1} \widehat{z}\frac{1}{(\widehat{y}-\widehat{z})^{2 \widetilde{\widehat{\Delta}^{\prime}}}}\widehat{\mathcal{O}}'(\widehat{z})\;,
\end{equation}
and the shadow dimension of a boundary operator with dimension $\widehat{\delta}$ is $\widetilde{\widehat{\delta}}\equiv d-1-\widehat{\delta}$. Equivalently, (\ref{bdyCPW}) can be interpreted as inserting two projectors $|\widehat{\mathcal{O}}|$ and $|\widehat{\mathcal{O}}'|$ into the three-point function $\langle\langle\widehat{\mathcal{O}}_1 \mathcal{O}_2 \mathcal{O}_3\rangle\rangle$. We will choose a different normalization
\begin{equation}
	\label{bdyCPWnorm}
\mathcal{N}_{\mathcal{\widehat{O}}}=\frac{\pi^{2h^{\prime}}\Gamma(\widehat{\Delta}-h^{\prime})\Gamma(\widehat{\Delta}'-h^{\prime})}{\Gamma(\widehat{\Delta})\Gamma(\widehat{\Delta}')}\;,
\end{equation}
which is determined by a different requirement compared to the one in (\ref{4ptCPWnorm}). We require that if we replace the shadow operators in (\ref{bdyCPW}) by the local operators with corresponding shadow dimensions the expression has unit coefficient
\begin{equation}
\label{bdycpw1}
\Psi^{\partial}_{\widehat{\Delta},\widehat{\Delta}^{\prime}}(x_i)= \int_{\mathbb{R}^{d-1}} d^{d-1}\widehat{x}\,d^{d-1}\widehat{y} \,\langle\langle \widehat{\mathcal{O}}_1 \widehat{\mathcal{O}}(\widehat{x}) \widehat{\mathcal{O}}'(\widehat{y}) \rangle \rangle \langle\langle \widehat{\mathcal{O}}_{\widetilde{\widehat{\Delta}}}(\widehat{x}) \mathcal{O}_2\rangle\rangle \langle\langle \widehat{\mathcal{O}}'_{\widetilde{\widehat{\Delta}^{\prime}}}(\widehat{y}) \mathcal{O}_3\rangle\rangle\;.
\end{equation}
Here we have used the conformal integral
\begin{equation}
    \langle\langle \widetilde{\widehat{\mathcal{O}}}(\widehat{x}) \mathcal{O}_2\rangle\rangle=\frac{\pi^{h^{\prime}}\Gamma(\widehat{\Delta}-h)}{\Gamma(\widehat{\Delta})}\langle\langle \widehat{\mathcal{O}}_{\widetilde{\widehat{\Delta}}}(\widehat{x}) \mathcal{O}_2\rangle\rangle\;,\quad \langle\langle\widetilde{\widehat{\mathcal{O}}'}(\widehat{y}) \mathcal{O}_3\rangle\rangle=\frac{\pi^{h^{\prime}}\Gamma(\widehat{\Delta}^{\prime}-h)}{\Gamma(\widehat{\Delta}^{\prime})} \langle\langle \widehat{\mathcal{O}}'_{\widetilde{\widehat{\Delta}^{\prime}}}(\widehat{y}) \mathcal{O}_3\rangle\rangle\;.
\end{equation}
From the definition (\ref{bdyCPW}), we find that the CPW $\Psi^{\partial}_{\widehat{\Delta},\widehat{\Delta}^{\prime}}(x_i)$ satisfies the same Casimir equations (\ref{bdyCasimir1}), (\ref{bdyCasimir2}) as the boundary conformal block $G^{\partial}_{\widehat{\Delta},\widehat{\Delta}^{\prime}}(x_i)$ 
\begin{eqnarray}
	&& \bigg(\frac{1}{2} \widehat{\mathbf{L}}_{2}^{2}+\widehat{\Delta}(\widehat{\Delta}-d+1)\bigg) \Psi_{\widehat{\Delta},\widehat{\Delta}^{\prime}}^{\partial}(x_i)=0\;, \\
	&& \bigg(\frac{1}{2} \widehat{\mathbf{L}}_{3}^{2}+\widehat{\Delta}^{\prime}(\widehat{\Delta}^{\prime}-d+1)\bigg) \Psi_{\widehat{\Delta},\widehat{\Delta}^{\prime}}^{\partial}(x_i)=0\;.
\end{eqnarray}
Similar to the four-point function case, the shadow symmetry also requires $\Psi^{\partial}_{\widehat{\Delta},\widehat{\Delta}^{\prime}}(x_i)$ to be a linear combination of the conformal block $G^{\partial}_{\widehat{\Delta},\widehat{\Delta}^{\prime}}(x_i)$ and all of its shadows
\begin{equation}
	\label{cpwboudrytoblock}
\Psi_{\widehat{\Delta},\widehat{\Delta}^{\prime}}^{\partial}(x_i)=\widetilde{\mathcal{K}}_{\widehat{\Delta},\widehat{\Delta}^{\prime}}^{\widehat{\Delta}_1} G_{\widehat{\Delta},\widehat{\Delta}^{\prime}}^{\partial}(x_i)+\mathcal{K}_{\widehat{\Delta},\widehat{\Delta}^{\prime}}^{\widehat{\Delta}_1} G_{\widetilde{\widehat{\Delta}},\widetilde{\widehat{\Delta}^{\prime}}}^{\partial}(x_i)+ \mathcal{S}_{\widehat{\Delta},\widehat{\Delta}^{\prime}}^{\widehat{\Delta}_1} G_{\widetilde{\widehat{\Delta}},\widehat{\Delta}^{\prime}}^{\partial}(x_i)+ \mathcal{S^{\prime}}_{\widehat{\Delta},\widehat{\Delta}^{\prime}}^{\widehat{\Delta}_1} G_{\widehat{\Delta},\widetilde{\widehat{\Delta}^{\prime}}}^{\partial}(x_i)\;. 
\end{equation}
In the following we compute these coefficients. We will exploit the same feature as in the four-point function example, namely different singularity behaviors can be singled out by using different orderings of the projector (\ref{shadowprojector}) in the definition (\ref{bdyCPW}).

The coefficient $\mathcal{K}^{\widehat{\Delta}_1}_{\widehat{\Delta},\widehat{\Delta}^{\prime}}$ can be obtained directly from the definition (\ref{bdycpw1}). It has the following singular behavior in boundary OPE limits of $\mathcal{O}_2$ and $\mathcal{O}_3$
\begin{equation}
\begin{aligned}
\label{asyconformalint}
\Psi^{\partial}_{\widehat{\Delta},\widehat{\Delta}^{\prime}}&(x_i) \\\to&\, x_{2\perp}^{\widetilde{\widehat{\Delta}}-\Delta_2} x_{3\perp}^{\widetilde{\widehat{\Delta}^{\prime}}-\Delta_3}\int_{\mathbb{R}^{d-1}} d^{d-1}\widehat{x}\,d^{d-1}\widehat{y} \,\langle\langle \widehat{\mathcal{O}}_1 \widehat{\mathcal{O}}(\widehat{x}) \widehat{\mathcal{O}}'(\widehat{y}) \rangle \rangle \langle \widehat{\mathcal{O}}_{\widetilde{\widehat{\Delta}}}(\widehat{x}) \widehat{\mathcal{O}}_{\widetilde{\widehat{\Delta}}}(\widehat{x}_2)\rangle\langle \widehat{\mathcal{O}}'_{\widetilde{\widehat{\Delta}^{\prime}}}(\widehat{y}) \widehat{\mathcal{O}}'_{\widetilde{\widehat{\Delta}'}}(\widehat{x}_3)\rangle\;.
\end{aligned}
\end{equation}
After evaluating the integral and comparing it with the asymptotic behavior of the boundary conformal block (\ref{asybdyblock}), we get $\mathcal{K}_{\Delta,\widehat{\Delta}^{\prime}}^{\widehat{\Delta}_1}$
\begin{equation}
\mathcal{K}_{\Delta,\widehat{\Delta}^{\prime}}^{\widehat{\Delta}_1}=\frac{\pi^{2 h^{\prime}} \Gamma(\widehat{\Delta}-h^{\prime})\Gamma(\widehat{\Delta}^{\prime}-h^{\prime})\Gamma\big(\frac{2h^{\prime}-\widehat{\Delta}-\widehat{\Delta}^{\prime}+\widehat{\Delta}_1}{2}\big)\Gamma\big(\frac{4h^{\prime}-\widehat{\Delta}-\widehat{\Delta}^{\prime}-\widehat{\Delta}_1}{2}\big)}{\Gamma(2h^{\prime}-\widehat{\Delta})\Gamma(2 h^{\prime}-\widehat{\Delta}^{\prime})\Gamma\big(\frac{\widehat{\Delta}+\widehat{\Delta}^{\prime}-\widehat{\Delta}_1}{2}\big)\Gamma\big(\frac{-2h^{\prime}+\widehat{\Delta}+\widehat{\Delta}^{\prime}+\widehat{\Delta}_1}{2}\big)}\;.
\end{equation}
To compute $\widetilde{\mathcal{K}}_{\widehat{\Delta},\widehat{\Delta}^{\prime}}^{\widehat{\Delta}_1}$, we should rewrite (\ref{bdyCPW}) as 
\begin{equation}
\Psi^{\partial}_{\widehat{\Delta},\widehat{\Delta}^{\prime}}(x_i)=	\frac{1}{\mathcal{N}_{\widehat{\mathcal{O}}}} \int_{\mathbb{R}^{d-1}} d^{d-1}\widehat{x}\,d^{d-1}\widehat{y} \,\langle\langle \widehat{\mathcal{O}}_1 \widetilde{\widehat{\mathcal{O}}}(\widehat{x}) \widetilde{\widehat{\mathcal{O}}'}(\widehat{y}) \rangle \rangle \langle\langle \widehat{\mathcal{O}}(\widehat{x}) \mathcal{O}_2\rangle\rangle \langle\langle\widehat{\mathcal{O}}'(\widehat{y}) \mathcal{O}_3\rangle\rangle\;,
\end{equation}
replacing the shadow operators by local operators introduces an overall factor
\begin{equation}
\label{bdyCPW2}
\Psi^{\partial}_{\widehat{\Delta},\widehat{\Delta}^{\prime}}(x_i)=\mathcal{A}_{\mathcal{\widehat{O}}} \int_{\mathbb{R}^{d-1}} d^{d-1}\widehat{x}\,d^{d-1}\widehat{y} \,\langle\langle \widehat{\mathcal{O}}_1 \widehat{\mathcal{O}}_{\widetilde{\widehat{\Delta}}}(\widehat{x}) \widehat{\mathcal{O}}'_{\widetilde{\widehat{\Delta}^{\prime}}}(\widehat{y}) \rangle \rangle \langle\langle \widehat{\mathcal{O}}(\widehat{x}) \mathcal{O}_2\rangle\rangle \langle\langle\widehat{\mathcal{O}}'(\widehat{y}) \mathcal{O}_3\rangle\rangle\;,
\end{equation}
where $\mathcal{A}_{\widehat{\mathcal{O}}}$ is computed by the conformal integral
\begin{equation}
\begin{aligned}
\mathcal{A}_{\widehat{\mathcal{O}}}&=\frac{1}{\mathcal{N}_{\widehat{\mathcal{O}}}\langle\langle \widehat{\mathcal{O}}_1 \widehat{\mathcal{O}}_{\widetilde{\widehat{\Delta}}}(\widehat{x}) \widehat{\mathcal{O}}'_{\widetilde{\widehat{\Delta}^{\prime}}}(\widehat{y}) \rangle \rangle }\int_{\mathbb{R}^{d-1}} d^{d-1}\widehat{w} d^{d-1} \widehat{z} \,\langle\langle \widehat{\mathcal{O}}_1 \widehat{\mathcal{O}}(\widehat{w}) \widehat{\mathcal{O}}'(\widehat{z}) \rangle \rangle (\widehat{w}-\widehat{x})^{-2\widetilde{\widehat{\Delta}}} (\widehat{z}-\widehat{y})^{-2\widetilde{\widehat{\Delta}^{\prime}}}\\&=\frac{\Gamma(\widehat{\Delta})\Gamma(\widehat{\Delta}^{\prime})\Gamma\big(\frac{2h^{\prime}-\widehat{\Delta}-\widehat{\Delta}^{\prime}+\widehat{\Delta}_1}{2}\big)\Gamma \big(\frac{4h^{\prime}-\widehat{\Delta}-\widehat{\Delta}^{\prime}-\widehat{\Delta}_1}{2}\big)}{ \Gamma(2h^{\prime}-\widehat{\Delta})\Gamma(2h^{\prime}-\widehat{\Delta}^{\prime})\Gamma\big(\frac{\widehat{\Delta}+\widehat{\Delta}^{\prime}-\widehat{\Delta}_1}{2}\big)\Gamma \big(\frac{-2h^{\prime}+\widehat{\Delta}+\widehat{\Delta}^{\prime}+\widehat{\Delta}_1}{2}\big)}\;.
\end{aligned}
\end{equation}
Then we can use the boundary OPE of $\mathcal{O}_2$ and $\mathcal{O}_3$ in (\ref{bdyCPW2}) and obtain 
\begin{equation}
\label{coeffkhat}
\widetilde{\mathcal{K}}_{\widehat{\Delta},\widehat{\Delta}^{\prime}}^{\widehat{\Delta}_1}=\frac{\pi^{2h^{\prime}}\Gamma(h^{\prime}-\widehat{\Delta})\Gamma(h^{\prime}-\widehat{\Delta}^{\prime})}{\Gamma(2h^{\prime}-\widehat{\Delta})\Gamma(2h^{\prime}-\widehat{\Delta}^{\prime})}\;.
\end{equation}
The remaining coefficients in (\ref{cpwboudrytoblock}) can be similarly obtained and read
\begin{eqnarray}
\nonumber&&\mathcal{S}_{\widehat{\Delta},\widehat{\Delta}^{\prime}}^{\widehat{\Delta}_1}= \frac{\pi^{2 h^{\prime}}\Gamma(\widehat{\Delta}-h^{\prime})\Gamma(h^{\prime}-\widehat{\Delta}^{\prime}) \Gamma\big(\frac{2h^{\prime}-\widehat{\Delta}-\widehat{\Delta}^{\prime}+\widehat{\Delta}_1}{2}\big) \Gamma\big(\frac{2h^{\prime}-\widehat{\Delta}+\widehat{\Delta}^{\prime}-\widehat{\Delta}_1}{2}\big)}{\Gamma(2h^{\prime}-\widehat{\Delta}) \Gamma(2h^{\prime}-\widehat{\Delta}^{\prime}) \Gamma\big(\frac{\widehat{\Delta}-\widehat{\Delta}^{\prime}+\widehat{\Delta}_1}{2}\big) \Gamma\big(\frac{\widehat{\Delta}+\widehat{\Delta}^{\prime}-\widehat{\Delta}_1}{2}\big)}\;,\\&&
\mathcal{S^{\prime}}_{\widehat{\Delta},\widehat{\Delta}^{\prime}}^{\widehat{\Delta}_1}= \mathcal{S}_{\widehat{\Delta},\widehat{\Delta}^{\prime}}^{\widehat{\Delta}_1} \Big|_{\widehat{\Delta} \leftrightarrow \widehat{\Delta}^{\prime}}\;.
\end{eqnarray}

\section{Spectral representation of exchange Witten diagrams}
\subsection{B$\partial\partial$ exchange diagram}
\label{appendixspectral}
\begin{figure}
    \centering
\includegraphics[width=0.71\linewidth]{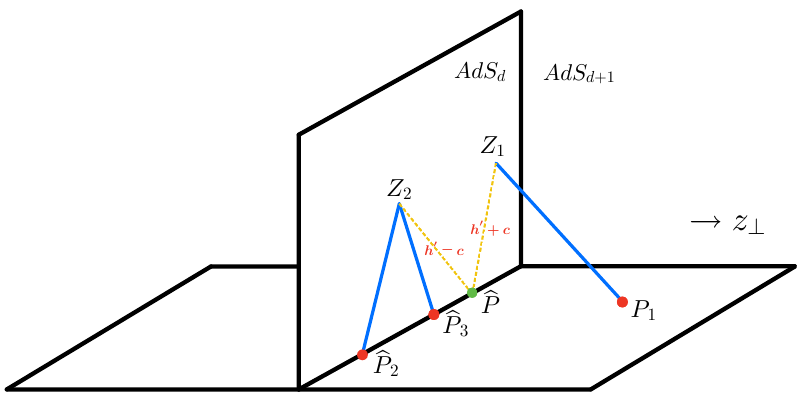}
    \caption{The split representation for the B$\partial\partial$ exchange Witten diagram. The yellow dash lines represent the auxiliary bulk-to-boundary propagators in $AdS_d$ with conformal dimensions $h^{\prime}\pm c$.}
    \label{Bppsplitfig}
\end{figure}
In this subsection we compute the conformal block decomposition of B$\partial\partial$ exchange Witten diagram using the method of spectral representation \cite{Penedones:2010ue,Costa:2014kfa}. The key formula is the split representation of the AdS bulk-to-bulk propagator
\begin{equation}
	\begin{aligned}
 \label{splitprop}
&G_{B B, (d)}^{\widehat{\Delta}}\left(Z_1,Z_2\right)=\int_{-i \infty}^{i \infty}\frac{d c}{2\pi i} \frac{\rho_{h^{\prime}}(c)}{(\widehat{\Delta}-h^{\prime})^{2}-c^{2}} \int D^{d-1}\widehat{P}\,G_{B\partial}^{h^{\prime}+c}(\widehat{P},Z_1)G_{B\partial}^{h^{\prime}-c}(\widehat{P},Z_2)\;,
\end{aligned}
\end{equation}
where
\begin{equation}
   \rho_{h^{\prime}}(c)\equiv \frac{\Gamma(h^{\prime}+c) \Gamma(h^{\prime}-c)}{2 \pi^{2 h^{\prime}} \Gamma(c) \Gamma(-c)}\;.
\end{equation}
Inserting it into the definition (\ref{exch}), we get the following expression pictorially represented in Fig. \ref{Bppsplitfig}
\begin{equation}
\begin{aligned}
\label{exch2}
&W_{\widehat{\Delta}}^{\text{B}\partial\partial,\text{exchange}}(P_1, \widehat{P}_{2}, \widehat{P}_{3})\\&=\int_{-i \infty}^{i \infty}\frac{d c}{2\pi i} \bigg[\frac{\rho_{h^{\prime}}(c)}{(\widehat{\Delta}-h^{\prime})^{2}-c^{2}} \int D^{d-1} \widehat{P}\langle \mathcal{O}_{\Delta_{1}}(P_1) \mathcal{\widehat{O}}_{h^{\prime}+c}(\widehat{P})\rangle_{\text {c}}\langle \mathcal{\widehat{O}}_{\widehat{\Delta}_{2}}(\widehat{P}_2) \mathcal{\widehat{O}}_{\widehat{\Delta}_{3}}(\widehat{P}_3)\mathcal{\widehat{O}}_{h^{\prime}-c}(\widehat{P})\rangle_{\text {c}} \bigg]\;,
\end{aligned}
\end{equation}
where inside the integrals we have a product of contact diagrams
\begin{eqnarray}
\nonumber&&\langle \mathcal{O}_{\Delta_{1}}(P_1) \mathcal{\widehat{O}}_{h^{\prime}+c}(\widehat{P})\rangle_{\text {c}}\equiv\int_{A d S_{d}} d Z_1\,(-2 P_{1} \cdot Z_1)^{-\Delta_{1}}(-2 \widehat{P} \cdot Z_{1})^{-(h^{\prime}+c)}\;,\\\nonumber&&\langle \mathcal{\widehat{O}}_{\widehat{\Delta}_{2}}(\widehat{P}_2) \mathcal{\widehat{O}}_{\widehat{\Delta}_{3}}(\widehat{P}_3)\mathcal{\widehat{O}}_{h^{\prime}-c}(\widehat{P})\rangle_{\text {c}}\equiv\int_{A d S_{d}} d Z_{2}\,(-2 \widehat{P}_{2} \cdot Z_2)^{-\widehat{\Delta}_{2}}(-2 \widehat{P}_{3} \cdot Z_2)^{-\widehat{\Delta}_{3}}(-2 \widehat{P} \cdot Z_{2})^{-(h^{\prime}-c)}\;.
\end{eqnarray}
These contact diagrams must be proportional to the BCFT two-point and three-point conformal invariant structures
\begin{eqnarray}
\nonumber\langle \mathcal{\widehat{O}}_{\widehat{\Delta}_{2}}(\widehat{P}_2) \mathcal{\widehat{O}}_{\widehat{\Delta}_{3}}(\widehat{P}_3)\mathcal{\widehat{O}}_{\widehat{\Delta}_i}(\widehat{P})\rangle_{\text {c}}&=&\frac{\pi ^{h^{\prime}} \Gamma (\widehat{\Delta}_{i2,3}) \Gamma(\widehat{\Delta}_{i3,2}) \Gamma(\widehat{\Delta}_{23,i}) \Gamma \big(\frac{\widehat{\Delta}_i+\widehat{\Delta}_2+\widehat{\Delta}_3-2h^{\prime}}{2} \big)}{2\, \Gamma (\widehat{\Delta}_i) \Gamma (\widehat{\Delta}_2) \Gamma (\widehat{\Delta} _3)}\langle\langle\widehat{O}_2 \widehat{O}_3 \widehat{O}_{\widehat{\Delta}_i}\rangle\rangle\;,\\\label{three}\\\label{two}
\langle \mathcal{O}_{\Delta_{1}}(P_1) \mathcal{\widehat{O}}_{\widehat{\Delta}}(\widehat{P})\rangle_{\text {c}}&=&\frac{\pi^{h^{\prime}} \Gamma\big(\frac{\Delta_{1}+\widehat{\Delta}-d+1}{2}\big) \Gamma\big(\frac{\Delta_{1}-\widehat{\Delta}}{2}\big)}{2\, \Gamma(\Delta_{1})}\langle\langle O_1\widehat{O}_{\widehat{\Delta}} \rangle\rangle\;.
\end{eqnarray}
The integral over $\widehat{P}$ can be performed and gives
\begin{equation}
\begin{aligned}
\label{Feyn}
\int& D^{d-1} \widehat{P} (-2 \widehat{P}\cdot P_{1})^{-(h^{\prime}+c)}(-2\widehat{P}\cdot \widehat{P}_2)^{\frac{-(h^{\prime}-c)-\widehat{\Delta}_2+\widehat{\Delta}_3}{2}}(-2\widehat{P}\cdot \widehat{P}_3)^{\frac{-(h^{\prime}-c)+\widehat{\Delta}_2-\widehat{\Delta}_3}{2}}=\\&\mathbf{A}\times
\int_{-i\infty}^{+i\infty} \frac{d\tau}{2\pi i}\xi^{\tau}\Gamma(-\tau)\Gamma(c-\tau)\Gamma\bigg(\frac{h^{\prime}-c+\widehat{\Delta}_2-\widehat{\Delta}_3}{2} + \tau \bigg) \Gamma\bigg(\frac{h^{\prime}-c-\widehat{\Delta}_2+\widehat{\Delta}_3}{2}+\tau\bigg) \;,
\end{aligned}
\end{equation}
where
\begin{equation}
\begin{gathered}
\mathbf{A}=\frac{\pi^{h^{\prime}}}{\Gamma(h^{\prime}+c)\Gamma\big(\frac{h^{\prime}-c+\widehat{\Delta}_2-\widehat{\Delta}_3}{2}\big)\Gamma\big(\frac{h^{\prime}-c-\widehat{\Delta}_2+\widehat{\Delta}_3}{2}\big)}\frac{(P_1\cdot N_b)^{-2c}}{(-2P_1\cdot \widehat{P}_{2})^{\frac{h^{\prime}-c+\widehat{\Delta}_2-\widehat{\Delta}_3}{2}}(-2P_1\cdot \widehat{P}_{3})^{\frac{h^{\prime}-c-\widehat{\Delta}_2+\widehat{\Delta}_3}{2}}}\;.
\end{gathered}
\end{equation}
Putting these ingredients together in (\ref{exch2}) and evaluating the $\tau$ integral, we get the following conformal partial wave decomposition as an integral over the unitary principal series of the boundary conformal group $SO(d,1)$
	\begin{equation}
	\begin{aligned}
&\mathcal{W}_{\widehat{\Delta}}^{\text{B}\partial\partial,\text{exchange}}(\xi)=\frac{\pi ^{h^{\prime}}}{8\, \Gamma(\Delta_{1}) \Gamma (\widehat{\Delta}_2) \Gamma(\widehat{\Delta} _3)}\\&\quad\times
	\int_{-i \infty}^{i \infty} \frac{d c}{2\pi i} \Bigg[\frac{ \Gamma\big(\frac{-h^{\prime}-c+\Delta_{1}}{2}\big)\Gamma\big(\frac{-h^{\prime}+c+\Delta_{1}}{2}\big) \Gamma \big(\frac{-h^{\prime}+c
			+\widehat{\Delta}_2+\widehat{\Delta}_3}{2} \big) \Gamma \big(\frac{-h^{\prime}-c +\widehat{\Delta}_2+\widehat{\Delta}_3}{2} \big)}{ \Gamma(c) \Gamma(-c)\left((\widehat{\Delta}-h^{\prime})^{2}-c^{2}\right)}\\&\quad\times\left(\Gamma (c) \Gamma \bigg(\frac{h'-c+\widehat{\Delta}_2-\widehat{\Delta}_3}{2} \bigg)\Gamma \bigg(\frac{h'-c-\widehat{\Delta}_2+\widehat{\Delta}_3}{2} \bigg)  g_{h^{\prime}-c}(\xi)+\left(c\leftrightarrow-c\right)\right) \Bigg]\;.
	\end{aligned}
	\end{equation}
Using the shadow symmetry $c\leftrightarrow-c$, we can rewrite it in terms of the $g_{h^{\prime}+c}(\xi)$ conformal blocks
\begin{equation}
\label{cpw}
\mathcal{W}_{\widehat{\Delta}}^{\text{B}\partial\partial,\text{exchange}}(\xi)=
\frac{\pi ^{h^{\prime}}}{4\,\Gamma(\Delta_{1}) \Gamma (\widehat{\Delta} _2) \Gamma(\widehat{\Delta} _3)}\int_{-i \infty}^{i \infty} \frac{d c}{2\pi i}\,F(c)\,g_{h^{\prime}+c}(\xi)\;,
\end{equation}
where
	\begin{equation}
	\begin{aligned}
&	F(c)=\\&\frac{ \Gamma\big(\frac{-h^{\prime}-c+\Delta_{1}}{2}\big)\Gamma\big(\frac{-h^{\prime}+c+\Delta_{1}}{2}\big) \Gamma \big(\frac{-h^{\prime}+c
			+\widehat{\Delta}_2+\widehat{\Delta}_3}{2} \big) \Gamma \big(\frac{-h^{\prime}-c +\widehat{\Delta}_2+\widehat{\Delta}_3}{2} \big)\Gamma \big(\frac{h'+c+\widehat{\Delta} _2-\widehat{\Delta}_3}{2} \big)\Gamma \big(\frac{h'+c-\widehat{\Delta}_2+\widehat{\Delta}_3}{2} \big)}{ \Gamma(c) \left((\widehat{\Delta}-h^{\prime})^{2}-c^{2}\right)}\;.
		\end{aligned}
	\end{equation}
We then close contour to the right. The poles in (\ref{cpw}) give the conformal dimension of conformal blocks and the residues give the OPE coefficients
\begin{equation}
\mathcal{W}_{\widehat{\Delta}}^{\text{B}\partial\partial,\text{exchange}}(\xi)=a^{\text{B}\partial\partial} g_{\widehat{\Delta}}(\xi)+\sum_{n=0}^{\infty}a_n^{\text{B}\partial\partial}\, g_{\Delta_1+2n}(\xi)+\sum_{n=0}^{\infty}b_n^{\text{B}\partial\partial}\, g_{\widehat{\Delta}_2+\widehat{\Delta}_3+2n}(\xi)\;,
\end{equation}
where the OPE coefficients are
\begin{equation}
 \begin{split}
	\label{coeffexchange1}
	&a^{\text{B}\partial\partial}=\\&\frac{\pi^{h^{\prime}}\Gamma \big(\frac{-\widehat{\Delta}+\Delta _1}{2}\big)\Gamma \big(\frac{-\widehat{\Delta} +\widehat{\Delta}_2+\widehat{\Delta}_3}{2}\big) \Gamma \big(\frac{\widehat{\Delta} +\widehat{\Delta}_2-\widehat{\Delta}_3}{2}
		\big) \Gamma \big(\frac{\widehat{\Delta} -\widehat{\Delta}
			_2+\widehat{\Delta} _3}{2} \big) \Gamma \big(\frac{-d+1+\widehat{\Delta} +\Delta _1}{2}\big) \Gamma \big(\frac{-d+1+\widehat{\Delta}+\widehat{\Delta}_2+\widehat{\Delta} _3}{2}
		\big)}{8\,\Gamma(\Delta_{1}) \Gamma (\widehat{\Delta} _2) \Gamma(\widehat{\Delta}_3) \Gamma \big(\frac{3-d}{2}+\widehat{\Delta}
		\big)}\;,
  \end{split}
  \end{equation}
\begin{equation}
 \begin{split}
	\label{coeffexchange2}
&a_n^{\text{B}\partial\partial}=\\&\frac{(-1)^n \pi^{h^{\prime}}\Gamma (n+\Delta_{12,3}) \Gamma
		(n+\Delta_{13,2}) \Gamma (-n+\Delta_{23,1}) \Gamma (-h^{\prime}+n+\Delta _1) \Gamma
		\big(-h^{\prime}+n+\frac{\sum_i\Delta_i}{2}\big)}{2n!\prod_i \Gamma(\Delta_i)\Gamma(-h^{\prime}+2 n+\Delta _1)\left(\widehat{\Delta}(\widehat{\Delta}-d+1)-(\Delta_1+2n)(\Delta_1+2n-d+1)\right)}\;,
  \end{split}
  \end{equation}
  \begin{equation}
 \begin{split}
	\label{coeffexchange3}
	&b_n^{\text{B}\partial\partial}=\frac{(-1)^n \pi^{h^{\prime}}}{2n!\prod_i \Gamma(\Delta_i)}\\&\times\frac{ \Gamma(n+\widehat{\Delta}_2)\Gamma (n+\widehat{\Delta}_3) \Gamma(-n-\Delta_{23,1})  \Gamma (-h^{\prime}+n+\widehat{\Delta}
		_2+\widehat{\Delta} _3) \Gamma \big(-h^{\prime}+n+\frac{\sum_i \Delta_i}{2} \big)}{\Gamma (-h^{\prime}+2 n+\widehat{\Delta} _2+\widehat{\Delta} _3)\left(\widehat{\Delta}(\widehat{\Delta}-d+1)-(\widehat{\Delta}_2+\widehat{\Delta}_3+2n)(\widehat{\Delta}_2+\widehat{\Delta}_3+2n-d+1)\right)}\;.
\end{split}
  \end{equation}
The above result for the double-trace OPE coefficients coincide exactly with (\ref{Bppexchdouble}).

\subsection{BB$\partial$ boundary exchange diagram}
\label{appendixspectral1}
\begin{figure}
    \centering
    \includegraphics[width=0.71\linewidth]{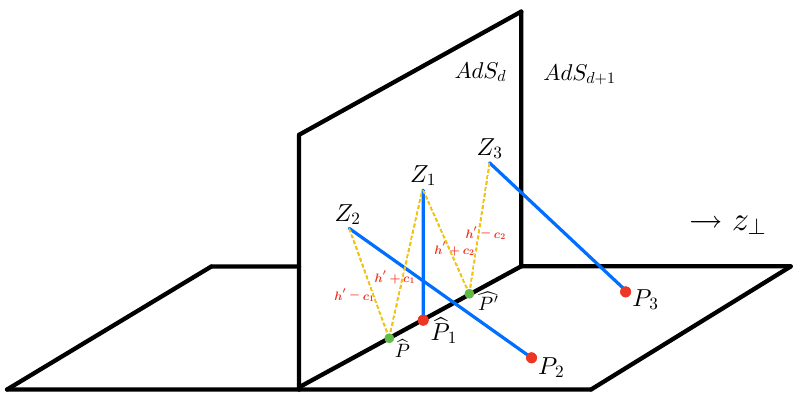}
    \caption{The split representation for the BB$\partial$ exchange Witten diagram. The split representation is applied to both bulk-to-bulk propagators.}
    \label{BBpbdysplitfig}
\end{figure}
From the BB$\partial$ boundary exchange Witten diagram (\ref{defbdyexchdiag}), we can use the split representation for both $AdS_{d}$ bulk-to-bulk propagators. This gives a two-fold spectral integral with spectral parameters $c_1$, $c_2$ and integrals over the $\mathbb{R}^{d-1}$ points $\widehat{P}$, $\widehat{P'}$ (Fig. \ref{BBpbdysplitfig})
\begin{equation}
\begin{split}
&W^{\partial}_{\widehat{\Delta},\widehat{\Delta}^{\prime}}(\widehat{P}_1,P_2,P_3)=\int_{-i \infty}^{i \infty} \int_{-i \infty}^{i \infty} \frac{d c_1 dc_2}{(2 \pi i)^2} \bigg[\frac{\rho_{h^{\prime}}(c_1) }{(\widehat{\Delta}-h^{\prime})^{2}-c_1^{2}} \frac{\rho_{h^{\prime}}(c_2) }{(\widehat{\Delta}^{\prime}-h^{\prime})^{2}-c_2^{2}} \int D^{d-1}\widehat{P}D^{d-1} \widehat{P^{\prime}}\\&\quad\quad\langle \mathcal{\widehat{O}}_{\Delta_{1}}(\widehat{P}_1) \mathcal{\widehat{O}}_{h^{\prime}+c_1}(\widehat{P})\mathcal{\widehat{O}}_{h^{\prime}+c_2}(\widehat{P^{\prime}})\rangle_{\text{c}} \langle \mathcal{O}_{\Delta_{2}}(P_2) \mathcal{\widehat{O}}_{h^{\prime}-c_1}(\widehat{P})\rangle_{\text{c}} \langle \mathcal{O}_{\Delta_{3}}(P_3) \mathcal{\widehat{O}}_{h^{\prime}-c_2}(\widehat{P^{\prime}})\rangle_{\text{c}}\bigg]\;.
\end{split}
\end{equation}
The contact diagrams in the integrand have already been computed in (\ref{three}) and (\ref{two}) in the last subsection. Recall the definition of boundary CPW (\ref{bdyCPW}), this gives automatically the CPW expansion of the boundary exchange diagram. Further using the symmetry of $c_1\to-c_1$, $c_2\to-c_2$, we can rewrite the exchange Witten diagram as 
\begin{equation}
\begin{aligned}
\label{cpwexpofbdyexch}
W^{\partial}_{\widehat{\Delta},\widehat{\Delta}^{\prime}}&=\frac{\pi ^{ h^{\prime}}}{8\, \Gamma (\widehat{\Delta}_1 ) \Gamma(\Delta_2)\Gamma (\Delta_3 )} \int_{-i \infty}^{i \infty} \int_{-i \infty}^{i \infty} \frac{d c_1 dc_2}{(2 \pi i)^2}\bigg[\frac{1}{(\widehat{\Delta}-h^{\prime})^{2}-c_1^{2}}\frac{1}{(\widehat{\Delta}^{\prime}-h^{\prime})^{2}-c_2^{2}}\frac{1}{\Gamma(c_1)\Gamma(c_2)}\\& \times\Gamma \bigg(\frac{c_1-c_2+\widehat{\Delta}_1}{2}\bigg) \Gamma \bigg(\frac{-c_1+c_2+\widehat{\Delta}_1}{2}
\bigg) \Gamma \bigg(\frac{c_1+c_2+\widehat{\Delta}
	_1}{2} \bigg)  \Gamma \bigg(\frac{2 h^{\prime}+c_1+c_2-\widehat{\Delta}_1}{2}
\bigg)\\&\times\Gamma \bigg(\frac{-h^{\prime}-c_1+\Delta_2}{2} \bigg) \Gamma \bigg(\frac{-h^{\prime}+c_1+\Delta _2}{2} \bigg) \Gamma
\bigg(\frac{-h^{\prime}-c_2+\Delta _3}{2} \bigg) \Gamma \bigg(\frac{-h^{\prime}+c_2+\Delta _3}{2}
\bigg)
\\&\times G_{h^{\prime}+c_1,h^{\prime}+c_2}^{\partial}\bigg]\;.
\end{aligned}
\end{equation}
To write it as a decomposition of conformal blocks, we should deform the $c_1$, $c_2$ contours to the right and sum over all the residues. If we pick up the poles of $c_1$, $c_2$  from Gamma functions in the middle and last lines, we obtain the conformal blocks of $W^{\partial}_{\widehat{\Delta},\widehat{\Delta}^{\prime}}$ that appear already in the boundary channel decomposition of contact Witten diagram (\ref{Boundarydecomofcontact}). The result agrees with the one obtained from the EOM relation in Section \ref{Boundary exchange Witten diagram}.

If at least one pole of $c_1$, $c_2$ is picked up from the factor $1/\big(\big((\widehat{\Delta}-h^{\prime})^2-c_1^2\big)\big((\widehat{\Delta}^{\prime}-h^{\prime})^2-c_2^2\big)\big)$ in the first line, we will find new conformal blocks that do not appear in contact diagram. For bulk exchange diagrams, this part only contains a single conformal block $g_{\Delta,J}^{\text{B}}$ dual to the exchanged field (we have seen this for scalar diagrams in Section \ref{Witten diagrams: The BBp case} and \ref{Mellin space representation}). But for boundary exchange diagrams it contains more conformal blocks in addition to the conformal block $g_{\widehat{\Delta},\widehat{\Delta}^{\prime}}^{\partial}$. They have the form
\begin{equation}
    g_{\widehat{\Delta},\widehat{\Delta}+\widehat{\Delta}_1+2m}^{\partial}\;,\quad g_{\widehat{\Delta},\Delta_3+2m}^{\partial}\;,\quad g_{\widehat{\Delta}^{\prime}+\widehat{\Delta}_1+2m,\widehat{\Delta}^{\prime}}^{\partial}\;,\quad g_{\Delta_2+2m,\widehat{\Delta}^{\prime}}^{\partial}\;,
\end{equation}
with $m=0,1,2,\dots$. The associated OPE coefficients of these conformal blocks can be read off from the residues at the corresponding poles in (\ref{cpwexpofbdyexch}) and we have
\begin{itemize}
    \item $g_{\widehat{\Delta},\widehat{\Delta}^{\prime}}^{\partial}$
\begin{equation}
    \begin{split}
    \label{ap}
       &a^{\partial}= \pi^{h^{\prime}} \Gamma\bigg(\frac{\Delta_2-\widehat{\Delta}}{2}\bigg)\Gamma\bigg(\frac{\Delta_3-\widehat{\Delta}^{\prime}}{2}\bigg)\times\\& \frac{ \Gamma\big(\frac{\widehat{\Delta}+\Delta_2-2h^{\prime}}{2}\big) \Gamma\big(\frac{\widehat{\Delta}^{\prime}+\Delta_3-2h^{\prime}}{2}\big)\Gamma\big(\frac{\widehat{\Delta}-\widehat{\Delta}^{\prime}+\widehat{\Delta}_1}{2}\big)\Gamma\big(\frac{\widehat{\Delta}+\widehat{\Delta}^{\prime}-
				\widehat{\Delta}_1}{2}\big)\Gamma\big(\frac{-\widehat{\Delta}+\widehat{\Delta}^{\prime}+\widehat{\Delta}_1}{2}\big) \Gamma\big(\frac{\widehat{\Delta}+\widehat{\Delta}^{\prime}+\widehat{\Delta}_1-2h^{\prime}}{2}\big)}{32\, \Gamma(\widehat{\Delta}_1)\Gamma(\Delta_2)\Gamma(\Delta_3) \Gamma(1-h^{\prime}+\widehat{\Delta})\Gamma(1-h^{\prime}+\widehat{\Delta}^{\prime})}\;,
    \end{split}
\end{equation}
    \item $g_{\widehat{\Delta},\widehat{\Delta}+\widehat{\Delta}_1+2m}^{\partial}$
\begin{equation}
    \begin{split}
    \label{bmp}
       &b_{m}^{\partial}= \frac{(-1)^{1+m} \pi^{h^{\prime}}\Gamma\big(\frac{\Delta_2-\widehat{\Delta}}{2}\big)\Gamma\big(\frac{\widehat{\Delta}+\Delta_2-2h^{\prime}}{2}\big)}{(2m+\widehat{\Delta}-\widehat{\Delta}^{\prime}+\widehat{\Delta}_1)(-2h^{\prime}+2m+\widehat{\Delta}+\widehat{\Delta}^{\prime}+\widehat{\Delta}_1)}\times\\& \frac{\Gamma(\widehat{\Delta}+m)\Gamma(\widehat{\Delta}_1+m)\Gamma(-h^{\prime}+m+\widehat{\Delta}+\widehat{\Delta}_1)\Gamma\big(\frac{-2m-\widehat{\Delta}-\widehat{\Delta}_1+\Delta_3}{2}\big)\Gamma\big(\frac{2m+\widehat{\Delta}+\widehat{\Delta}_1+\Delta_3-2h^{\prime}}{2}\big)}{8m!\Gamma(\widehat{\Delta}_1)\Gamma(\Delta_2)\Gamma(\Delta_3)\Gamma(1-h^{\prime}+\widehat{\Delta})\Gamma(-h^{\prime}+2m+\widehat{\Delta}+\widehat{\Delta}_1)}\;,
    \end{split}
\end{equation}
\item $g_{\widehat{\Delta},\Delta_3+2m}^{\partial}$
\begin{equation}
    \begin{split}
    \label{cmp}
       &c_m^{\partial}=\frac{(-1)^{1+m} \pi^{h^{\prime}} \Gamma\big(\frac{\Delta_2-\widehat{\Delta}}{2}\big)\Gamma\big(\frac{\widehat{\Delta}+\Delta_2-2h^{\prime}}{2}\big)\Gamma(m+\Delta_3-h^{\prime})}{(2m+\Delta_3-\widehat{\Delta}^{\prime})(-2h^{\prime}+2m+\Delta_3+\widehat{\Delta}^{\prime})}\\&\times\frac{ \Gamma\big(\frac{-2m+\widehat{\Delta}+\widehat{\Delta}_1-\Delta_3}{2}\big)\Gamma\big(\frac{2m+\widehat{\Delta}-\widehat{\Delta}_1+\Delta_3}{2}\big)\Gamma\big(\frac{2m-\widehat{\Delta}+\widehat{\Delta}_1+\Delta_3}{2}\big)\Gamma\big(\frac{2m-2h^{\prime}+\widehat{\Delta}+\widehat{\Delta}_1+\Delta_3}{2}\big)}{8m!\Gamma(\widehat{\Delta}_1)\Gamma(\Delta_2)\Gamma(\Delta_3)\Gamma(1-h^{\prime}+\widehat{\Delta})\Gamma(-h^{\prime}+2m+\Delta_3)}\;,
    \end{split}
\end{equation}
\item
$g_{\widehat{\Delta}^{\prime}+\widehat{\Delta}_1+2m,\widehat{\Delta}^{\prime}}^{\partial}$ 
\begin{equation}
\label{00}
 b^{\prime\partial}_m=b^{\partial}_m \big|_{\widehat{\Delta} \leftrightarrow \widehat{\Delta}^{\prime}, \Delta_2 \leftrightarrow \Delta_3}\;,
\end{equation}
\item $g_{\Delta_2+2m,\widehat{\Delta}^{\prime}}^{\partial}$
\begin{equation}
\begin{aligned}
\label{01}
 c^{\prime\partial}_m=c^{\partial}_m \big|_{\widehat{\Delta} \leftrightarrow \widehat{\Delta}^{\prime}, \Delta_2 \leftrightarrow \Delta_3}\;.
\end{aligned}
\end{equation}
\end{itemize}

\section{Integrated vertex identity for vector exchange}
\label{Integrated vertex identity for vector exchange}
In Section \ref{Mellin space representation} we defined the integral $A^{\widehat{\mu}}(w,x_2,x_3)$
\begin{equation}
	\label{Aintegral}
	A^{\widehat{\mu}}(w,x_2,x_3)=\int_{A d S_{d+1}} \frac{d^{d+1} z}{z_0^{d+1}} G_{BB,(d+1)}^{\Delta,\widehat{\mu} \nu}(w, z)\left( G_{B \partial}^{\Delta_{\mathcal{O}}}(x_{2}, z) \stackrel{\leftrightarrow}{\nabla}_{\nu} G_{B \partial}^{\Delta_{\mathcal{O}}}(x_{3}, z)\right)\;.
\end{equation}
We extend the method of \cite{DHoker:1999mqo} to calculate the integral for general conformal dimensions. The initial steps are identical and are briefly reviewed below. To simplify (\ref{Aintegral}), we can perform a translation followed by a conformal inversion
\begin{equation}
	x_2 \rightarrow 0\;, \quad x_3 \rightarrow x_{32} \equiv x_3-x_2\;,
\end{equation}
\begin{equation}
	x_{23}^{\prime}=\frac{x_{23}}{(x_{23})^2}\;, \quad z^{\prime}=\frac{z}{z^2}\;, \quad w^{\prime}=\frac{w}{w^2} \;.
\end{equation}
The integral $A^{\widehat{\mu}}(w,x_2,x_3)$ becomes  \cite{Freedman:1998tz,DHoker:1999mqo}
\begin{equation}
	A^{\widehat{\mu}}(w, x_2, x_3)=(x_{23})^{-2 \Delta_{\mathcal{O}}} \frac{1}{w^2} J^{\widehat{\mu}}{}_{\nu}(w) I^\nu(w^{\prime}-x_{23}^{\prime})\;,
\end{equation}
where
\begin{equation}
	J^{\widehat{\mu}}{}_{\nu}(w)=\delta^{\widehat{\mu}}{}_{\nu}-2 \frac{w^{\widehat{\mu}} w_{\nu}}{w^2}\; ,
\end{equation}
and
\begin{equation}
	I^\mu(w)=\int \frac{d^{d+1} z}{z_0^{d+1}} G^{\Delta,\mu\nu}_{BB,(d+1)}(w, z) z_0^{\Delta_{\mathcal{O}}} \stackrel{\leftrightarrow}{\nabla}_{\nu}\left(\frac{z_0}{z^2}\right)^{\Delta_{\mathcal{O}}}\;.
\end{equation}
Scaling symmetry and Poincar\'e symmetry require $I^\mu(w)$ to have the form \cite{DHoker:1999mqo} 
\begin{equation}
	I^\mu(w)=\frac{w^\mu}{w^2} f(t)\;,
\end{equation}
where 
\begin{equation}
	t= \frac{w_0^2}{w^2}\;.
\end{equation}
Using the equation of motion of the vector field propagator $G^{\Delta,\mu\nu}_{BB,(d+1)}(w, z)$, we obtain a differential equation for $f(t)$
\begin{equation}
	\label{diffeq}
	4 t^2(t-1) f^{\prime \prime}(t)+4 t\left(2 t+\frac{d-4}{2}\right) f^{\prime}(t)+M^2 f(t)=-2 \Delta_{\mathcal{O}} t^{\Delta_{\mathcal{O}}}\;,
\end{equation}where $M^2=(\Delta-1)(\Delta-d+1)$. Moreover, $f(t)$ should satisfy the following two conditions
\begin{enumerate}
    \item $f(t)$ has the behavior of $f(t) \sim t^{\frac{\Delta-1}{2}}$ as $t \rightarrow 0$, which follows from the OPE.
    \item $f(t)$ is smooth at $t=1$, which follows from its integral definition \cite{DHoker:1998ecp}.
\end{enumerate}
The differential equation (\ref{diffeq}) has a special solution
\begin{equation}
f_0(t)=C_0 \, t^{\Delta_{\mathcal{O}}} {}_3 F_{2} \bigg(1,\Delta_{\mathcal{O}},\Delta_{\mathcal{O}}+1;\frac{3-\Delta}{2}+\Delta_{\mathcal{O}}, \frac{3-d+\Delta}{2}+\Delta_{\mathcal{O}};t\bigg)\;,
\end{equation}
where
\begin{equation}
	C_0=\frac{2 \Delta_{\mathcal{O}}}{(1-\Delta+2\Delta_{\mathcal{O}})(1-d+\Delta+2\Delta_{\mathcal{O}})}\;.
\end{equation}
But it does not satisfy the boundary conditions. On the other hand, considering the linear combination with one of the  homogeneous solutions 
 \begin{equation}
f_1(t)=t^{\frac{\Delta-1}{2}} {}_2F_1\bigg(\frac{\Delta-1}{2},\frac{\Delta+1}{2};1-\frac{d}{2}+\Delta;t\bigg)\;,
 \end{equation}
it is easy to see both conditions are satisfied by
\begin{equation}
	\label{ftotal}
	f_{\text{total}}(t)=f_0(t)+C_1 f_1(t)\;,
\end{equation}
where
\begin{equation}
C_1=	-\frac{\Gamma\big(\frac{\Delta-1}{2}\big) \Gamma\big(\frac{\Delta+1}{2}\big)\Gamma\big(\frac{1-\Delta}{2}+\Delta_{\mathcal{O}}\big)\Gamma\big(\frac{1-d+\Delta}{2}+\Delta_{\mathcal{O}}\big)}{2\Gamma(1-\frac{d}{2}+\Delta)\Gamma(\Delta_{\mathcal{O}})^2}\;.
\end{equation}
It is more convenient to write $f_{\text{total}}(t)$ as a series expansion
\begin{equation}
	\label{ftotalexpan}
	f_{\text{total}}(t)=t^{\Delta_{\mathcal{O}}} \sum_{k=0}^{\infty} \widetilde{P}_k t^k+ t^{\frac{\Delta-1}{2}} \sum_{k=0}^{\infty} \widetilde{Q}_k t^k\;,
\end{equation}
where
\begin{equation}
	\widetilde{P}_k=\frac{2(\Delta_{\mathcal{O}}+k)(\Delta_{\mathcal{O}})_k ^2 }{(1-\Delta+2\Delta_{\mathcal{O}})(1-d+\Delta+2\Delta_{\mathcal{O}})\left(\frac{3-\Delta}{2}+\Delta_{\mathcal{O}}\right)_k \left(\frac{3-d+\Delta}{2}+\Delta_{\mathcal{O}}\right)_k}\;,
\end{equation}
\begin{equation}
\widetilde{Q}_k=-\frac{\Gamma\big(\frac{\Delta-1}{2}+k\big)\Gamma\big(\frac{\Delta+1}{2}+k\big)\Gamma\big(\frac{1-\Delta}{2}+\Delta_{\mathcal{O}}\big)\Gamma\big(\frac{1-d+\Delta}{2}+\Delta_{\mathcal{O}}\big)}{2k! \Gamma(1-\frac{d}{2}+k+\Delta)\Gamma(\Delta_{\mathcal{O}})^2}\;.
\end{equation}
Undoing the inversion and translation, we find each power $t^a$ in (\ref{ftotalexpan}) contributes to $A^{\widehat{\mu}}(w,x_2,x_3)$ as a contact vertex 
\begin{equation}
	- \frac{(x_{23})^{2(a-\Delta_{\mathcal{O}})}}{2 a} g^{\widehat{\mu} \nu}(w)\left(G_{B \partial}^a(w, x_2) \stackrel{\leftrightarrow}{\nabla}_{\nu} G_{B \partial}^a (w, x_3)\right)\;.
\end{equation}
Therefore the bulk spin-$1$ exchange diagram $\left(\ref{spin1exch}\right)$ can be replaced by the following infinite sums of contact diagrams with derivative interactions
\begin{equation}\label{WvecexchangeinWcon2der}
	W^{\text{B}}_{\Delta,1}(x_i)=-\sum_{k=0}^{\infty}P_k(x_{23}^2)^{k} W^{\text{contact,2-der}}_{\widehat{\Delta}_1,\Delta_{\mathcal{O}}+k,\Delta_{\mathcal{O}}+k}- \sum_{k=0}^{\infty} Q_k  (x_{23}^2)^{\left(\frac{\Delta-1}{2}+k-\Delta_{\mathcal{O}}\right)} W^{\text{contact,2-der}}_{\widehat{\Delta}_1,\frac{\Delta-1}{2}+k,\frac{\Delta-1}{2}+k}\;,
\end{equation}
where $W^{\text{contact,2-der}}_{\widehat{\Delta}_1,\Delta_a,\Delta_b}$ are two-derivative contact Witten diagrams defined in (\ref{contact1derivative}) and the coefficients are given by
\begin{equation}
	\label{Pk}
	P_k=\frac{(\Delta_{\mathcal{O}})_k ^2 }{(1-\Delta+2\Delta_{\mathcal{O}})(1-d+\Delta+2\Delta_{\mathcal{O}})\left(\frac{3-\Delta}{2}+\Delta_{\mathcal{O}}\right)_k \left(\frac{3-d+\Delta}{2}+\Delta_{\mathcal{O}}\right)_k}\;,
\end{equation}
\begin{equation}
	\label{Qk}
	Q_k=-\frac{\Gamma\big(\frac{\Delta-1}{2}+k\big)^2\Gamma\big(\frac{1-\Delta}{2}+\Delta_{\mathcal{O}}\big)\Gamma\big(\frac{1-d+\Delta}{2}+\Delta_{\mathcal{O}}\big)}{4k! \Gamma(1-\frac{d}{2}+k+\Delta)\Gamma(\Delta_{\mathcal{O}})^2}\;.
\end{equation}
It is useful to note the identity
\begin{equation}
	\begin{aligned}
 \nabla_{\widehat{\mu}} G_{B \partial}^{\widehat{\Delta}_{1}}(\widehat{x}_{1}, w) \nabla^{\widehat{\mu}}G_{B\partial}^{\Delta_i}(x_i,w)=&\widehat{\Delta}_1 \Delta_i\left( G_{B \partial}^{\widehat{\Delta}_{1}}(\widehat{x}_{1}, w) G_{B\partial}^{\Delta_i}(x_i,w)\right.\\&\left.-2(\widehat{x}_1-x_i)^2 G_{B \partial}^{\widehat{\Delta}_{1}+1}(\widehat{x}_{1}, w) G_{B\partial}^{\Delta_i+1}(x_i,w)\right)\;.
 \end{aligned}
\end{equation}
The two-derivative contact Witten diagram can then be written in terms of the zero-derivative contact Witten diagrams
\begin{equation}
\label{spin1contactasDfunction}
W^{\text{contact,2-der}}_{\widehat{\Delta}_1,\Delta_{\mathcal{O}},\Delta_{\mathcal{O}}}=2\widehat{\Delta}_1 \Delta_{\mathcal{O}} \left((\widehat{x}_1-x_2)^2 W^{\text{contact}}_{\widehat{\Delta}_1+1,\Delta_{\mathcal{O}}+1,\Delta_{\mathcal{O}}}-(\widehat{x}_1-x_3)^2 W^{\text{contact}}_{\widehat{\Delta}_1+1,\Delta_{\mathcal{O}},\Delta_{\mathcal{O}}+1}\right)\;.
\end{equation}
Using it in (\ref{WvecexchangeinWcon2der}), we find the upshot is that the BB$\partial$ bulk channel spin-$1$ exchange Witten diagram can be written as infinite sums of contact Witten diagrams with shifted dimensions as in (\ref{spin1exchangeascontact}).

\bibliography{refs} 
\bibliographystyle{utphys}
\end{document}